\documentclass{article}
\usepackage{PRIMEarxiv}

\usepackage[utf8]{inputenc} 
\usepackage[T1]{fontenc}    

\usepackage{amsmath,amsfonts}
\usepackage{algorithmic}
\usepackage{algorithm}
\usepackage{array}
\usepackage[caption=false,font=normalsize,labelfont=sf,textfont=sf]{subfig}
\usepackage{textcomp}
\usepackage{stfloats}
\usepackage{url}
\usepackage{verbatim}
\usepackage{graphicx}
\usepackage{cite}

\usepackage{booktabs}
\usepackage{multicol}
\usepackage{multirow}
\usepackage{xcolor}
\usepackage{colortbl}
\usepackage{balance}
\usepackage{lscape}
\usepackage{float}
\usepackage[export]{adjustbox}
\usepackage{tikz}
\usepackage[colorlinks=true,linkcolor=green,citecolor=green,urlcolor=blue]{hyperref}
\usepackage[capitalise,noabbrev]{cleveref}

\begin{document}

\title{A Survey of Deep Learning-based Point Cloud Denoising}

\author{
    Jinxi~Wang,
    Ben~Fei,
    Dasith~de~Silva~Edirimuni,
    Zheng~Liu,
    Ying~He,
    and~Xuequan~Lu
\thanks{
J. Wang, D. de Silva Edirimuni and X. Lu are with the Department of Computer Science and Software Engineering, University of Western Australia, Crawley, WA, 6009, Australia. 
B. Fei is with Multimedia Lab, Department of Information Engineering, the Chinese University of Hong Kong, Hong Kong SAR. 
Z. Liu is with the School of Computer Science, China University of Geosciences (Wuhan), Wuhan 430079, China. 
Y. He is with the College of Computing and Data Science, Nanyang Technological University, 639798, Singapore.}%
}

\maketitle
\thispagestyle{empty}
\pagestyle{empty}

\begin{abstract}
Accurate 3D geometry acquisition is essential for a wide range of applications, such as computer graphics, autonomous driving, robotics, and augmented reality. However, raw point clouds acquired in real-world environments are often corrupted with noise due to various factors such as sensor, lighting, material, environment etc, which reduces geometric fidelity and degrades downstream performance. 
Point cloud denoising is a fundamental problem, aiming to recover clean point sets while preserving underlying structures. 
Classical optimization-based methods, guided by hand-crafted filters or geometric priors, have been extensively studied but struggle to handle diverse and complex noise patterns. Recent deep learning approaches leverage neural network architectures to learn distinctive representations and demonstrate strong outcomes, particularly on complex and large-scale point clouds.
Provided these significant advances, this survey provides a comprehensive and up-to-date review of deep learning-based point cloud denoising methods up to August 2025. We organize the literature from two perspectives: (1) supervision level (supervised vs. unsupervised), and (2) modeling perspective, proposing a functional taxonomy that unifies diverse approaches by their denoising principles. We further analyze architectural trends both structurally and chronologically, establish a unified benchmark with consistent training settings, and evaluate methods in terms of denoising quality, surface fidelity, point distribution, and computational efficiency. Finally, we discuss open challenges and outline directions for future research in this rapidly evolving field.
\end{abstract}

\section{Introduction}\label{sec:introduction}

Imagine scanning a finely crafted sculpture displayed in a well-lit exhibition hall using a handheld 3D device.
The sensor captures not only the delicate curves and surface patterns of the sculpture but also portions of the surrounding space, producing a dense point cloud rich in geometric detail yet inevitably imperfect.
Some points are displaced due to sensor jitter or surface reflectivity, while others are lost due to occlusion.
Similar challenges arise in scanning cultural artifacts, product prototypes, and both indoor and outdoor scenes
—domains where high accuracy is critical for applications such as high-fidelity 3D reconstruction~\cite{mildenhall2021nerf,kerbl20233d,barron2023zip}, virtual or augmented reality~\cite{xu2023vr,chen2023mobilenerf,wu20244d}, autonomous navigation~\cite{bescos2018dynaslam, prakash2021multi, hu2023planning} and robotics~\cite{sundermeyer2021contact,xu2022fast,ke20243d}.

\begin{figure}[t]
    \centering
    \includegraphics[width=0.6\linewidth]{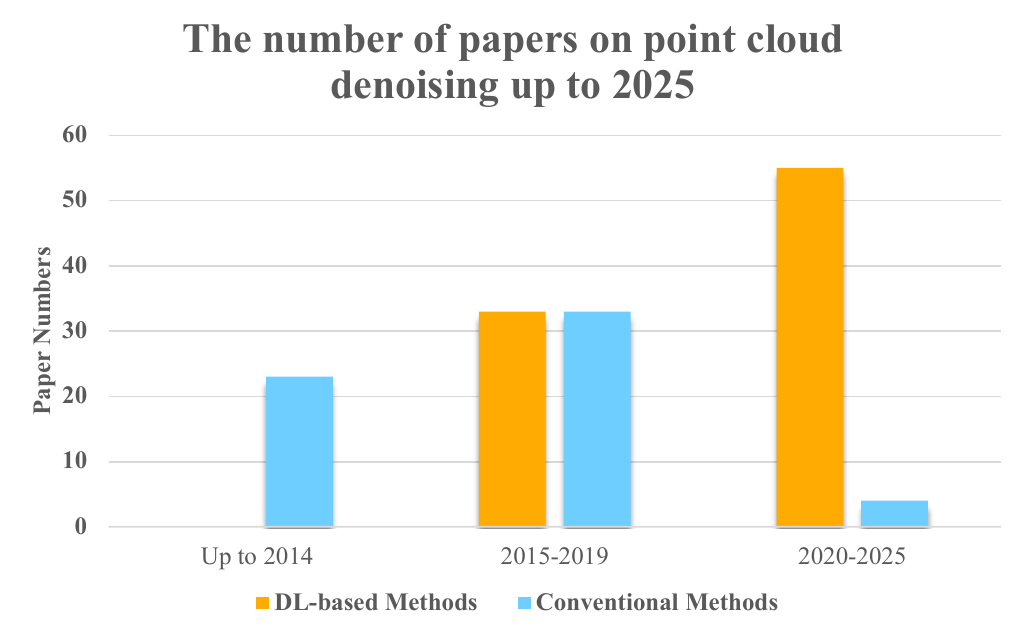}
    \caption{Number of publications on point cloud denoising up to August 2025. The results show that deep learning-based methods have rapidly overtaken conventional approaches and now dominate the field.}

    \label{fig:paper-number}
\end{figure}

This is the challenge addressed by point cloud denoising, a fundamental problem which aims to restore a noise-free point cloud while preserving the underlying geometric structures.
Over the past two decades, a rich body of point cloud denoising techniques has emerged.
Early progress was primarily driven by traditional approaches, which laid the foundations for the field through explicit geometric modeling. 
Representative examples include Moving Least Squares (MLS)~\cite{alexa2003computing_mls,fleishman2005robust_rmls,oztireli2009feature_rimls,xu2019anisotropic_mls},
Locally Optimal Projection (LOP)~\cite{lipman2007parameterization_lop,huang2009consolidation_wlop,huang2013edge_ear,preiner2014continuous_clop,liao2013efficient_flop}, and non-local filtering~\cite{zeng20193d_nonlocal,chen2019multi_nonlocal,lu2020low_nonlocal,zhou2021point_nonlocal,zhu2022nonlocal}. 
More recently, deep learning has revolutionized this field, enabling data-driven models to learn complex geometric priors directly from raw points.
As shown in Fig.~\ref{fig:paper-number}, the number of learning-based methods has surged since 2015, and today they account for the vast majority of state-of-the-art methods. 
This rapid growth reflects not only the strength of deep learning in capturing geometric structures but also the increasing diversity of problem formulations and architectural designs.

Deep learning-based denoising methods distinguish themselves from traditional counterparts in several aspects.
First, they are capable of learning complex geometric priors directly from raw points, rather than relying on hand-crafted assumptions such as local smoothness or explicit surface fitting.
Second, the design space of these methods is considerably diverse: some approaches treat denoising as a pointwise regression task, others as a probabilistic inference problem, while more recent works leverage generative models or diffusion processes to capture higher-order structures.
Third, the rapid evolution of architectures—from early PointNet-style encoders to graph neural networks and transformers—demonstrates a shift toward increasingly expressive and scalable frameworks.

Building on these advances, only a few surveys have contributed to consolidating knowledge in point cloud denoising.
For example, Han et al.~\cite{han2017review} provide a thorough review of classical techniques such as MLS, bilateral filtering, and projection-based approaches, while Zhou et al.~\cite{zhou2022point} summarize early learning-based pipelines and categorize them alongside traditional methods.
Nevertheless, existing taxonomies often rely on network architectural distinctions, which—while useful for tracing the evolution of networks—can obscure the fact that methods with similar architectures may adopt fundamentally different problem formulations.

These observations highlight the need for a dedicated and systematic analysis of learning-based point cloud denoising. 
To this end, we present the first survey exclusively focused on deep learning-based approaches, introducing a functional taxonomy grounded in problem modeling, an unified benchmarking and an in-depth analysis of both performance and computational efficiency.
Our contributions are threefold:
\begin{itemize}
    \item \textbf{Comprehensive and structured review.} 
    We systematically analyze existing learning-based denoising approaches, organizing them according to their underlying problem formulations and architectural designs.
    \item \textbf{Principled functional taxonomy.} 
    Unlike earlier surveys that mainly organize deep learning-based methods by network architectures or input scales, we propose a functional taxonomy grounded in modeling strategies, as these strategies are the essential factor distinguishing different approaches.
    \item \textbf{Unified benchmark and insights.} 
    We retrain representative methods under consistent settings and evaluate both denoising performance and computational efficiency, highlighting current challenges and future research opportunities, including the potential for large-scale visual foundation models.
\end{itemize}

The structure of this survey is organized as follows: 
Section~\ref{sec:Background} introduces the background of point cloud denoising.
Sections~\ref{sec:supervised} and~\ref{sec:unsupervised} review supervised and unsupervised methods, respectively.
Section~\ref{sec:benchmark} provides a unified experimental benchmark and ablation studies for performance and efficiency analysis.
Finally, Section~\ref{sec:Discussion} discusses open challenges and future directions.
\section{Background}
\label{sec:Background}

\subsection{Common Noise Types}
Point cloud corruption generally falls into three categories: synthetic perturbations, outliers, and real-scan artifacts.
Synthetic noise, commonly used in controlled evaluations, includes isotropic and anisotropic Gaussian noise, Laplacian noise with heavy-tailed deviations, Uniform-ball noise with spherically bounded displacements in arbitrary directions, and discrete perturbations that impose irregular displacements on a small subset of points.
Outliers, such as impulse noise or randomly scattered background points, represent spurious measurements unrelated to the true surface and typically require explicit removal.  
Point clouds captured from real sensors (e.g., Kinect, LiDAR) exhibit complex, mixed noise, including spatial jitter, occlusion-induced gaps, and non-uniform sampling densities, posing significant challenges to real-world generalization.

\subsection{Problem Formulation}
Point cloud denoising aims to recover a clean and geometrically faithful point cloud from a set of noisy 3D measurements.
Let
\begin{equation}
\mathcal{P} = \{\pmb{p}_i = (x_i, y_i, z_i)\}_{i=1}^n \subset \mathbb{R}^3,
\end{equation}
denote the observed point cloud, which is a noisy sampling of some underlying surface.
The goal is to estimate a denoised version: 
\begin{equation}
\mathcal{\hat{P}} = \{\pmb{\hat{p}}_i\}_{i=1}^n,
\end{equation}
that better reflects the true surface geometry while removing noise and outliers.

Traditionally, this task has been formulated as an optimization problem that minimizes point-wise deviation from the input while enforcing geometric priors such as smoothness or manifold consistency.
However, these hand-crafted priors often struggle to generalize to complex surfaces or unseen noise, and typically require carefully tuned parameters to perform well across diverse scenarios.

In deep learning-based methods, point cloud denoising is typically cast as a regression problem, where a neural network \( f_\theta \) with learnable parameters \( \theta \) predicts denoised positions from noisy inputs:
\begin{equation}
    \mathcal{\hat{P}} = f_\theta(\mathcal{P}).
\end{equation}
When paired ground-truth clean point clouds \( P^c \) are available, the network is trained in a supervised manner by minimizing a discrepancy loss:
\begin{equation}
    \mathcal{L}_{\text{supervised}}(\theta) = \ell(\mathcal{\hat{P}}, \mathcal{P}^c), 
\end{equation}
where \( \ell(\cdot, \cdot) \) is commonly an \(\ell_2\) distance, Chamfer Distance, or Earth Mover’s Distance (EMD).

Beyond directly predicting coordinates, some methods learn per-point displacement vectors, latent implicit fields, or surface-aware corrections, depending on the architectural design and task formulation. The supervision strategies also vary, ranging from fully supervised settings to weakly supervised and unsupervised paradigms.

\begin{figure*}[htbp]
    \centering

        \includegraphics[width=\linewidth]{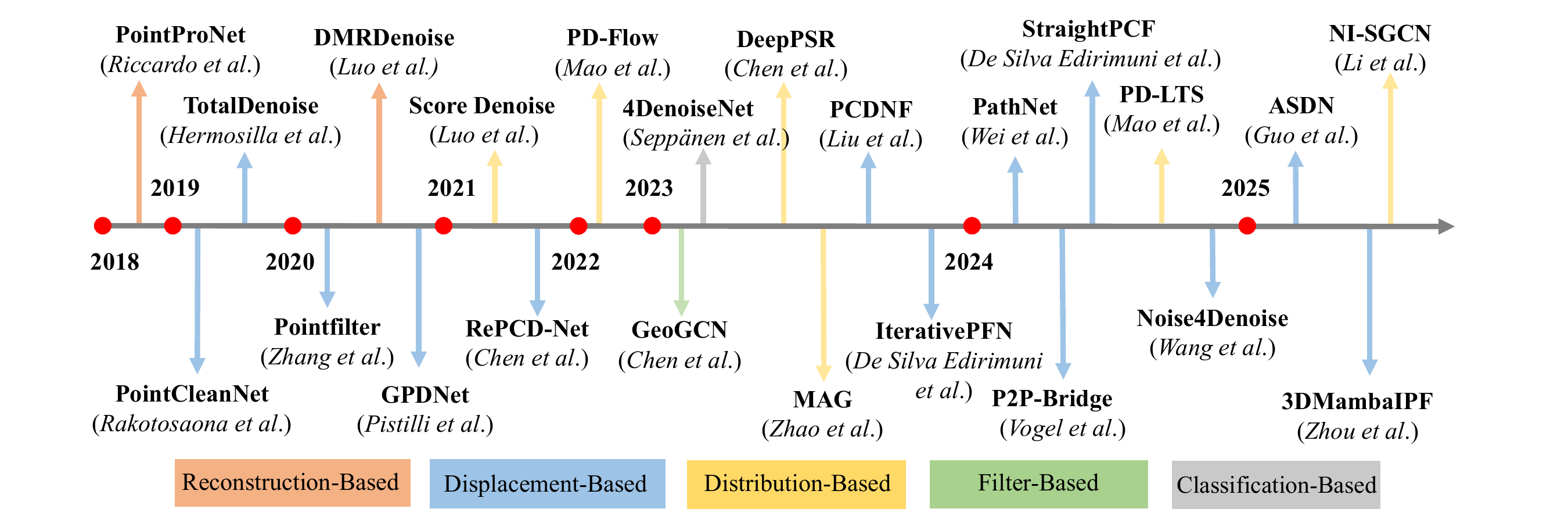}
        \caption{Timeline of representative learning-based point cloud denoising methods from 2018 to 2025. Methods are categorized based on their underlying principles, including reconstruction-based, displacement-based, distribution-based, filter-based, and classification-based approaches.}
    
    \label{fig:timeline}
\end{figure*}

In the following sections, we review representative deep learning-based denoising methods, starting with those that rely on paired supervision during training.
Fig.~\ref{fig:timeline} presents the timeline of representative learning-based point cloud denoising methods.

\begin{table*}[!t]
\caption{Overview of representative deep learning-based point cloud denoising methods arranged chronologically. Each entry is categorized by its underlying denoising principle and network architecture, with a brief highlight of the key innovation that sets it apart from prior work. Abbreviations: \textbf{Reconst.} = Reconstruction-based, \textbf{Displ.} = Displacement-based, \textbf{Distrib.} = Distribution-based, \textbf{Filter} = Filter-based, \textbf{Classif.} = Classification-based.}
\centering
\scriptsize   
\setlength{\tabcolsep}{2.5pt}
\renewcommand{\arraystretch}{1.3} 
\begin{tabular}{
    >{\centering\arraybackslash}p{1.3cm}|  
    >{\centering\arraybackslash}p{1.5cm}|  
    >{\centering\arraybackslash}m{0.7cm}|  
    >{\centering\arraybackslash}m{2cm}|    
    >{\raggedright\arraybackslash}m{8cm}   
}

\hline
\textbf{Principle} & \textbf{Architecture} & \textbf{Year} & \textbf{Method} & \textbf{Highlights} \\
\hline
\multicolumn{1}{c|}{}                                       & \multicolumn{1}{c|}{Convolution-based}                 & \multicolumn{1}{c|}{2018} & PointProNet~\cite{Roveri2018PointProNet}   & First converts 3D to 2D and leverages CNN for denoising. \\ \cline{2-5}
\multicolumn{1}{c|}{}                                       & \multicolumn{1}{c|}{Graph-based}                       & \multicolumn{1}{c|}{2020} & DMRDenoise~\cite{luo2020differentiable}    & First to use downsampling-upsampling strategy for denoising. \\ \cline{2-5}
\multicolumn{1}{c|}{\multirow{-4}{*}{Reconst.}}             & \multicolumn{1}{c|}{Transformer-based}                 & \multicolumn{1}{c|}{2022} & TDNet~\cite{xu2022tdnet}                   & First to employ a transformer-based architecture for point cloud denoising, leveraging global self-attention to capture long-range dependencies and enhance reconstructive accuracy. \\ \hline

\multicolumn{1}{c|}{}                                       & \multicolumn{1}{c|}{}                                  & \multicolumn{1}{c|}{2019} & PointCleanNet~\cite{rakotosaona2020pointcleannet} & First to use PointNet-like architecture for denoising. \\
\multicolumn{1}{c|}{}                                       & \multicolumn{1}{c|}{}                                  & \multicolumn{1}{c|}{2020} & Pointfilter~\cite{zhang2020pointfilter}   & First to propose bilateral projection loss and sharp edge preservation. \\
\multicolumn{1}{c|}{}                                       & \multicolumn{1}{c|}{}                                  & \multicolumn{1}{c|}{2022} & MODNet~\cite{huang2022modnet}             & First to propose multi-offset learning on multi-scale patches for structure-aware denoising. \\
\multicolumn{1}{c|}{}                                       & \multicolumn{1}{c|}{\multirow{-4}{*}{Point MLP-based}} & \multicolumn{1}{c|}{2024} & Pathnet~\cite{wei2024pathnet}             & First to introduce a path-selective attention mechanism that adaptively aggregates multi-scale neighborhood features for enhanced denoising performance. \\ \cline{2-5}
\multicolumn{1}{c|}{}                                       & \multicolumn{1}{c|}{}                                  & \multicolumn{1}{c|}{2020} & GPDNet~\cite{pistilli2020learning}        & First graph convolutional network for point cloud denoising. \\
\multicolumn{1}{c|}{}                                       & \multicolumn{1}{c|}{}                                  & \multicolumn{1}{c|}{2023} & IterativePFN~\cite{de2023iterativepfn}    & First to utilize iteratively refined ground truth during training, enabling progressive learning of denoising. \\
\multicolumn{1}{c|}{}                                       & \multicolumn{1}{c|}{}                                  & \multicolumn{1}{c|}{2024} & StraightPCF~\cite{de2024straightpcf}      & Achieves the most uniform point distribution among recent methods. \\
\multicolumn{1}{c|}{}                                       & \multicolumn{1}{c|}{}                                  & \multicolumn{1}{c|}{2025} & ASDN~\cite{guo2025you}                    & First to introduce an adaptive stopping mechanism that learns when to halt denoising dynamically. \\
\multicolumn{1}{c|}{}                                       & \multicolumn{1}{c|}{\multirow{-5}{*}{Graph-based}}     & \multicolumn{1}{c|}{2025} & 3DMambaIPF~\cite{zhou20253dmambaipf}      & First to introduce a Mamba-based state space model for iterative point cloud filtering. \\ \cline{2-5}
\multicolumn{1}{c|}{}                                       & \multicolumn{1}{c|}{RNN-based}                         & \multicolumn{1}{c|}{2021} & RePCDNet~\cite{chen2022repcd}             & First multi-scale feature learning using RNN modules. \\ \cline{2-5}
\multicolumn{1}{c|}{\multirow{-10}{*}{Displ.}}              & \multicolumn{1}{c|}{Transformer-based}                 & \multicolumn{1}{c|}{2023} & MSaD-Net~\cite{zhu2023msad}               & First to integrate multiple self-attention mechanisms into a unified network, enhancing feature interactions across mixed receptive fields for improved denoising performance. \\ \hline

\multicolumn{1}{c|}{}                                       & \multicolumn{1}{c|}{}                                  & \multicolumn{1}{c|}{2021} & ScoreDenoise~\cite{luo2021score}          & First to learn gradient field for denoising using score matching. \\
\multicolumn{1}{c|}{}                                       & \multicolumn{1}{c|}{}                                  & \multicolumn{1}{c|}{2023} & MAG~\cite{zhao2023mag}                    & First to introduce a momentum ascent strategy within gradient fields for progressive point cloud denoising. \\
\multicolumn{1}{c|}{}                                       & \multicolumn{1}{c|}{}                                  & \multicolumn{1}{c|}{2024} & P2P-Bridge~\cite{vogel2024p2p}            & First to formulate point cloud denoising as a conditional diffusion bridging process between noisy and clean point distributions, enabling continuous refinement with strong generative priors. \\
\multicolumn{1}{c|}{}                                       & \multicolumn{1}{c|}{\multirow{-4}{*}{Diffusion-based}} & \multicolumn{1}{c|}{2025} & NI-SGCN~\cite{li2025nisgcn}               & First to integrate spiking neural networks (SNNs) with graph convolution for energy-efficient denoising. \\
\cline{2-5}
\multicolumn{1}{c|}{}                                       & \multicolumn{1}{c|}{Point MLP-based}                   & \multicolumn{1}{c|}{2022} & SVCNet~\cite{zhao2022noise}               & First to learn self-variation in noise distribution and reverse the corruption process for denoising. \\
\cline{2-5}
\multicolumn{1}{c|}{}                                       & \multicolumn{1}{c|}{}                                  & \multicolumn{1}{c|}{2022} & PDFlow~\cite{mao2022pd}                   & First formulates point cloud denoising as a noise-disentanglement problem in a latent space modeled by invertible normalizing flows. \\
\multicolumn{1}{c|}{\multirow{-7}{*}{Distrib.}}             & \multicolumn{1}{c|}{\multirow{-2}{*}{Flow-based}}      & \multicolumn{1}{c|}{2024} & PD-LTS~\cite{mao2024denoising}            & Extends flow-based denoising by introducing a latent transformation strategy (LTS) to enhance noise disentanglement. \\ \hline

\multicolumn{1}{c|}{}                                       & \multicolumn{1}{c|}{Convolution-based}                 & \multicolumn{1}{c|}{2021} & DFPN~\cite{lu2020deep}                    & First to utilize a CNN-based network to predict guided normals for feature-aware denoising. \\ \cline{2-5}
\multicolumn{1}{c|}{\multirow{-2}{*}{Filter}}               & \multicolumn{1}{c|}{Graph-based}                       & \multicolumn{1}{c|}{2023} & GeoGCN~\cite{geogcn2023}                  & First dual-domain graph convolutional framework that jointly denoises point positions and surface normals. \\ \hline

\multicolumn{1}{c|}{}                                       & \multicolumn{1}{c|}{Convolution-based}                 & \multicolumn{1}{c|}{2020} & WeatherNet~\cite{heinzler2020cnn}         & First to apply CNN denoising on LiDAR data under adverse weather. \\ \cline{2-5}
\multicolumn{1}{c|}{}                                       & \multicolumn{1}{c|}{Transformer-based}                 & \multicolumn{1}{c|}{2022} & RFTNet~\cite{gao2022reflective}           & First to incorporate reflectivity cues and transformer backbone for filtering reflective noise in large-scale point clouds. \\ \cline{2-5}
\multicolumn{1}{c|}{\multirow{-3}{*}{Classif.}}             & \multicolumn{1}{c|}{Point MLP-based}                   & \multicolumn{1}{c|}{2023} & RocketPCD~\cite{zuo2023attention}         & First applies attention-based MLP network for denoising scanned rocket tank panels under real-world noise. \\ \hline
\end{tabular}
\label{tab:repre_methods}
\end{table*}

\begin{table}
\centering
\caption{Abbreviations used for denoising principle.}
\renewcommand{\arraystretch}{1.2}
\begin{tabular}{p{1cm} p{5.5cm}}
\toprule
\textbf{Abbr.} & \textbf{Full Term and Meaning} \\
\midrule
Reconst.  & Reconstruction-based: reconstruct clean geometry via projection or depth regression. \\
Displ.   & Displacement-based: regress per-point offsets to restore clean positions. \\
Distrib. & Distribution-based: model and refine the statistical distribution of clean points. \\
Filter   & Filter-based: update point positions via optimization, guided by geometric priors (e.g., normals) estimated by learning-based methods. \\
Classif. & Classification-based: detect and remove noisy or outlier points. \\
\bottomrule
\end{tabular}
\label{tab:abbr_explained}
\end{table}

\section{Supervised Learning}
\label{sec:supervised}
Supervised point cloud denoising methods learn from paired noisy–clean data to remove noise while preserving fine geometric structures.
Whereas previous works often categorize methods by network architecture~\cite{han2017review,zhou2022point}, such taxonomies may overlook that methods with similar backbones can embody fundamentally different problem formulations.
Since the modeling strategy dictates how noise is represented and removed, a strategy-oriented organization offers a more faithful view of the underlying principles and limitations of each approach.
Accordingly, we group existing works into five categories: reconstruction-, displacement-, distribution-, filter-, and classification-based approaches.
Representative methods from each category are listed in Table~\ref{tab:repre_methods}, while the definitions of these categories are summarized in Table~\ref{tab:abbr_explained}.

\subsection{Reconstruction-based Methods}
Reconstruction-based denoising methods do not regress displacement vectors or filter noise directly. Instead, they generate new 
point sets that represent a cleaner and more complete version of the surface, thereby reconstructing the underlying geometry.

The application of deep learning to point cloud denoising did not initially follow the same trajectory as general point cloud understanding. While PointNet~\cite{qi2017pointnet} pioneered direct learning on unstructured point sets for classification and segmentation, its architecture was not designed to capture fine-grained geometric details, which are crucial for tasks such as denoising and surface reconstruction. As a result, early denoising methods were more influenced by image processing pipelines. PointProNet~\cite{Roveri2018PointProNet} projects unordered points onto 2D height maps, allowing the reuse of CNN-based techniques and better preserving geometric details. 
Although this design enables the reuse of established 2D techniques, it also introduces geometric distortions due to the repeated transformations between 2D and 3D spaces.

Subsequent approaches shifted toward explicitly reconstructing the latent manifold of noisy point clouds, with Luo et al.~\cite{luo2020differentiable} introducing a framework that leverages differentially subsampled points with minimal perturbation and their neighborhood features.
The network follows an autoencoder-like architecture: the encoder extracts both local and non-local features for each point and performs adaptive point selection via a differentiable pooling operation, while the decoder reconstructs the manifold by transforming each sampled point and its neighborhood into a local surface patch. 
The denoised point cloud is then obtained by resampling from this manifold.
This method also supports unsupervised training, in addition to the standard supervised setting. However, the downsampling process may still compromise fine geometric structures, leading to oversmoothing.

Another representative example is SSPCN (Single-Stage Point Cloud Cleaning Network) proposed by Li and Sheng~\cite{li2023single}, which unifies outlier removal and denoising within a single model. The network also follows a downsampling–upsampling paradigm: it first selects a prefiltered subset using a density-based farthest point sampling (DBFPS) strategy, compensates noisy features, restores point density through upsampling, and finally reconstructs the cleaned point cloud from refined features. By integrating these steps into an end-to-end pipeline, SSPCN achieves robust denoising while maintaining uniformity in the reconstructed point set.

In parallel, transformer-based designs have also emerged.
Xu et al.~\cite{xu2022tdnet} introduced a transformer-based architecture named TDNet for point cloud denoising, designed to capture global dependencies and reconstruct clean point distributions from noisy inputs.
The network adopts an encoder–decoder structure inspired by NLP Transformers. The encoder treats each point as a token and extracts features through multi-head self-attention layers, allowing the model to capture global contextual relationships among points. In the decoder, TDNet reconstructs a denoised point cloud by learning to regress local surface patches around each sampled point. An adaptive sampling module is introduced to select surface-representative points dynamically, ensuring that the decoder focuses on meaningful regions and avoids outlier-dominated neighborhoods.

\subsection{Displacement-based Methods}
Displacement-based methods take a direct approach by explicitly estimating the noise component of each point. Specifically, the network learns a displacement vector for every noisy point, which is then used to shift the point back toward the underlying clean surface. This strategy allows for point-wise correction and is often guided by local geometric context extracted through neighborhood features. By regressing these displacement vectors, the network effectively denoises the point cloud while maintaining spatial coherence.

\textbf{Point-wise MLP Methods.}
To overcome the limitations of indirect 2D projection, later methods focus on directly learning denoising mappings in 3D space.
Qi et al.~\cite{qi2017pointnet} established a foundational framework by applying shared multi-layer perceptrons (MLPs) to process unordered point sets, treating each point independently before aggregating global features. Its hierarchical extension, PointNet++~\cite{qi2017pointnet++}, further captures local structures by applying PointNet recursively on neighborhoods of increasing granularity. 
While originally designed for high-level semantic tasks such as classification, these architectures also laid the foundation for low-level geometric processing tasks like denoising. 
PCPNet~\cite{Guerrero2018PCPNet} and ECNet~\cite{Yu2018ECNet} were among the first to adapt them for low-level geometric tasks, such as normal estimation and point cloud consolidation. By constructing local patches from neighboring points and focusing on geometric features like surface orientation and edge saliency, these methods demonstrated the feasibility of learning fine-grained structural cues from unstructured point sets. 
They provided valuable insights and design paradigms that later inspired point cloud denoising frameworks.

Building upon PCPNet, PointCleanNet~\cite{rakotosaona2020pointcleannet} presents one of the earliest frameworks that directly operates on raw point clouds specifically for denoising.
It directly operates in the 3D domain and formulates denoising as a two-stage process: first detecting and removing outliers, and then refining the remaining points through point-wise displacement prediction.
Zhang et al.~\cite{zhang2020pointfilter} introduced Pointfilter, an encoder-decoder architecture that processes local point neighborhoods and predicts displacement vectors to reconstruct clean point clouds. This method addresses a key challenge in learning-based point cloud denoising: how to preserve sharp geometric features. 
To this end, they introduce a bilateral projection-based loss function that leverages surface normal information during training,
effectively enhancing edge preservation. Notably, although the projection loss is defined with respect to surface normals during training, no normal information is required at inference time, making it more practical and widely applicable in real-world scenarios.

Huang et al.~\cite{huang2022modnet} proposed MODNet, a  
multi-scale point cloud denoising network designed to mitigate surface degradation issues such as residual noise and loss of geometric details. Unlike conventional approaches that naively aggregate multi-scale features, MODNet employs a multi-scale perception module to adaptively weight features based on local geometric complexity. The method consists of three key stages: (1) multi-scale patch feature extraction, (2) multi-scale weight regression to guide denoising, and (3) multi-offset displacement prediction for refined point adjustment.
De Silva Edirimuni et al.~\cite{desilva2023contrastive} proposed to jointly estimate surface normals and predict point-wise displacements for denoising, leveraging contrastive learning to improve feature robustness. The method first pretrains a patch-level feature encoder using contrastive learning, where noise corruption serves as the augmentation strategy. The pretrained features are then used to regress both normals and displacements in a unified network. To preserve fine structures and sharp edges under high noise levels, the method enforces geometric consistency between the predicted normals and displacement directions through a joint loss function. This is the first framework to integrate contrastive representation learning with joint normal estimation and geometry-guided displacement prediction, offering improved resilience against challenging noise conditions.

Wang et al.~\cite{wang2023fcnet} proposed FCNet, a framework that explicitly cleans feature noise during training through a teacher-student learning model and two key modules: non-local self-similarity (NSS) and weighted average pooling (WAP). The NSS module exploits the inherent non-local similarities in point clouds to smooth features, while WAP suppresses noise from outliers using statistical outlier removal.
Wei et al.~\cite{wei2024pathnet} introduced a reinforcement learning (RL)-based framework for point cloud denoising, which dynamically selects the most appropriate denoising path for each point. Unlike conventional models that apply a single denoising network to all points, PathNet models the per-point variability in noise levels and geometric complexity by learning a routing agent that assigns points to specialized denoising branches. The agent is trained via a noise- and geometry-aware reward function, encouraging better adaptation to multi-scale surface structures and varying noise intensities. The routing policy and denoising modules are trained jointly to avoid over- or under-smoothing.

To address computational overhead, Sheng and Li~\cite{sheng2024denoising} proposed LPCDNet, a lightweight network designed to address the trade-off between denoising efficacy and computational efficiency. Their method employs a trigonometric-based feature extraction module, a nonparametric feature aggregation module, and a decoder for surface realignment, enabling high-quality denoising with minimal parameters.

\textbf{Graph-based.}
Graph convolutional networks have become a core backbone for point cloud learning, with DGCNN~\cite{wang2019dynamic} introducing dynamic graph updates through EdgeConv layers. 
While effective for classification and segmentation, the DGCNN is less suited for denoising: the global spatial transformer adds unnecessary overhead, max-based aggregation is unstable under noise, and fixed weights reduce adaptability~\cite{pistilli2020learning}.
Building on these ideas, Pistilli et al.~\cite{pistilli2020learning} proposed GPDNet, which simultaneously denoises all points in the graph using residual blocks composed of graph convolutional layers. 
By progressively refining feature-based neighborhood graphs and integrating adaptive weighting with edge attention, GPDNet enhances local structure modeling and achieves improved robustness to noise.

Extending the effectiveness of graph-based residual refinement, subsequent works explored more explicit iterative strategies to further improve denoising efficiency and accuracy.
De Silva Edirimuni et al.~\cite{de2023iterativepfn} introduced IterativePFN that stacks multiple IterationModules, each designed to perform a single, learnable denoising step. By supervising each module with progressively cleaner targets, the network learns to iteratively filter point clouds in just a single forward pass. Patch-level directed graphs and graph convolutions enable geometry-aware displacement predictions. This integrated iterative learning leads to faster convergence and more accurate denoising compared to prior displacement‑based methods.

Liu et al.~\cite{liu2023pcdnf} introduced PCDNF, which formulates point cloud denoising as a multitask learning problem by jointly optimizing normal filtering and point position restoration. Instead of relying solely on direct denoising or filtering-based approaches, the method introduces an auxiliary normal filtering task to guide the denoising process.
It integrates two key modules: a shape-aware selector that constructs latent tangent spaces using point and normal features along with geometric priors, and a feature refinement module that fuses point and normal features to leverage their complementary strengths—point features for local detail and normal features for structural cues such as edges and corners. This design enables more accurate noise removal while preserving geometric features.
Such normal-guided denoising methods are particularly effective in preserving 3D geometric structures and fine-scale details. Although initial normals can be estimated from noisy point clouds using techniques such as PCA, they are often unreliable under high noise levels, which limits the effectiveness of normal-based guidance in such scenarios.

Recently,
De Silva Edirimuni et al.~\cite{de2024straightpcf} proposed StraightPCF, which models the denoising process as straight-line motion in 3D space. Unlike prior approaches that rely on iterative or stochastic refinement, StraightPCF directly learns a constant flow direction and distance for each point to recover the clean geometry in a single forward pass. Specifically, the network comprises two key components: a VelocityModule that predicts the flow direction, and a DistanceModule that estimates the magnitude needed to move each noisy point back to the surface. By enforcing straight, deterministic trajectories, the method avoids overshooting and mitigates discretization errors common in stepwise updates.
Notably, it achieves uniform point distributions without relying on any explicit regularization.

Chen et al.~\cite{chen2024progressive} introduced C2AENet, a progressive point cloud denoising framework designed to address structural degradation caused by noise. The method operates in multiple stages, where each stage refines point positions through a learned displacement. To enhance denoising consistency and effectiveness, the authors propose a Cross-Stage Cross-Coder architecture that integrates features both across stages and between encoder-decoder pairs. Additionally, to overcome the limitations of purely semantic graph construction in prior graph-based approaches, C2AENet incorporates a geometric graph convolution module with adaptive edge attention, enabling the network to capture both local and global contextual information.

Guo et al.~\cite{guo2025you} introduced the Adaptive Stop Denoising Network (ASDN).
This method addresses a common issue in traditional approaches: the tendency to over-process already clean points while insufficiently correcting noisy ones, which often leads to artifacts.
To overcome this problem, ASDN adopts a U-Net architecture complemented by an adaptive classifier constructed with EdgeConv layers from DGCNN. This classifier leverages a recoverability factor to evaluate the denoising progress of each point and make dynamic, point-wise decisions on when to halt the process. By adaptively controlling the denoising depth, ASDN effectively prevents over-smoothing and better preserves geometric details.

Building on the iterative refinement paradigm of IterativePFN~\cite{de2023iterativepfn}, Zhou et al.~\cite{zhou20253dmambaipf} proposed 3DMambaIPF, which incorporates Selective State Space Models (SSMs) to efficiently process dense and large-scale point clouds. The authors observed that existing methods struggle with scalability and denoising quality when processing large point clouds (e.g., 50K points), often generating noisy outliers post-denoising. 
To address this, 3DMambaIPF leverages the selective input handling and long-range dependency modeling of SSMs, enabling robust denoising of highly dense point clouds with up to 500K points. Moreover, a differentiable rendering loss is introduced to enforce surface consistency, thereby enhancing geometric fidelity and improving boundary alignment with real-world objects.

\textbf{RNN-based.}
RePCD-Net~\cite{chen2022repcd} adopts a recurrent architecture to predict point-wise displacements for point cloud denoising progressively. The network is composed of shared modules—including MLPs, attention-enhanced RNNs, bidirectional RNNs (BRNNs), and fully connected layers—that operate recurrently across multiple denoising stages. A key design is the recurrent feature propagation layer, which enables the reuse and refinement of deep geometric features across stages. Furthermore, the method introduces a feature-aware loss function that promotes the preservation of multi-scale geometric structures. By jointly learning feature dynamics and spatial regularity, RePCD-Net effectively balances noise removal and detail preservation.

\textbf{Transformer-based.}
Transformer architectures have been actively explored for point cloud denoising, with different methods emphasizing local-global feature integration, multi-scale modeling, or domain-specific robustness.
Zhu et al.~\cite{zhu2023msad} proposed MSaD-Net, a transformer-based, single-stage network that overcomes the limitations of prior PointNet-style offset prediction methods by incorporating a hybrid attention mechanism to better capture both local and global contextual information.
Specifically, it introduces:
(1) a lightweight GAAS module, which combines probabilistic sampling with k-NN grouping to extract meaningful local neighborhoods while reducing input redundancy;
(2) transformer layers for modeling long-range dependencies across local patches, and a channel-attention module to enhance inter-channel feature interactions; and
(3) a final offset regression stage, where an MLP predicts point-wise displacement vectors that are added to the noisy input to obtain denoised coordinates.

Concurrently, Wang and Li~\cite{wang2023transformer} proposed TPCDN network that integrates both local geometric context and global dependencies for enhanced denoising. The network is built upon three key modules: a local feature extraction module that adaptively encodes neighborhood geometry, a point cloud Transformer module that captures long-range spatial interactions, and a feature fusion module that seamlessly integrates local and global features. By framing denoising as a per-point displacement prediction task, TPCDN refines the positions of noisy points while preserving fine structural details.

Recently, Ran et al.~\cite{ran2025tpdnet} introduced TPDNet tailored for offshore drilling platforms, which combines local feature aggregation and Transformer-style self-attention to identify and correct noise. The network first applies a feature abstraction module to extract multi-scale local features from k-NN neighborhoods, followed by a self-attention-based feature extraction module that captures both local and non-local dependencies. Finally, a displacement prediction head calculates per-point corrections to restore clean geometry. A new offshore drilling platform point cloud dataset is also released, demonstrating TPDNet's practical efficacy in reconstructing 3D surfaces under real-world marine conditions. 

Liu et al.~\cite{liu2025pyramidpcd} proposed a pyramid network named PyramidPCD, addressing the challenge of restoring high-quality point clouds from noisy inputs. Their key contribution lies in a UNet-shaped Transformer architecture incorporating structure-aware and detail-preserving units at multiple scales. The method leverages a feature pyramid, where coarser scales capture primary structures and finer scales retain detailed shape characteristics. By progressively refining features across scales, the approach effectively removes noise while preserving structural integrity.

\subsection{Distribution-based Methods}
Distribution-based methods formulate point cloud denoising from a probabilistic perspective. Rather than regressing clean coordinates or explicit displacements, they assume noisy observations follow some probabilistic corruption of the clean distribution. Denoising can then proceed in different ways: score-based approaches (e.g., diffusion models) interpret the gradient of the log-probability as a denoising direction, while flow-based approaches learn invertible mappings that disentangle noise from structure in latent space. 
Such methods offer a principled framework for uncertainty modeling and often support unsupervised or self-supervised learning.

\textbf{Diffusion-based Methods.} 
Diffusion-based point cloud denoising methods are closely related to score-based approaches, as both share the same theoretical foundation of learning the gradient of the log-probability density $\nabla_x \log p(x)$, also known as the score function. 
In this view, classical score matching~\cite{hyvarinen2005estimation}, its extension to high-dimensional generative modeling~\cite{song2019generative}, and subsequent diffusion formulations (e.g., DDPM~\cite{ho2020denoising}, DDIM~\cite{song2020denoising}) can all be understood within a unified score-based diffusion framework, differing mainly in how the forward noisifying process and the reverse denoising dynamics are defined. 
Thus, earlier gradient-based denoisers for point clouds can naturally be seen as instances of diffusion-based modeling.

Building on these,
ScoreDenoise~\cite{luo2021score} is the first to model noisy point clouds as samples from a convolved distribution \((p * n)(x)\), where \(p(x)\) denotes the clean surface distribution and \(n(x)\) is the noise model. Instead of directly predicting the noise displacement, the method estimates the score function \(\nabla_x \log(p * n)(x)\), i.e., the 
gradient-log of the noise convolved probability distribution.
Denoising is performed by iteratively applying gradient ascent to move points toward the high-density regions of the distribution. 
This probabilistic formulation improves robustness against noise artifacts such as outliers and shrinkage.
Chen et al.~\cite{chen2022deep} built upon ScoreDenoise by introducing a more robust and flexible resampling framework. While inheriting the score-based modeling paradigm, this method enhances the pipeline in three key aspects: 
(1) learning a global, smoother gradient field;
(2) incorporating a context-aware feature extractor for better generalization across varying point structures; 
and (3) integrating iterative regularization (e.g., Laplacian smoothing) to further preserve geometric fidelity. These improvements lead to significantly improved denoising and upsampling performance, especially in challenging or unstructured regions.
Furthermore,
MAG~\cite{zhao2023mag} introduces a momentum-enhanced gradient ascent scheme. A neural network first estimates the gradient field \(\nabla_x \log p(x)\), similar to prior score-based methods. Instead of using raw gradient ascent, MAG applies a momentum term that accumulates past update directions when moving each point:
\[
v_{t+1} = \beta\,v_t + \alpha\,g_t,\quad
x_{t+1} = x_t + v_{t+1},
\]
where \(g_t\) is the estimated gradient, \(v_t\) is the momentum, and \(\beta, \alpha\) are hyper-parameters. This approach reduces oscillations and accelerates convergence, cutting inference time by 25–40\% and reducing artifacts such as outliers or shrinkage.  

Yang et al.~\cite{yang20243d} incorporated global feature guidance to enhance robustness and structure preservation, which consists of three modules:
(1) Adaptive Feature Extraction (AFE): Extracts global features from the noisy point cloud to dynamically guide the embedding of local and non-local features;
(2) Gradient Field Estimation (GFE): Uses these adaptive features to estimate the gradient of the log-probability function \(\nabla_x \log (p * n)(x)\) via a densely connected MLP;
(3) Further Accelerated Gradient Ascent (FAGA): Applies an accelerated gradient ascent update to move points toward the clean surface more efficiently and stably.
This combination of global-conditioning and refined optimization allows the method to outperform prior score-based denoisers, effectively removing noise while preserving geometric details.

Hu and Hu~\cite{hu2025dynamic} extended to dynamic scenarios by leveraging temporal consistency across frames. The method follows a four-stage pipeline. First, it estimates the gradient of the log-probability density \(\nabla_x \log p(x)\) for each frame using a local feature-based neural network. Second, it establishes temporal correspondences by treating local patches as rigid bodies that move within the gradient field of adjacent frames until reaching equilibrium, thereby revealing aligned structures. Third, it refines the gradient estimation by aggregating gradients from temporally corresponding patches, producing a smoother and more stable field. Finally, denoising is performed by iteratively updating point positions along the aggregated gradient directions via gradient ascent. 

Li et al.~\cite{li2025nisgcn} present NI‑SGCN, an energy-efficient network that introduces spiking neurons into the ScoreDenoise framework~\cite{luo2021score}. The network is composed of two modules: a spiking feature extraction module, which utilizes noise-injected Integrate-and-Fire neurons within a graph convolutional structure to extract local features, and a score estimation module, which predicts the gradient of the underlying data distribution (\(\nabla \log p(x)\)) to guide denoising.

More recently, Vogel et al.~\cite{vogel2024p2p} introduced P2P-Bridge by adapting Diffusion Schrödinger Bridges optimal paths for point cloud denoising.
This work formulates denoising as an optimal transport problem, learning a transport plan between noisy and clean point clouds. 
Further advancing this direction, Wang et al.~\cite{wang2025adaptive} introduced an adaptive and iterative method based on a score-based diffusion model. The key contribution lies in addressing the challenge of efficiently handling different noise levels and patterns in point cloud data. Unlike traditional deep learning approaches that empirically repeat denoising steps, the authors propose an adaptive denoising schedule that estimates noise variation and adjusts step sizes accordingly. Their method employs a two-stage sampling strategy during training to enable feature fusion and gradient fusion, which enhances iterative denoising performance.
The most comprehensive diffusion-based approach is presented by Du et al.~\cite{du2025superpc} through their SuperPC framework. Unlike task-specific solutions that may suffer from error accumulation, SuperPC introduces a unified diffusion model capable of joint completion, upsampling, denoising, and colorization. The method's effectiveness stems from its three-level-conditioned diffusion framework, augmented by a spatial-mix-fusion strategy that explicitly models inter-task correlations. This holistic approach demonstrates superior performance by simultaneously addressing multiple point cloud defects through shared feature learning.

\textbf{Point MLP Methods.}
Zhao et al.~\cite{zhao2022noise} proposed a Self-Variation Capture Network (SVCNet) built on the assumption that noise is relatively small compared to the underlying structure. By perturbing the latent features of a noisy point cloud, SVCNet generates self-variation representations that differ in noise distribution but preserve structural consistency. The network then captures their commonality through feature aggregation and directly regresses denoised coordinates. Additionally, an edge constraint module is introduced to suppress over-smoothing and preserve sharp geometric features. Unlike prior methods that rely on explicit noise priors, SVCNet adopts a noise-agnostic approach and demonstrates strong performance, particularly in filtering surface drift noise and producing uniformly distributed point clouds.

\textbf{Flow-based generative models.}
Normalizing flows define invertible transformations that map data into tractable distributions, and are typically categorized into discrete variants~\cite{kingma2018glow} and continuous ones~\cite{grathwohl2018ffjord}. 
In the context of 3D point clouds, PointFlow~\cite{yang2019pointflow} was the first flow-based method, leveraging continuous normalizing flows to model a two-level distribution hierarchy for point cloud generation.
As for point cloud denoising, PD-Flow~\cite{mao2022pd} formulates point cloud denoising as a distribution learning and feature disentanglement problem. Specifically, noisy point clouds are treated as samples from a joint distribution of clean structures and noise. By disentangling the noise component from the latent representation, the clean points can be reconstructed. The mapping between the Euclidean space and the latent space is modeled via a normalizing flow, enabling invertible transformations and exact likelihood estimation.
Hu et al.~\cite{hu2023ndpuflow} introduced a noising‑denoising framework, ND‑PUFlow, for point cloud upsampling using Continuous Normalizing Flows (CNF). In the noising stage, the method generates dense noisy point sets by adding Gaussian perturbations to sparse inputs. During the denoising stage, a CNF model learns to transform each noisy point back towards the underlying surface distribution via invertible mappings. This process is trained in both supervised and self‑supervised settings, enabling flexible, arbitrary-scale upsampling without relying on explicit high-resolution ground truth. ND‑PUFlow achieves competitive performance while providing principled likelihood-based modeling of point distributions.
Building on PD-Flow~\cite{mao2022pd}, Mao et al.~\cite{mao2024denoising} proposed a significantly enhanced framework named PD-LTS, which further explores point cloud denoising in the latent space. While PD-Flow relies on a simple feature extractor, PD-LTS introduces a deeper multi-level graph convolution (MLGC) module to capture both local and global geometric features inspired by IterativePFN~\cite{de2023iterativepfn}. More importantly, it explicitly disentangles clean structure and noise within the latent representation: noise-related channels are zeroed out before the clean latent code is reconstructed via a more stable invertible neural network based on monotone operator theory.

\subsection{Filter-based Methods.}
These methods estimate or learn auxiliary information (e.g., normals) and apply a traditional
filtering rule to update point positions. 

Lu et al.~\cite{lu2020deep} proposed a feature-preserving point cloud filtering framework that combines deep normal estimation with an iterative filtering process. 
The method first classifies each point as a feature or non-feature point, and then predicts surface normals using two separate CNN-based networks. To better leverage the structured nature of convolutional operations, the method projects local 3D point neighborhoods onto a 2D tangent image plane, enabling the use of standard 2D CNNs for normal estimation. In the testing phase, the denoising process is performed via a deterministic filtering rule~\cite{lu2020low_nonlocal} that updates each point's position based on its neighborhood and the estimated normals. This filtering formula encourages the updated points to align with locally estimated tangent planes, effectively removing noise while preserving sharp geometric structures. 
Unlike purely learning-based approaches, the network in this method does not directly predict the denoised coordinates, but rather guides a traditional filter through learned geometric priors.
GeoDualCNN~\cite{wei2021geodualcnn} presents a geometry-supporting dual-branch CNN designed for joint surface normal estimation and point cloud denoising. The key insight is that accurate normal estimation requires local surface smoothness, and denoising relies on reliable normals. The method first defines homogeneous neighborhoods (HoNe) to isolate locally smooth patches, then computes initial normals via PCA on these patches. GeoDualCNN refines both normals and point positions through two parallel branches: one processes a homogeneous height map to denoise positionally, while the other leverages a homogeneous normal map to preserve surface features.

Different from predicting the displacement, Wang et al.~\cite{wang2022pointfilternet} proposed PointFilterNet (PFN), which integrates classical filtering principles with neural networks to perform denoising.
PFN learns optimal filtering coefficients through an outlier recognizer and a denoiser. The outlier recognizer mitigates noise by assigning small weights to outliers, while the denoiser progressively refines the point cloud, mimicking human visual perception.
It allows for interpretable and robust point aggregation, where each neighbor’s contribution is adaptively controlled.

Chen et al.~\cite{geogcn2023} proposed a dual-domain graph convolutional framework by estimating two types of normals: real normals (RNs) for local geometry refinement and virtual normals (VNs) for global structure regularization. 
The first stage applies a Spatial GCN (S-GCN) to regress initial denoised points. Then, Principal Component Analysis (PCA) is performed on these preliminary results to estimate initial real normals (RNs). These normals are concatenated with the denoised points and passed into a Normal-domain GCN (N-GCN), which refines the RNs. Finally, the denoised points are updated using an optimised normal-guided filtering module, resulting in the final clean point cloud.

\subsection{Classification-based Methods}
These methods treat point cloud denoising as a point-wise classification or segmentation task. Instead of directly modifying point positions, they aim to identify and remove noisy or outlier points by labeling each point as either valid or corrupted. Such approaches are particularly effective in scenarios where noise is sparse but structurally disruptive, such as LiDAR scans under adverse weather conditions.

\textbf{Convolution-based Methods.}
A representative method is WeatherNet~\cite{heinzler2020cnn}, which uses a CNN to classify weather-induced noise in LiDAR data. The authors propose a CNN-based method that performs point-wise weather segmentation, distinguishing between valid points and those affected by adverse weather. This segmentation enables the removal of noise while preserving essential structural details in the point cloud. The network is trained on a dataset collected under controlled weather conditions, providing point-wise ground truth annotations for various weather scenarios.
LiSnowNet~\cite{yu2022lisnownet} presents a real-time, CNN-based method for unsupervised snow removal from LiDAR point clouds. The network first projects raw LiDAR scans into range–intensity images and processes them via a deep residual convolutional architecture interleaved with wavelet transform layers to predict a residual map. Unlike point-displacement or reconstruction methods, LiSnowNet performs point classification: points associated with predicted snow noise are removed rather than adjusted. The model is trained without labels, using a sparsity-promoting loss applied in the wavelet domain. 
4DenoiseNet~\cite{seppanen20224denoisenet} introduces a real-time deep learning framework that exploits spatial–temporal context for LiDAR point cloud denoising in adverse weather. By jointly processing consecutive frames \(P_t\) and \(P_{t-1}\), the network extracts spatio-temporal features through a k‑nearest‑neighbor convolution followed by a motion‑guided attention mechanism.

Gao et al.~\cite{gao2022reflective} addressed the challenge of reflective noise in large-scale LiDAR point clouds, such as virtual points caused by glass surfaces. The method first projects the 3D point cloud into a 2D range-intensity image and processes it with a Transformer-based autoencoder to classify points as either valid or reflective noise. To improve reliability, the classification results are further cross-validated using data from multiple sensor positions. Finally, points identified as noise are removed from the original point cloud.

Zuo et al.~\cite{zuo2023attention} proposed a classification-based denoising framework tailored to scanned point clouds of rocket tank panels. The method leverages an attention-enhanced deep learning model to identify and remove noisy points, rather than predicting displacements. By incorporating attention mechanisms into the feature extraction pipeline and stabilizing the learning process with layer normalization, the model achieves high accuracy in distinguishing surface points from noise, making it suitable for precise industrial applications.

\section{Unsupervised Learning}
\label{sec:unsupervised}
Unsupervised learning methods for point cloud denoising are particularly valuable in scenarios where labeled data is scarce or expensive to obtain. These methods aim to improve the quality of point clouds by identifying and reducing noise through patterns discovered in the data itself, without relying on ground truth labels. Here, we discuss several innovative approaches in the field.

Total Denoising~\cite{hermosilla2019total} is the first approach to point cloud denoising that does not rely on labeled training data. Instead, it utilizes the inherent structures and distributions within the point clouds to identify and remove noise. The technique leverages geometric constraints and statistical models to differentiate between noise and valid data points, optimizing the point cloud's overall structure without prior knowledge of the noise characteristics.

Regaya et al.~\cite{regaya2021point} proposed Point-Denoise, an unsupervised outlier detection method for enhancing 3D point clouds. Point-Denoise leverages unsupervised learning to detect and remove outliers, regardless of their type or level. 
They conducted experiments on their new dataset, acquired using the Riegl VZ-400 3D Terrestrial Laser Scanner.

Wang et al.~\cite{wang2024noise4denoise} proposed Noise4Denoise, an unsupervised framework that removes the dependency on ground-truth clean point clouds.
The method generates two noisy versions of each input point cloud, with the second containing twice the noise level of the first, and trains the network to predict their displacement as an approximation of the underlying clean surface. This strategy enables effective learning directly from noisy data and achieves performance comparable to supervised approaches, offering greater practicality for real-world scenarios where clean references are unavailable. However, this method rely on the modeling of nosie, as twice noise is a ideal assumption.

Most recently, inspired by unsupervised score-based image denoising~\cite{kim2021noise2score}, Wei et al.~\cite{wei2025noise2score3d} introduced Noise2Score3D, an unsupervised score-based denoising framework that leverages Tweedie’s~\cite{efron2011tweedie} formula to estimate the posterior mean of clean points from noisy observations. By directly learning the score function without requiring paired supervision, this method offers a principled probabilistic alternative to displacement regression, further advancing the applicability of unsupervised denoising in real-world settings.

\section{Benchmark}
\label{sec:benchmark}
\begin{figure}[htbp]
    \centering
        \includegraphics[width=0.95\linewidth]{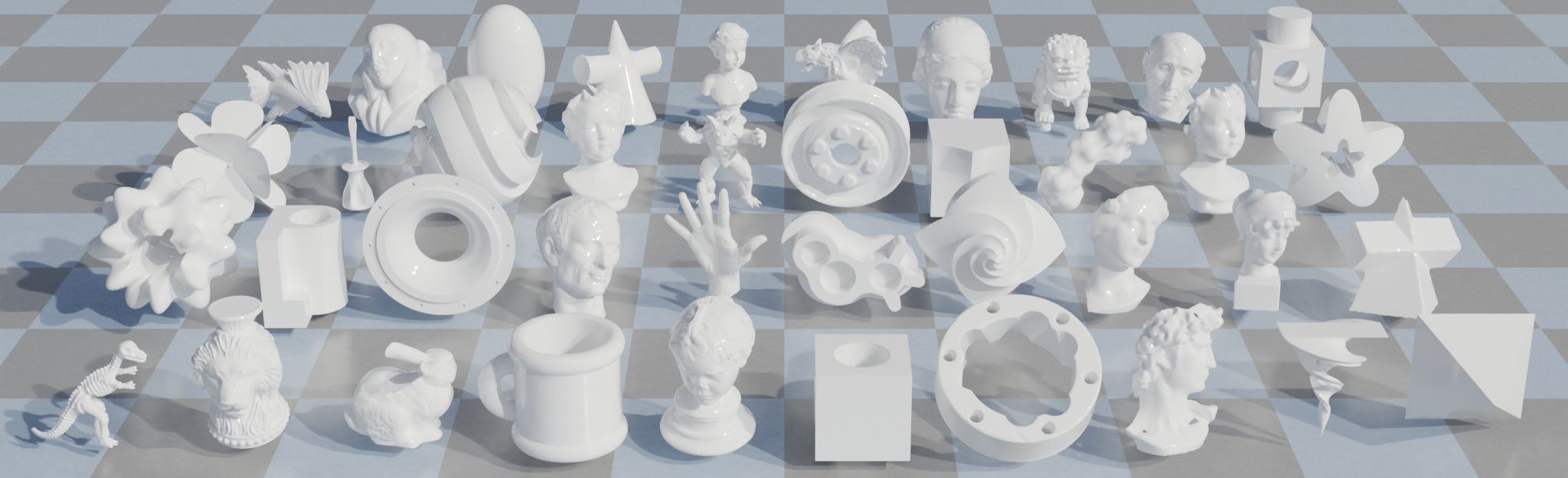}%
        \label{fig:trainset}
    \\[0.5em]
        \includegraphics[width=0.95\linewidth]{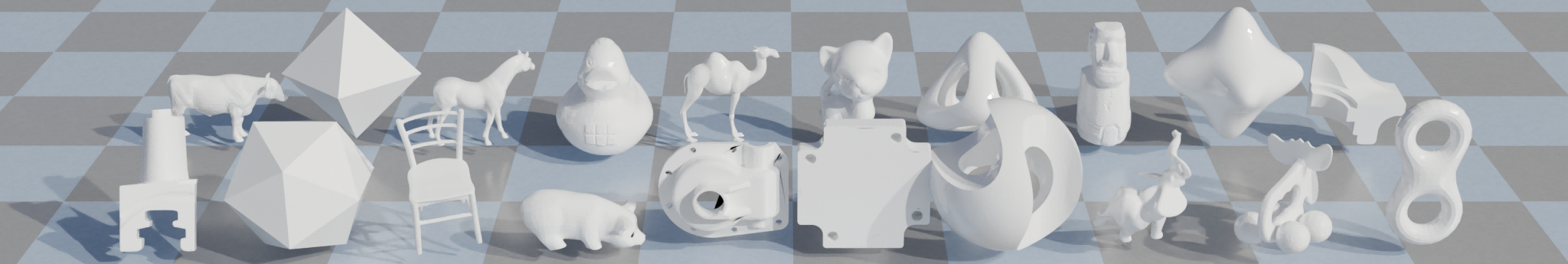}%
        \label{fig:testset}
    \caption{Visualizations of training (top) and testing (bottom) datasets.}
    \label{fig:train_test}
\end{figure}

\subsection{Dataset Setup}
\label{sec:dataset}
To fairly evaluate the performance of various point cloud denoising methods, we adopt a standardized set of datasets for training and testing.

\textbf{Training data.} 
We adopt the training dataset provided by PUNet~\cite{yu2018punet}, which has become the de facto standard for learning-based point cloud denoising in recent state-of-the-art works~\cite{zhou20253dmambaipf,luo2021score}. Its widespread use facilitates fair and reproducible comparisons across different methods.
The dataset comprises $40$ synthetic 3D meshes for training, encompassing both CAD and non-CAD shapes, providing a diverse range of geometric structures.  
For each shape, we follow previous works and uniformly sample point clouds at three resolutions: 10K, 30K, and 50K points using Poisson disk sampling. Before adding random noise, all point clouds are normalized to lie within a unit sphere centered at the origin. To simulate noisy observations, Gaussian noise is added to each point cloud during training, with the standard deviation randomly sampled from the interval $[0.005, 0.02]$. 
This noise variability enriches the training distribution, enabling the dataset to cover a broader range of corruption levels and reducing the risk of overfitting to any specific noise configuration. The multi-resolution setup also encourages the model to adapt to different input densities.
Fig.~\ref{fig:train_test} provides an overview of the training and testing datasets.

\textbf{Testing data.}  
For testing, we adopt two widely used datasets to evaluate denoising performance under both synthetic and real-world noise.
The first is the PUNet benchmark, which contains $20$ synthetic 3D models corrupted with Gaussian noise at three levels ($\sigma = 0.01$, $0.02$, $0.03$) and two resolutions ($10K$ and $50K$ points).
The second is the \textit{Paris-rue-Madame} dataset~\cite{ruemadame2014}, captured using a laser scanner in a real urban environment. It is used to evaluate generalization under realistic sensor noise, and features diverse object categories with varying surface geometries and noise. Since no ground truth is available for Paris-rue-Madame, evaluation on this dataset is limited to qualitative visual comparisons.

All input point clouds are normalized using the same preprocessing pipeline, and the same ground-truth meshes are used across methods to ensure a fair comparison. Additionally, supervised and unsupervised approaches are evaluated separately under the same settings.

\subsection{Evaluation Metrics}

To quantitatively assess the denoising performance, we adopt three commonly used metrics: Chamfer Distance (CD), Point-to-Mesh Distance (P2M), and Hausdorff Distance (HD). CD measures the average bidirectional point-wise distance between the denoised point cloud and the ground-truth. P2M computes the shortest distance from each denoised point to the reference mesh surface, providing a surface-aware evaluation. HD captures the maximum deviation between the predicted and ground-truth shapes, highlighting outlier sensitivity.

\vspace{1mm}
\noindent \textbf{1) Point-to-Point Metrics.} \\
Let the ground-truth point cloud be $S_c = \{x_i\}_{i=1}^{|S_c|}$ and the denoised point cloud be $S_p = \{y_i\}_{i=1}^{|S_p|}$. The number of points $|S_c|$ and $|S_p|$ may vary across methods due to filtering operations.

Here, $\|x - y\|_2$ denotes the Euclidean distance between two points $x$ and $y$ in $\mathbb{R}^3$. The operator $\min_{y \in S_p} \|x - y\|_2$ computes the distance from a point $x \in S_c$ to its nearest neighbor in $S_p$, while $\sup$ denotes the supremum (i.e., the maximum) over all such values.

\begin{itemize}
    \item[a)]
    \textit{Chamfer Distance (CD):}
    \begin{equation}\label{eq:chamfer}
        \begin{aligned}
            e_{\mathrm{CD}}(S_c, S_p) = \frac{1}{|S_c|} \sum_{x \in S_c} \min_{y \in S_p} \|x - y\|_2^2 \\+ 
            \frac{1}{|S_p|} \sum_{y \in S_p} \min_{x \in S_c} \|y - x\|_2^2
        \end{aligned}
    \end{equation}

    \item[b)]
    \textit{Hausdorff Distance (HD):}
    \begin{equation}\label{eq:hausdorff}
        \begin{aligned}
            e_{\mathrm{HD}}(S_c, S_p) = \max \big\{ 
            &\sup_{x \in S_c} \min_{y \in S_p} \|x - y\|_2, \\
            &\sup_{y \in S_p} \min_{x \in S_c} \|y - x\|_2 
            \big\}
        \end{aligned}
    \end{equation}
\end{itemize}

\vspace{1mm}
\noindent \textbf{2) Point-to-Mesh Metric.} \\
The Point-to-Mesh Distance (P2M) evaluates how well the denoised point cloud conforms to the surface of the ground-truth mesh. 
It is defined in a bi-directional manner: (1) for each predicted point $y \in S_p$, the closest point on the reference mesh surface $M$ is found; and (2) for each face $f \in M$, the closest point in the predicted set $S_p$ is identified. The final distance is the average of these two terms:
\begin{equation}
\begin{aligned}
\mathrm{P2M}(\mathcal{M}, S_p)
&= \frac{1}{|S_p|}
   \sum_{y \in S_p}
   \min_{f \in \mathcal{M}}
   \left\lVert y - \operatorname{proj}_{f}(y) \right\rVert_2^{2} \\
&\quad + \frac{1}{|\mathcal{M}|}
   \sum_{f \in \mathcal{M}}
   \min_{y \in S_p}
   \left\lVert \operatorname{proj}_{f}(y) - y \right\rVert_2^{2}
\end{aligned}
\end{equation}

All denoised point clouds are normalized using the same transformation as their corresponding ground-truth meshes. This ensures consistent alignment in the original coordinate space for fair comparison. All metrics are computed directly between the normalized point cloud and the ground-truth. we use PyTorch3D~\cite{ravi2020accelerating} implementations of these metrics, and lower values indicate better denoising performance.

\subsection{Benchmark Settings}

Ensuring a fair comparison across point cloud denoising methods is non-trivial due to variations in training configurations, input resolutions, noise settings, and normalization pipelines. Such inconsistencies can lead to misleading conclusions if not carefully controlled. To address this, we implement a unified evaluation protocol designed for both comparability and reproducibility.

All compared methods are evaluated under the same training conditions using the PUNet dataset, with consistent input resolutions and Gaussian noise augmentation, where noise levels are randomly sampled from $[0.005, 0.02]$, as described in Section~\ref{sec:dataset}.
When official pretrained models are available and consistent with our setup (e.g., \textit{3DMambaIPF}, \textit{ASDN}), we use the released weights directly. 
For methods that rely on precomputed geometric feature weights for training supervision (e.g., \textit{RePCD}, \textit{Pathnet}), we retrain the models using the datasets and hyperparameters provided in the original papers, as the required labels are not derivable from raw input alone.
For methods requiring input normals in the training stage (e.g., \textit{Pointfilter}, \textit{PCDNF}), we estimate normals using Open3D with default settings.

Notably, deep networks are often highly sensitive to input scale. Instead of enforcing a universal normalization pipeline, we preserve each method's original centering and scaling strategy during inference to respect its design. Nevertheless, all output point clouds are re-normalized using the ground-truth transformation prior to metric computation (CD, P2M, HD) to ensure consistent evaluation.
For iterative methods, we follow the iteration counts recommended in the original papers.
We moderately increase the number of iterations under high-noise conditions to improve denoising performance, while keeping the configuration aligned with the method’s original design.

\begin{table*}[th]
\caption{Quantitative comparison on the PUNet test set with 10K input points under Gaussian noise levels of 1\%, 2\%, and 3\%. 
Chamfer Distance (CD) and Point-to-Mesh Distance (P2M) are multiplied by $10^4$, while Hausdorff Distance (HD) is multiplied by $10^3$. Lower values indicate better performance. The lowest value in each column is shown in \textbf{bold}, and the second- and third-lowest values are \underline{underlined}.}
\label{tab:punet_10k}
\resizebox{\textwidth}{!}{%
\begin{tabular}{lccccccccccc}\toprule
\multirow{2}{*}{Method} & \multirow{2}{*}{Year} & \multirow{2}{*}{TrainSet} & \multicolumn{3}{c}{1\% noise} & \multicolumn{3}{c}{2\% noise} & \multicolumn{3}{c}{3\% noise} \\
\cmidrule(lr){4-6}\cmidrule(lr){7-9}\cmidrule(lr){10-12}
 &  &  & CD$\downarrow$ & P2M$\downarrow$ & HD$\downarrow$ & CD$\downarrow$ & P2M$\downarrow$ & HD$\downarrow$ & CD$\downarrow$ & P2M$\downarrow$ & HD$\downarrow$ \\
\midrule
\multicolumn{12}{c}{\textbf{Supervised}} \\
3DMambaIPF~\cite{zhou20253dmambaipf}       & 2025 & PUNet  & 1.989 & \underline{0.477} & 0.807 & 2.995 & \underline{0.803} & 1.583 & 3.719 & \underline{1.323} & \textbf{2.385} \\
ASDN~\cite{guo2025you}            & 2025 & PUNet  & \underline{1.836} & \underline{0.490} & \underline{0.722} & \underline{2.509} & \underline{0.776} & \underline{1.329} & \textbf{3.115} & \textbf{1.282} & 2.848 \\
P2P-Bridge~\cite{vogel2024p2p}      & 2024 & PUNet  & 2.283 & 0.685 & 5.012 & 3.204 & 1.115 & 4.156 & 3.994 & 1.758 & 4.565 \\
Pathnet~\cite{wei2024pathnet}         & 2024 & PU-GAN & 3.184 & 1.307 & 1.680  & 4.107 & 1.707 & 3.220  & 4.860  & 2.257 & 5.560 \\
StraightPCF~\cite{de2024straightpcf}     & 2024 & PUNet  & \underline{1.905} & 0.546 & \underline{0.791} & \underline{2.673} & 0.929 & \underline{1.475} & \underline{3.263} & \underline{1.429} & \underline{2.786} \\
PD-LTS~\cite{mao2024denoising}          & 2024 & PUNet  & \textbf{1.783} & \textbf{0.468} & \textbf{0.692} & \textbf{2.449} & \textbf{0.765} & \textbf{1.234} & \underline{3.431} & 1.500 & 2.856 \\
MAG~\cite{zhao2023mag}             & 2023 & PUNet  & 2.522 & 0.753 & 4.233 & 3.685 & 1.381 & 5.569 & 4.708 & 2.261 & 9.172 \\
PCDNF~\cite{liu2023pcdnf}           & 2023 & PUNet  & 2.557 & 0.773 & 1.226 & 3.750  & 1.347 & 3.070  & 4.784 & 2.212 & 6.469 \\
MODNet~\cite{huang2022modnet}          & 2023 & PUNet  & 2.442 & 0.669 & 1.085 & 3.868 & 1.383 & 3.935 & 6.976 & 3.818 & 12.94 \\
IterativePFN~\cite{de2023iterativepfn}    & 2023 & PUNet  & 2.055 & 0.501 & 0.929 & 3.043 & 0.843 & 1.666 & 4.242 & 1.689 & \underline{2.587} \\
DeepPSR~\cite{chen2022deep}         & 2022 & PUNet  & 3.425& 1.200 & 5.844& 3.761& 1.21& 2.765& 5.291& 2.234& 4.094\\
RePCD~\cite{chen2022repcd}           & 2022 & PU-GAN & 2.751 & 0.905 & 1.101 & 4.068 & 1.557 & 2.966 & 6.552 & 3.521 & 6.549 \\
PDFlow~\cite{mao2022pd}          & 2022 & PUNet  & 2.133 & 0.681 & 1.066 & 3.252 & 1.330  & 2.969 & 5.208 & 2.917 & 6.445 \\
ScoreDenoise~\cite{luo2021score}    & 2021 & PUNet  & 2.597 & 0.827 & 1.721 & 3.704 & 1.404 & 5.227 & 4.896& 2.397& 6.584\\
DMRDenoise~\cite{luo2020differentiable}      & 2020 & PUNet  & 3.220  & 1.003 & 2.794 & 4.105 & 1.679 & 3.039 & 6.450  & 3.594 & 5.605 \\
GPD~\cite{pistilli2020learning}             & 2020 & PUNet  & 2.310  & 0.714 & 0.905 & 4.283 & 1.855 & 3.203 & 8.089 & 4.936 & 9.207 \\
Pointfilter~\cite{zhang2020pointfilter}     & 2020 & PUNet  & 2.602 & 0.843 & 1.406 & 3.655 & 1.242 & 2.489 & 5.099 & 2.187 & 6.252 \\
PointCleanNet~\cite{rakotosaona2020pointcleannet}   & 2019 & PUNet  & 3.687 & 1.600   & 1.571 & 7.931 & 4.764 & 6.062 & 13.370 & 9.586 & 13.49 \\
\midrule
\multicolumn{12}{c}{\textbf{Un-supervised}} \\
Noise4Denoise~\cite{wang2024noise4denoise}   & 2024 & PUNet  & \textbf{2.205} & \textbf{0.644} & \textbf{0.882} & \textbf{3.448} & \textbf{1.228} & \textbf{2.326} & \textbf{4.154} & \textbf{1.861} & \textbf{4.113} \\
ScoreDenoise-U~\cite{luo2021score}  & 2021 & PUNet  & \underline{3.558} & \underline{1.069} & \underline{5.573} & \underline{5.231} & \underline{2.020} & \underline{7.475} & \underline{7.249} & \underline{3.523} & \underline{12.236} \\
DMRDenoise-U~\cite{luo2020differentiable}    & 2020 & PUNet  & 5.234 & 2.280 & 7.171 & \underline{6.225} & \underline{2.935} & 14.131 & 8.842 & \underline{5.065} & 22.390 \\
TotalDenoising~\cite{hermosilla2019total}  & 2019 & PUNet  & \underline{3.068} & \underline{0.925} & \underline{1.362} & 6.241 & 2.947 & \underline{5.933} & 11.452 & 7.117 & \underline{13.447} \\
\bottomrule
\end{tabular}
}
\end{table*}

\begin{table*}[h]
\caption{Quantitative comparison on the PUNet test set with 50K input points under Gaussian noise levels of 1\%, 2\%, and 3\%. 
Chamfer Distance (CD) and Point-to-Mesh Distance (P2M) are multiplied by $10^4$, while Hausdorff Distance (HD) is multiplied by $10^3$. Lower values indicate better performance. The lowest value in each column is shown in \textbf{bold}, and the second- and third-lowest values are \underline{underlined}.}
\label{tab:punet_50k}
\resizebox{\textwidth}{!}{%
\begin{tabular}{lccccccccccc}\toprule
\multirow{2}{*}{Method} & \multirow{2}{*}{Year} & \multirow{2}{*}{TrainSet} & \multicolumn{3}{c}{1\% noise} & \multicolumn{3}{c}{2\% noise} & \multicolumn{3}{c}{3\% noise} \\
\cmidrule(lr){4-6}\cmidrule(lr){7-9}\cmidrule(lr){10-12}
 &  &  & CD$\downarrow$ & P2M$\downarrow$ & HD$\downarrow$ & CD$\downarrow$ & P2M$\downarrow$ & HD$\downarrow$ & CD$\downarrow$ & P2M$\downarrow$ & HD$\downarrow$ \\
\midrule
\multicolumn{12}{c}{\textbf{Supervised}} \\
3DMambaIPF~\cite{zhou20253dmambaipf}     & 2025 & PUNet & 0.589 & \textbf{0.291} & 0.375 & \underline{0.755} & \textbf{0.405} & \textbf{0.880}  & \underline{1.455} & \underline{0.924} & \textbf{2.294} \\
ASDN~\cite{guo2025you}                   & 2025 & PUNet & \underline{0.499} & 0.308 & \underline{0.303} & \underline{0.687} & 0.448 & 1.124 & \underline{1.397} & \underline{0.974} & 13.679 \\
P2P-Bridge~\cite{vogel2024p2p}           & 2024 & PUNet & \underline{0.586} & 0.330  & 2.619 & 0.902 & 0.580  & 6.964 & 1.554 & 1.124 & 8.863 \\
Pathnet~\cite{wei2024pathnet}            & 2024 & PU-GAN& 0.729 & 0.386 & 0.849 & 1.160  & 0.685 & 3.022 & 2.612 & 1.842 & 8.338 \\
StraightPCF~\cite{de2024straightpcf}     & 2024 & PUNet & 0.620  & 0.389 & \underline{0.479} & 0.812 & 0.551 & \underline{1.074} & \textbf{1.227} & \textbf{0.856} & 3.222 \\
PD-LTS~\cite{mao2024denoising}           & 2024 & PUNet & \textbf{0.472} & \underline{0.296} & \textbf{0.302} & \textbf{0.658} & \underline{0.432} & 1.115 & 1.854 & 1.341 & 5.064 \\
MAG~\cite{zhao2023mag}                   & 2023 & PUNet & 0.717 & 0.400   & 2.296 & 1.288 & 0.832 & 4.902 & 1.928 & 1.334 & 9.342 \\
PCDNF~\cite{liu2023pcdnf}                                      & 2023 & PUNet & 0.671 & 0.348 & 0.617 & 1.017 & 0.577 & 2.525 & 1.583 & 1.002 & 5.370  \\
MODNet~\cite{huang2022modnet}            & 2023 & PUNet & 0.711 & 0.375 & 0.806 & 1.100   & 0.645 & 3.508 & 2.840  & 2.086 & 15.864 \\
IterativePFN~\cite{de2023iterativepfn}   & 2023 & PUNet & 0.605 & \underline{0.302} & \underline{0.374} & 0.803 & \underline{0.436} & \underline{1.08} & 1.970  & 1.302 & \underline{2.986} \\
DeepPSR~\cite{chen2022deep}                                      & 2022 & PUNet & 0.768 & 0.390  & 1.680  & 0.970  & 0.508 & 1.202 & 2.610  & 1.675 & \underline{2.802} \\
RePCD~\cite{chen2022repcd}               & 2022 & PU-GAN& 0.769 & 0.406 & 0.801 & 2.216 & 1.593 & 3.904 & 5.527 & 4.583 & 8.980  \\
PDFlow~\cite{mao2022pd}                  & 2022 & PUNet & 0.652 & 0.417 & 1.113 & 1.427 & 1.056 & 3.814 & 3.911 & 3.245 & 7.776 \\
ScoreDenoise~\cite{luo2021score}         & 2021 & PUNet & 0.747 & 0.426 & 1.861 & 1.274 & 0.821 & 4.562 & 2.033 & 1.381 & 8.783 \\
DMRDenoise~\cite{luo2020differentiable}  & 2020 & PUNet & 0.838 & 0.486 & 1.165 & 1.722 & 1.191 & 2.798 & 4.355 & 3.514 & 6.953 \\
GPD~\cite{pistilli2020learning}          & 2020 & PUNet & 1.078 & 0.648 & 1.483 & 3.905 & 3.045 & 6.890  & 8.454 & 7.289 & 16.545 \\
Pointfilter~\cite{zhang2020pointfilter}  & 2020 & PUNet & 0.896 & 0.537 & 0.873 & 1.036 & 0.595 & 1.831 & 1.590  & 0.982 & 7.260  \\
PointCleanNet~\cite{rakotosaona2020pointcleannet} & 2019 & PUNet & 1.071 & 0.626 & 1.868 & 2.036 & 1.426 & 10.270 & 5.710  & 4.769 & 24.664 \\
\midrule
\multicolumn{12}{c}{\textbf{Un-supervised}} \\
Noise4Denoise~\cite{wang2024noise4denoise}                              & 2024 & PUNet & \textbf{0.699} & \textbf{0.385} & \textbf{0.698} & \textbf{1.056} & \textbf{0.636} & \textbf{2.185} & \textbf{1.284} & \textbf{0.829} & \textbf{4.045} \\
ScoreDenoise-U~\cite{luo2021score}        & 2021 & PUNet & \underline{1.046} & \underline{0.467} & \underline{2.693} & \underline{2.216} & \underline{1.304} & \underline{6.529} & \underline{4.413} & \underline{3.140} & \underline{8.741} \\
DMRDenoise-U~\cite{luo2020differentiable} & 2020 & PUNet & 1.579 & 0.927 & 3.024 & 3.330 & 2.356 & \underline{5.684} & 7.076 & 5.715 & 16.364 \\
TotalDenoising~\cite{hermosilla2019total}                              & 2019 & PUNet & \underline{0.790} & \underline{0.418} & \underline{1.057} & \underline{3.030} & \underline{2.226} & 7.030 & \underline{6.593} & \underline{5.349} & \underline{16.618} \\
\bottomrule
\end{tabular}%
}
\end{table*}

\subsection{Quantitative Results}
We report the quantitative performance of all compared methods on the synthetic datasets, under varying Gaussian noise levels (1\%, 2\%, 3\%) and two input resolutions (10K and 50K points). 
Tables~\ref{tab:punet_10k} and~\ref{tab:punet_50k} summarize the results using Chamfer Distance (CD), Point-to-Mesh Distance (P2M), and Hausdorff Distance (HD). 
To facilitate a clearer comparison, we group methods into two categories: supervised and unsupervised, and analyze their performance separately. Within each group, methods are arranged chronologically by publication year to highlight progress over time. Lower values indicate better denoising quality across all metrics.

\textbf{Supervised methods.}
Recent supervised methods, particularly those proposed in 2024–2025 such as \textit{3DMambaIPF}, \textit{ASDN}, \textit{PD-LTS}, and \textit{StraightPCF}, consistently achieve strong performance across all metrics and noise levels. For instance, under 3\% noise at 50K resolution, 
\textit{PD-LTS} achieves a CD of 1.227 and P2M of 0.856, while \textit{ASDN} follows closely with a CD of 1.397.
These models also maintain low CD and P2M values under lighter noise, demonstrating robust generalization across varying input conditions.

Earlier methods such as \textit{P2P-Bridge} and \textit{Pathnet} remain competitive under low-noise settings but tend to exhibit elevated HD values at higher noise levels (e.g., \textit{Pathnet} reaches a HD of 8.338 under 3\% noise at 50K), suggesting greater sensitivity to outliers or structural distortions. Similarly, legacy architectures like \textit{PointCleanNet} and \textit{DMRDenoise} generally underperform in high-resolution or high-noise scenarios.

Resolution also plays a key role in model performance. Most methods show noticeable improvement when moving from 10K to 50K inputs. For example, \textit{StraightPCF} reduces its CD from 3.263 to 1.227 at 3\% noise, indicating its ability to benefit from denser point distributions.
These observations highlight differences in design objectives: some methods excel at average-case recovery, as reflected by low CD values, while others place greater emphasis on structural fidelity, captured by HD and P2M.

In summary, supervised denoising methods over the past several years have adopted increasingly diverse architectures—from point-wise refinement to structure-aware networks—each with unique strengths under specific noise and resolution settings. Our unified benchmark allows for a consistent and transparent comparison of these developments.

\textbf{Unsupervised methods.}
Among unsupervised methods, \textit{Noise4Denoise} consistently demonstrates competitive performance across different noise levels and resolutions. 
For example, under 3\% noise at 50K resolution, it achieves a CD of 1.284 and HD of 4.045, outperforming several other unsupervised approaches in all metrics.
While overall denoising quality remains behind the top supervised models, several unsupervised methods perform reasonably well under moderate noise. 
For instance, \textit{TotalDenoising} achieves a CD of 0.79 under 1\% noise at 50K, and maintains acceptable HD and P2M scores across other settings. 
This suggests that with suitable architectural design and training strategies, unsupervised models can handle low to medium level corruption even without ground-truth supervision.

However, as noise intensity increases, unsupervised methods exhibit performance degradation, especially in terms of HD. 
For example, \textit{ScoreDenoise-U} reaches an HD of 12.236 under 3\% noise at 10K resolution, 
indicating challenges of suppressing outliers in unsupervised settings with severe noise.
Overall, unsupervised methods offer a promising direction for label-free denoising, and their performance in low-noise scenarios shows encouraging potential.

\subsection{Visual Comparison}
\begin{figure*}[htbp]
\centering
\vspace{0.3em}
\textbf{Supervised}\par\vspace{0.1em}

\begin{flushleft}
\begin{minipage}[b]{0.27\linewidth}\centering
  \includegraphics[width=\linewidth]{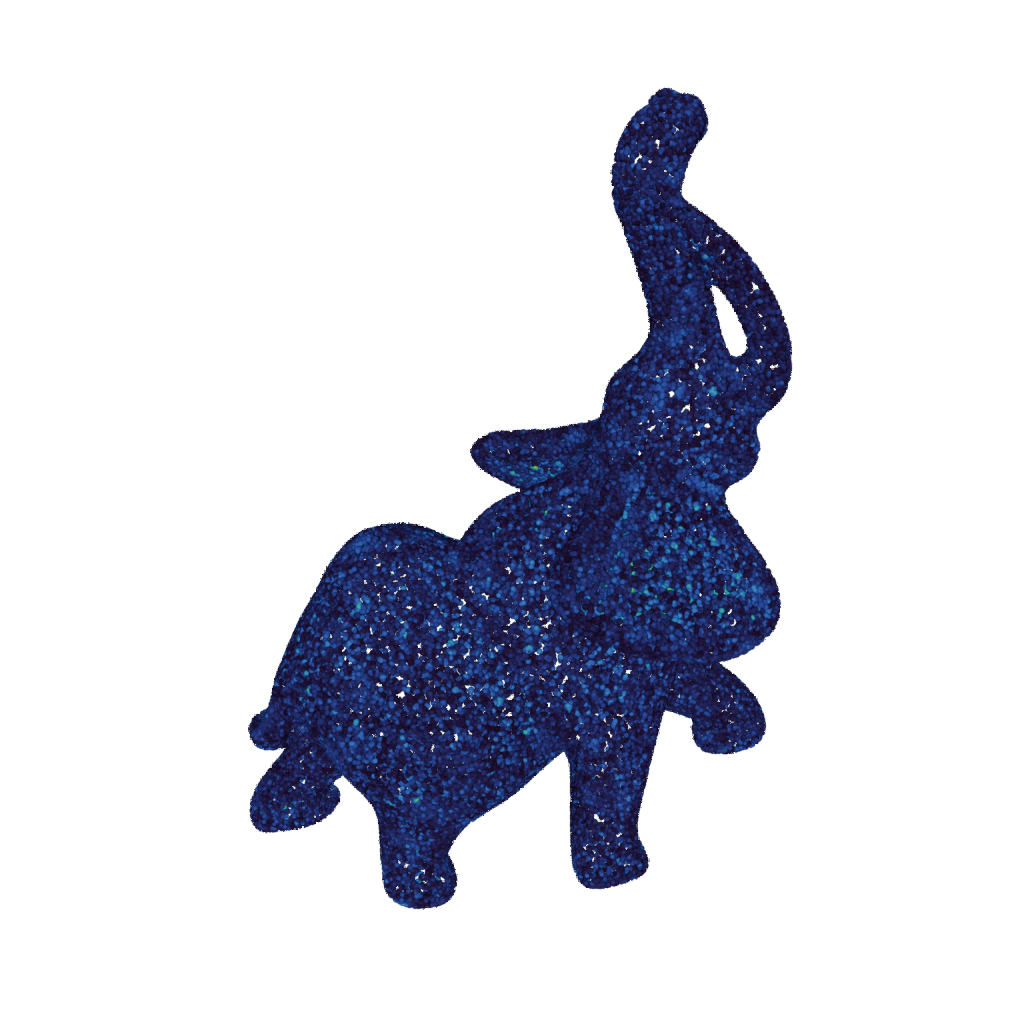}\\[-0.2em]
  \footnotesize 3DMambaIPF
\end{minipage}\hspace{-23mm}%
\begin{minipage}[b]{0.27\linewidth}\centering
  \includegraphics[width=\linewidth]{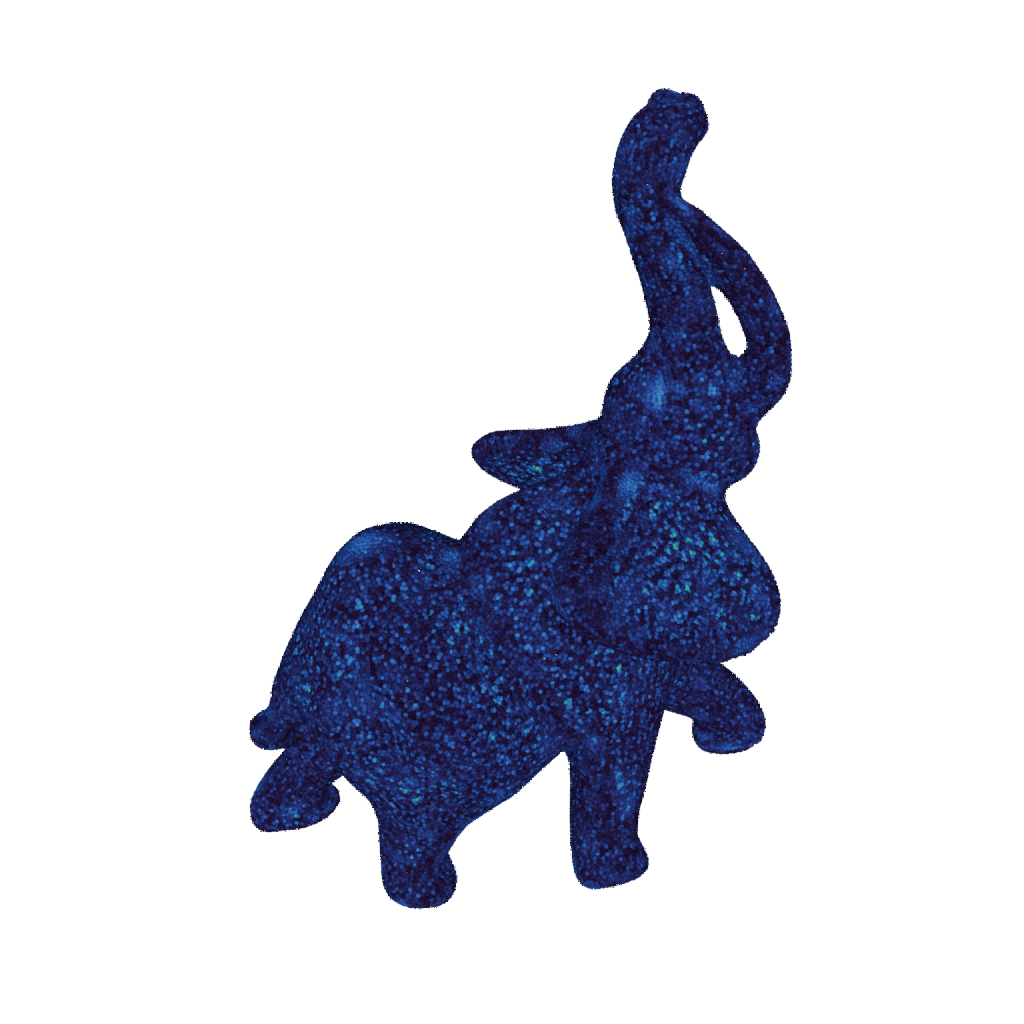}\\[-0.2em]
  \footnotesize ASDN
\end{minipage}\hspace{-23mm}%
\begin{minipage}[b]{0.27\linewidth}\centering
  \includegraphics[width=\linewidth]{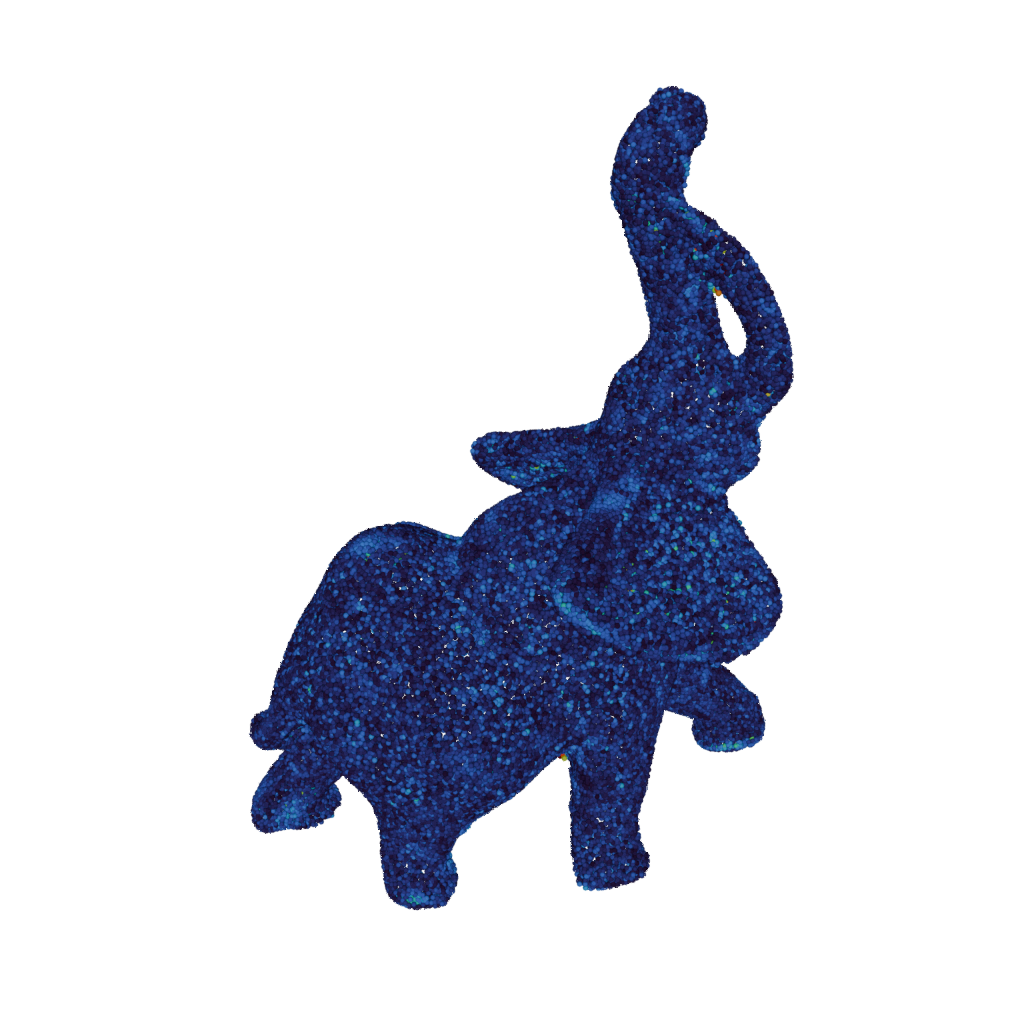}\\[-0.2em]
  \footnotesize P2P-Bridge
\end{minipage}\hspace{-23mm}%
\begin{minipage}[b]{0.27\linewidth}\centering
  \includegraphics[width=\linewidth]{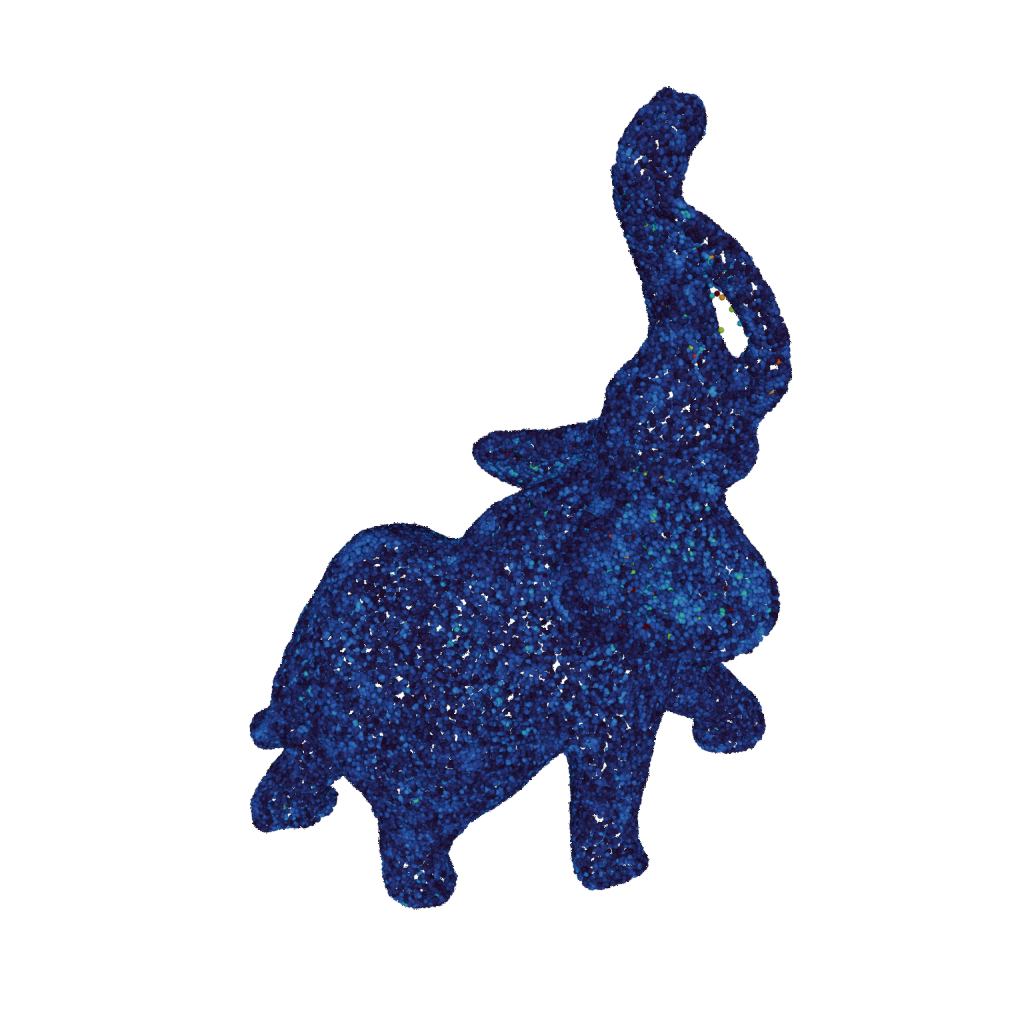}\\[-0.2em]
  \footnotesize Pathnet
\end{minipage}\hspace{-23mm}%
\begin{minipage}[b]{0.27\linewidth}\centering
  \includegraphics[width=\linewidth]{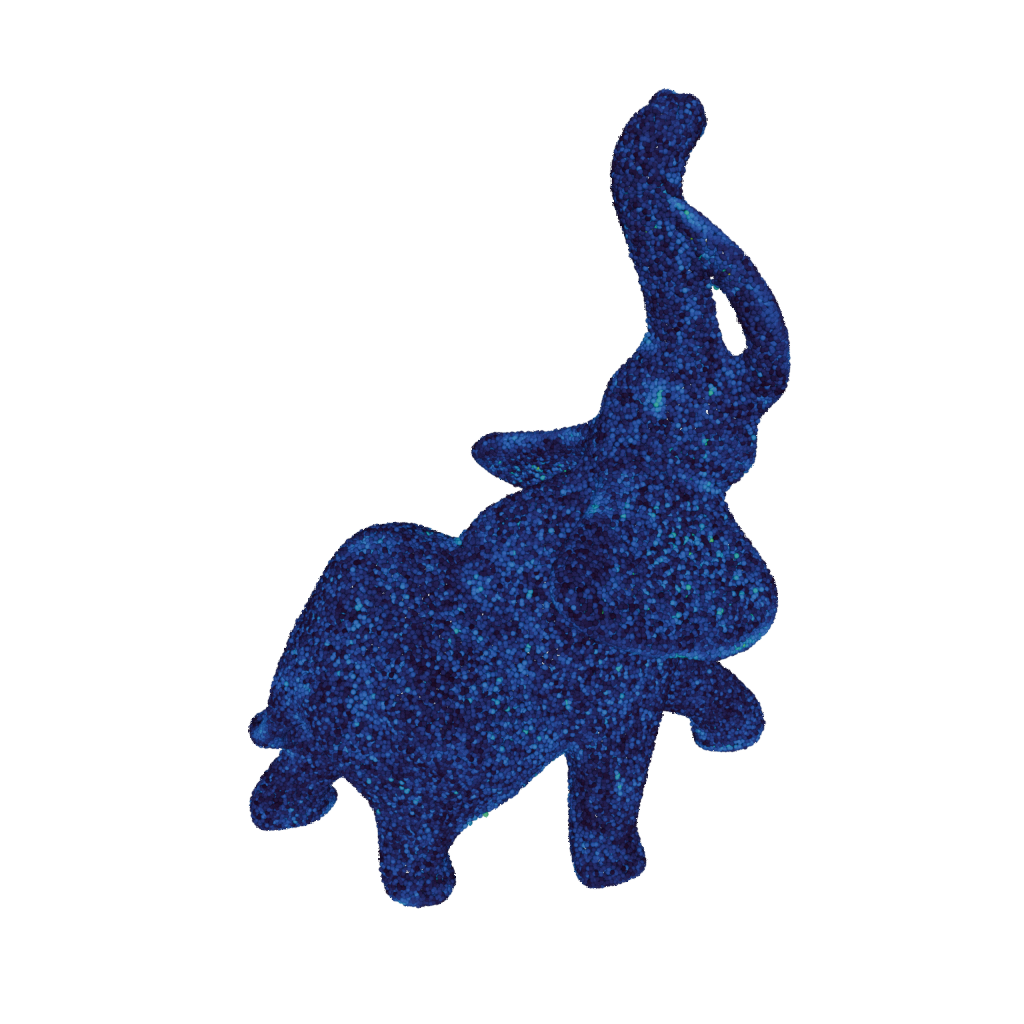}\\[-0.2em]
  \footnotesize StraightPCF
\end{minipage}\hspace{-23mm}%
\begin{minipage}[b]{0.27\linewidth}\centering
  \includegraphics[width=\linewidth]{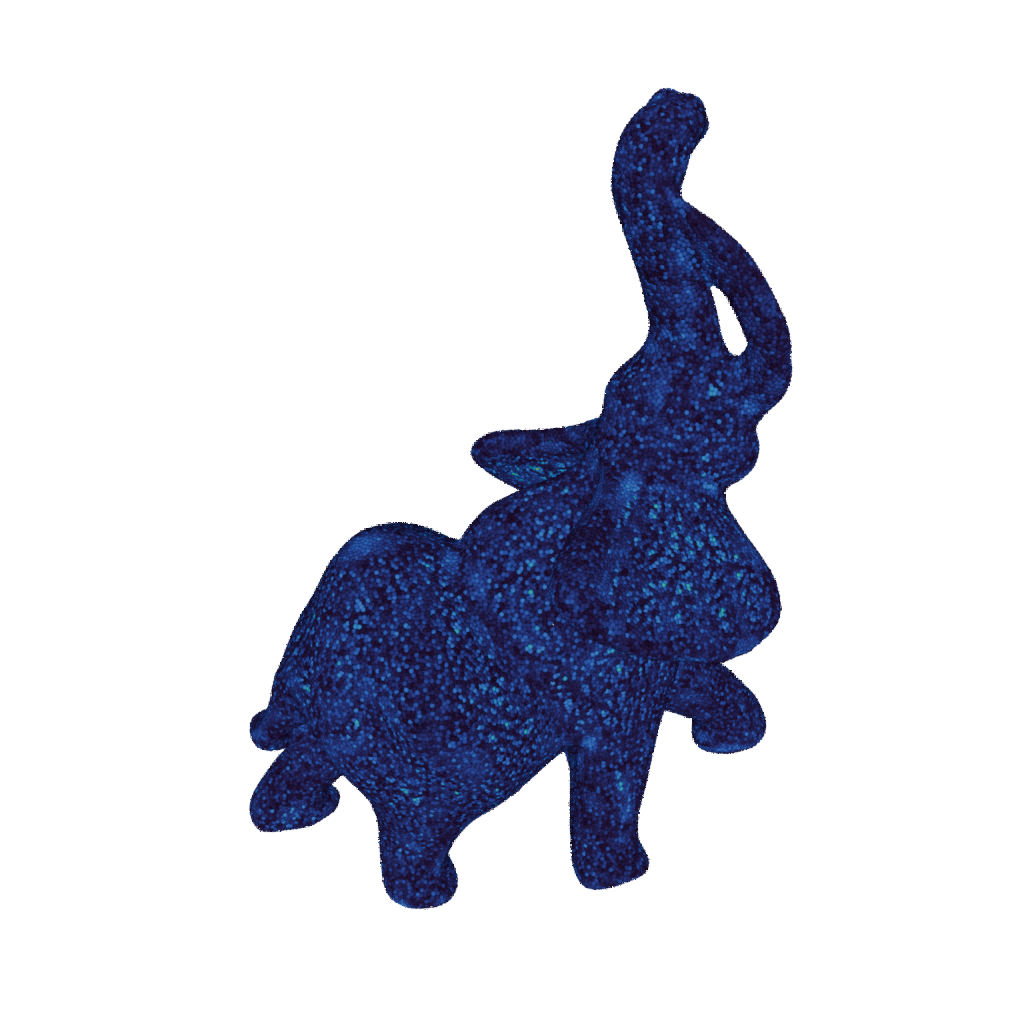}\\[-0.2em]
  \footnotesize PD-LTS
\end{minipage}%
\end{flushleft}

\begin{flushleft}
\begin{minipage}[b]{0.27\linewidth}\centering
  \includegraphics[width=\linewidth]{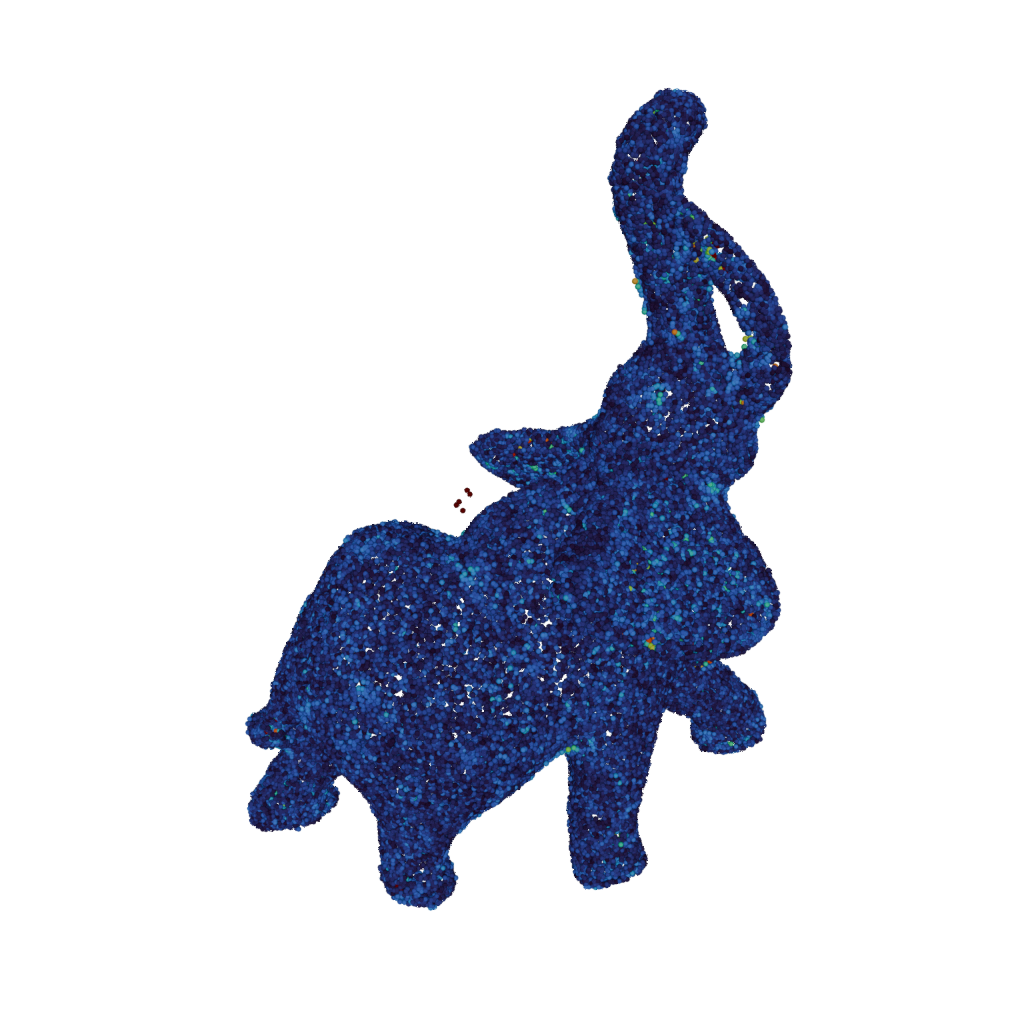}\\[-0.2em]
  \footnotesize MAG
\end{minipage}\hspace{-23mm}%
\begin{minipage}[b]{0.27\linewidth}\centering
  \includegraphics[width=\linewidth]{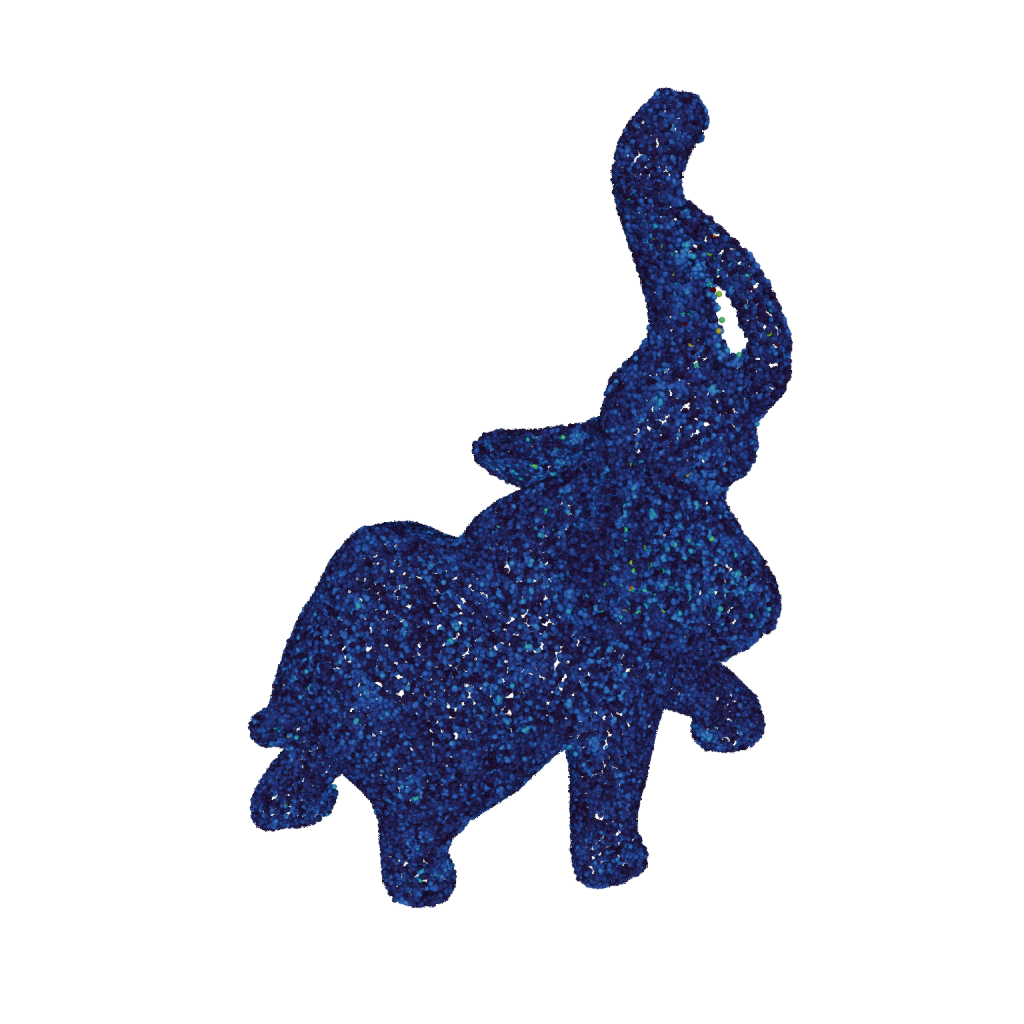}\\[-0.2em]
  \footnotesize PCDNF
\end{minipage}\hspace{-23mm}%
\begin{minipage}[b]{0.27\linewidth}\centering
  \includegraphics[width=\linewidth]{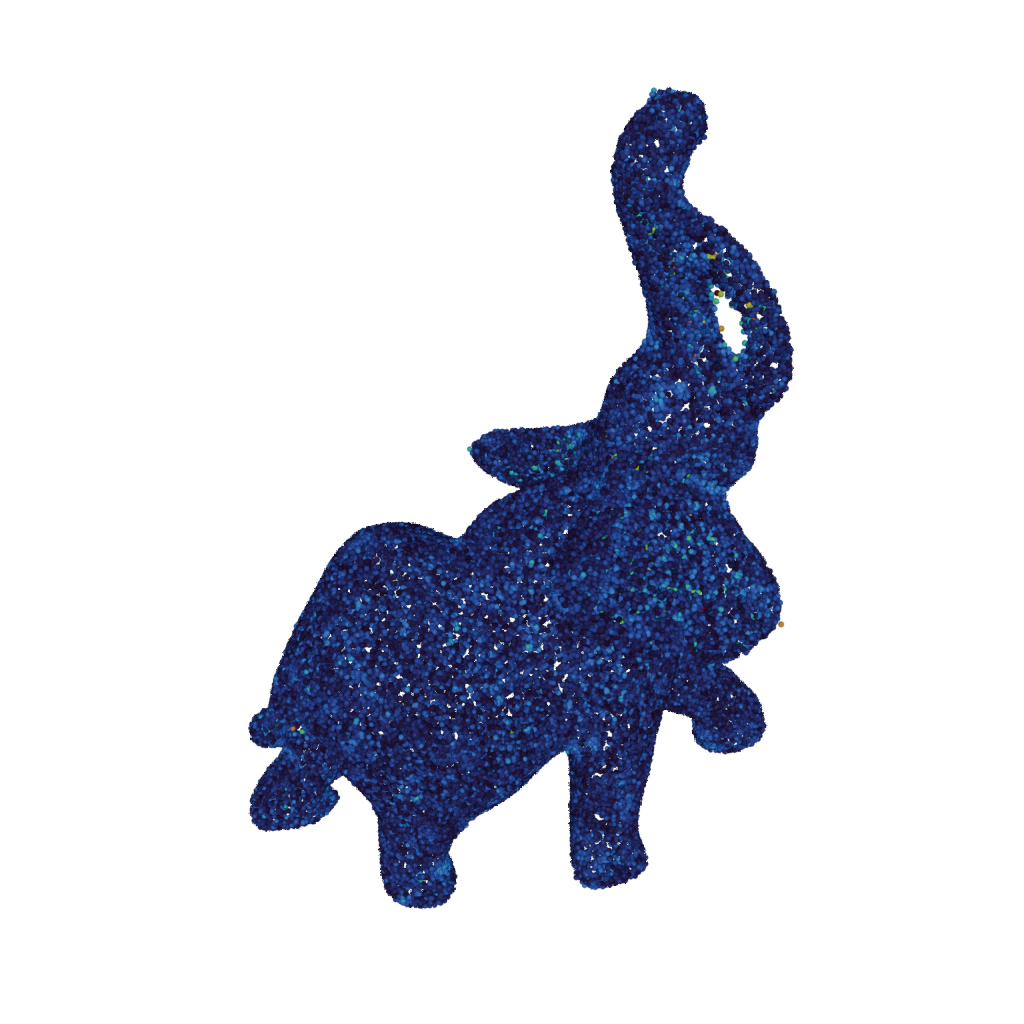}\\[-0.2em]
  \footnotesize MODNet
\end{minipage}\hspace{-23mm}%
\begin{minipage}[b]{0.27\linewidth}\centering
  \includegraphics[width=\linewidth]{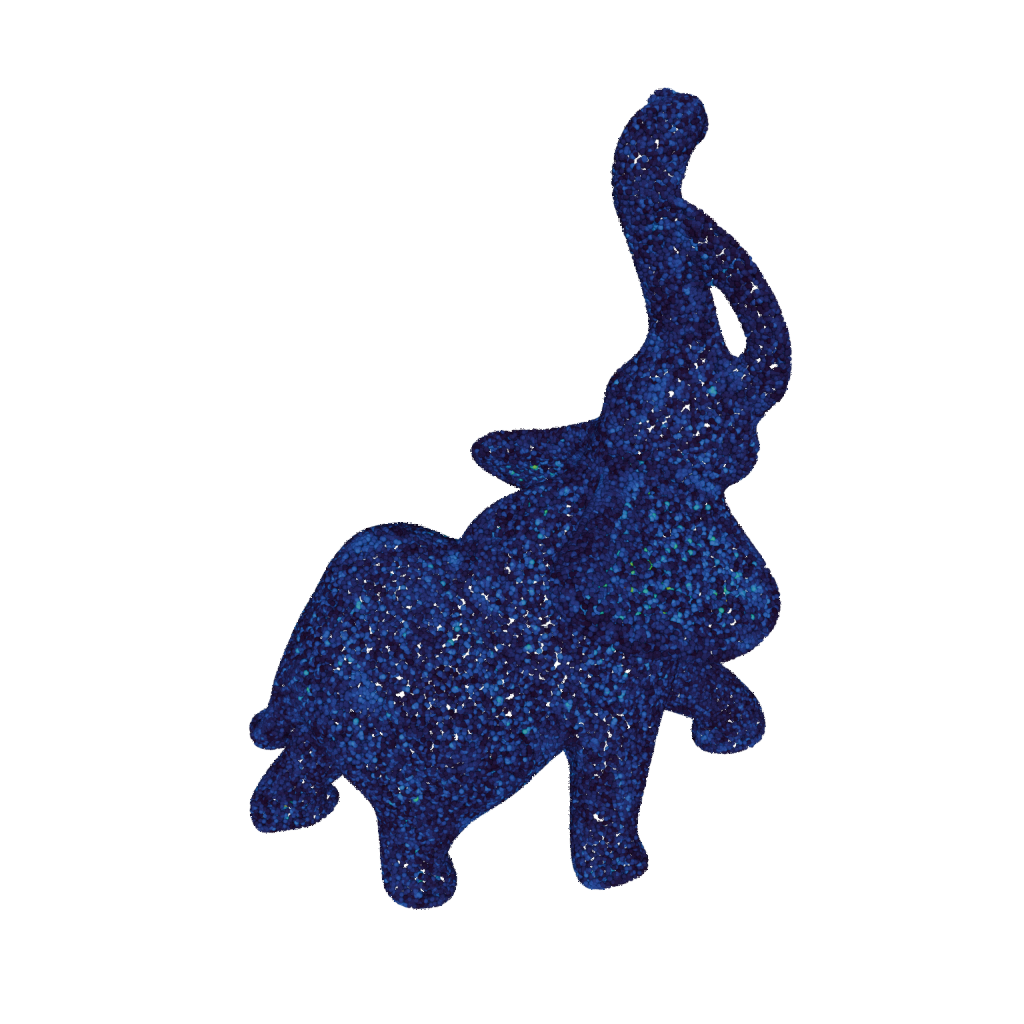}\\[-0.2em]
  \footnotesize IterativePFN
\end{minipage}\hspace{-23mm}%
\begin{minipage}[b]{0.27\linewidth}\centering
  \includegraphics[width=\linewidth]{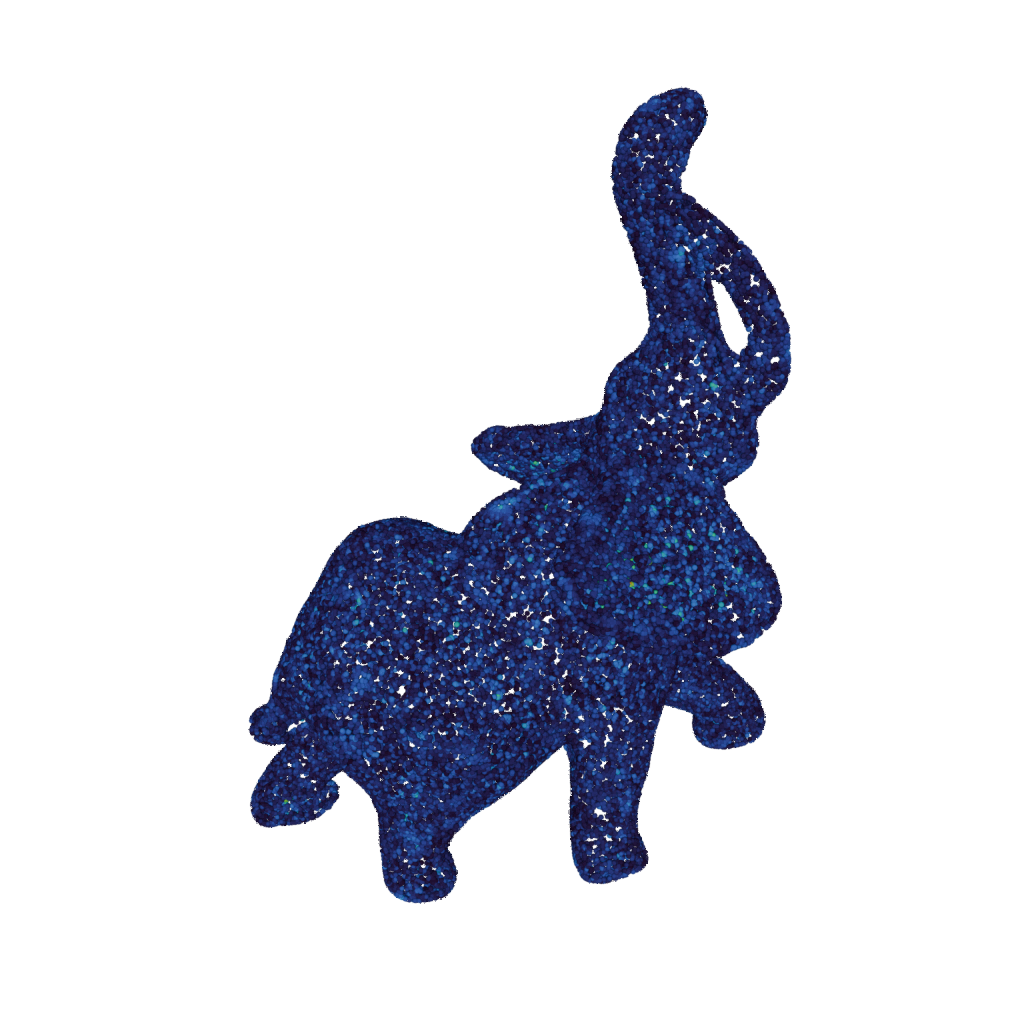}\\[-0.2em]
  \footnotesize DeepPSR
\end{minipage}\hspace{-23mm}%
\begin{minipage}[b]{0.27\linewidth}\centering
  \includegraphics[width=\linewidth]{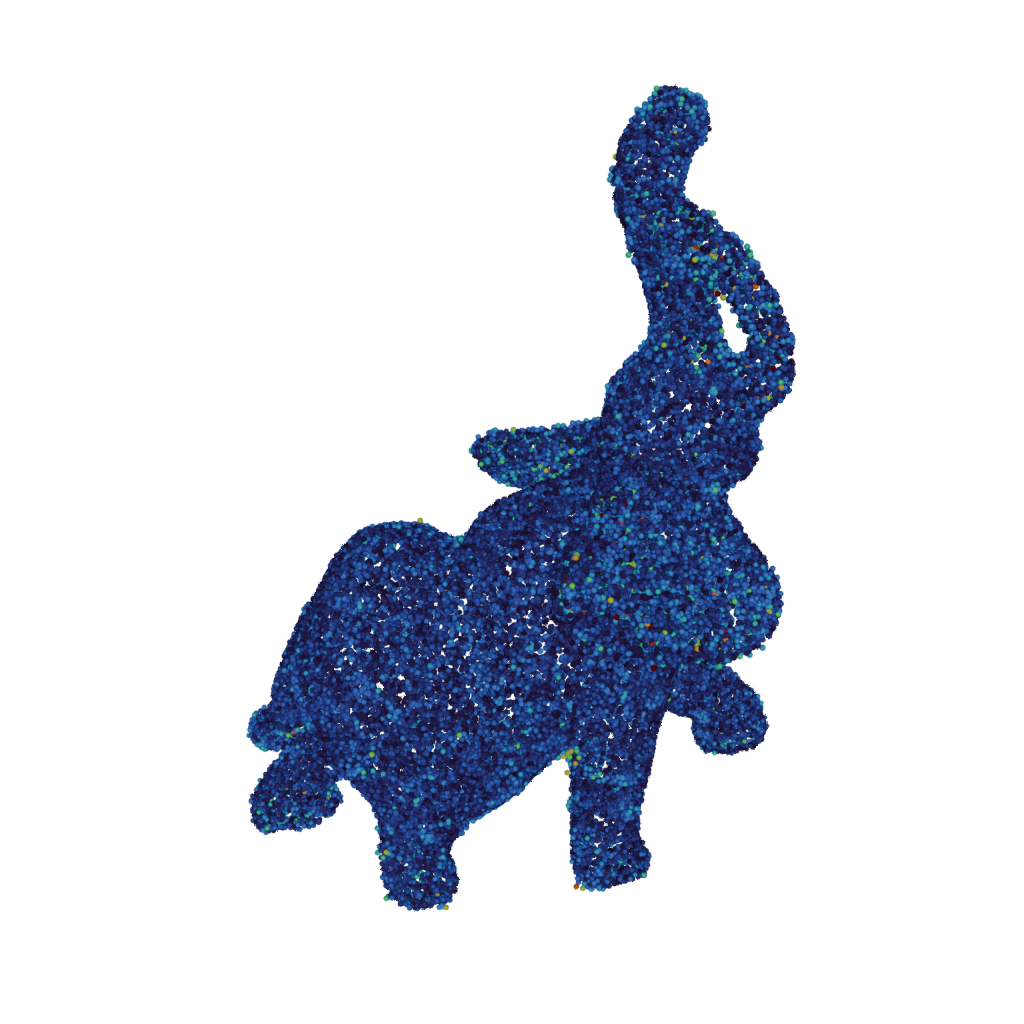}\\[-0.2em]
  \footnotesize RePCD
\end{minipage}%
\end{flushleft}

\begin{flushleft}
\begin{minipage}[b]{0.27\linewidth}\centering
  \includegraphics[width=\linewidth]{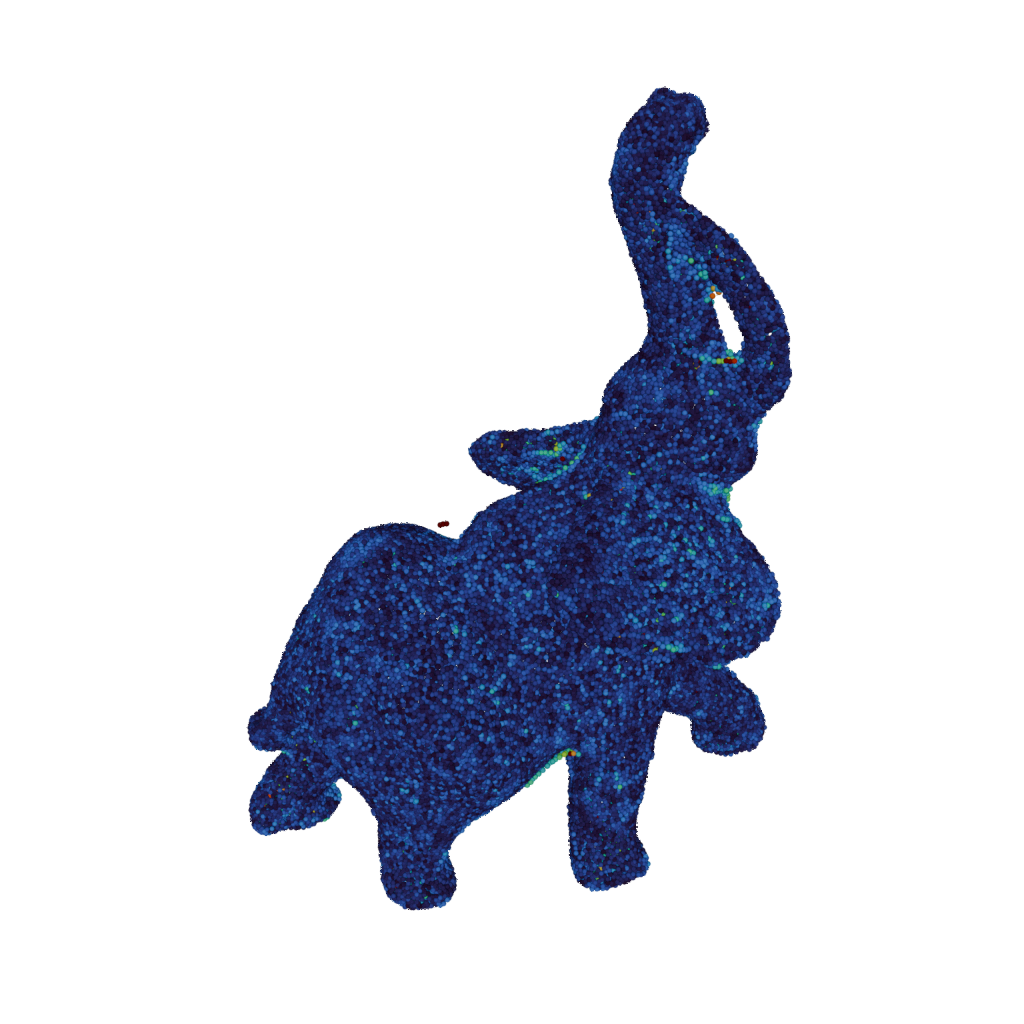}\\[-0.2em]
  \footnotesize PDFlow
\end{minipage}\hspace{-23mm}%
\begin{minipage}[b]{0.27\linewidth}\centering
  \includegraphics[width=\linewidth]{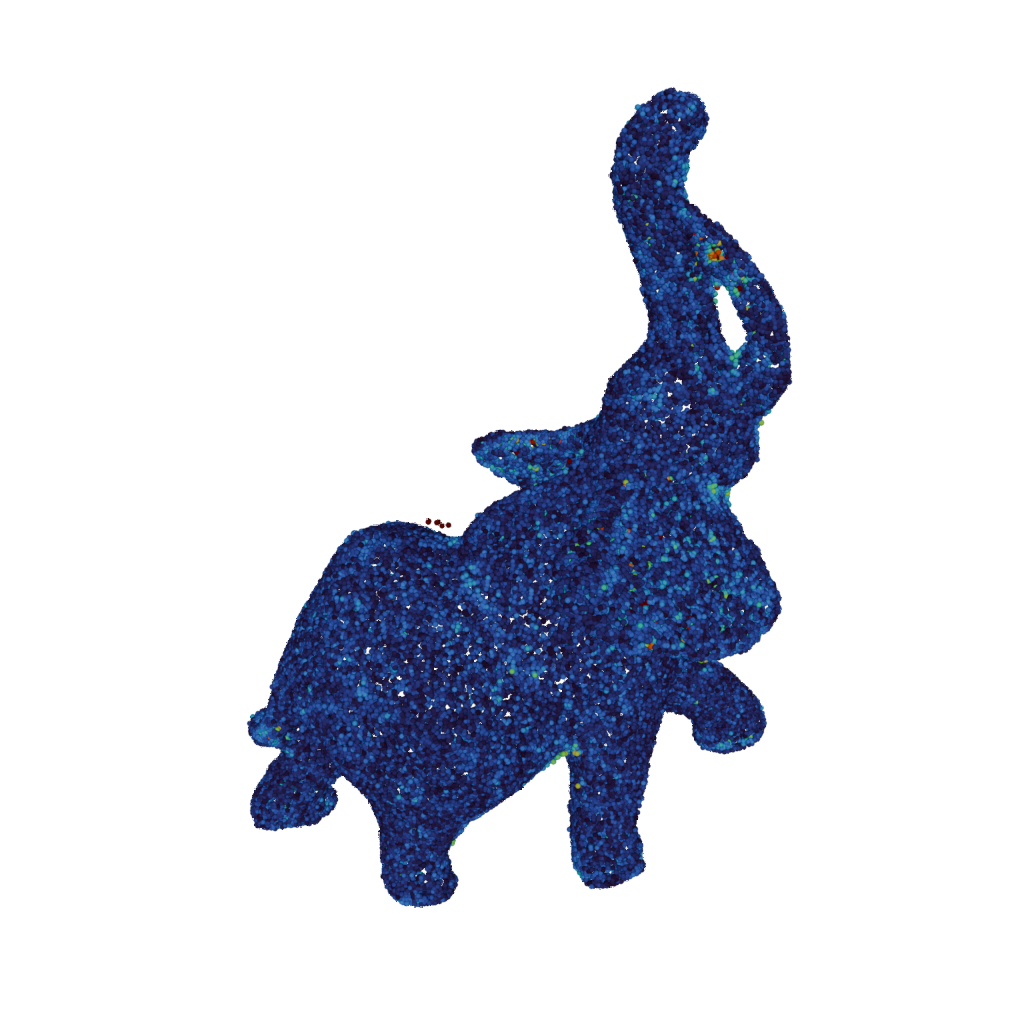}\\[-0.2em]
  \footnotesize ScoreDenoise
\end{minipage}\hspace{-23mm}%
\begin{minipage}[b]{0.27\linewidth}\centering
  \includegraphics[width=\linewidth]{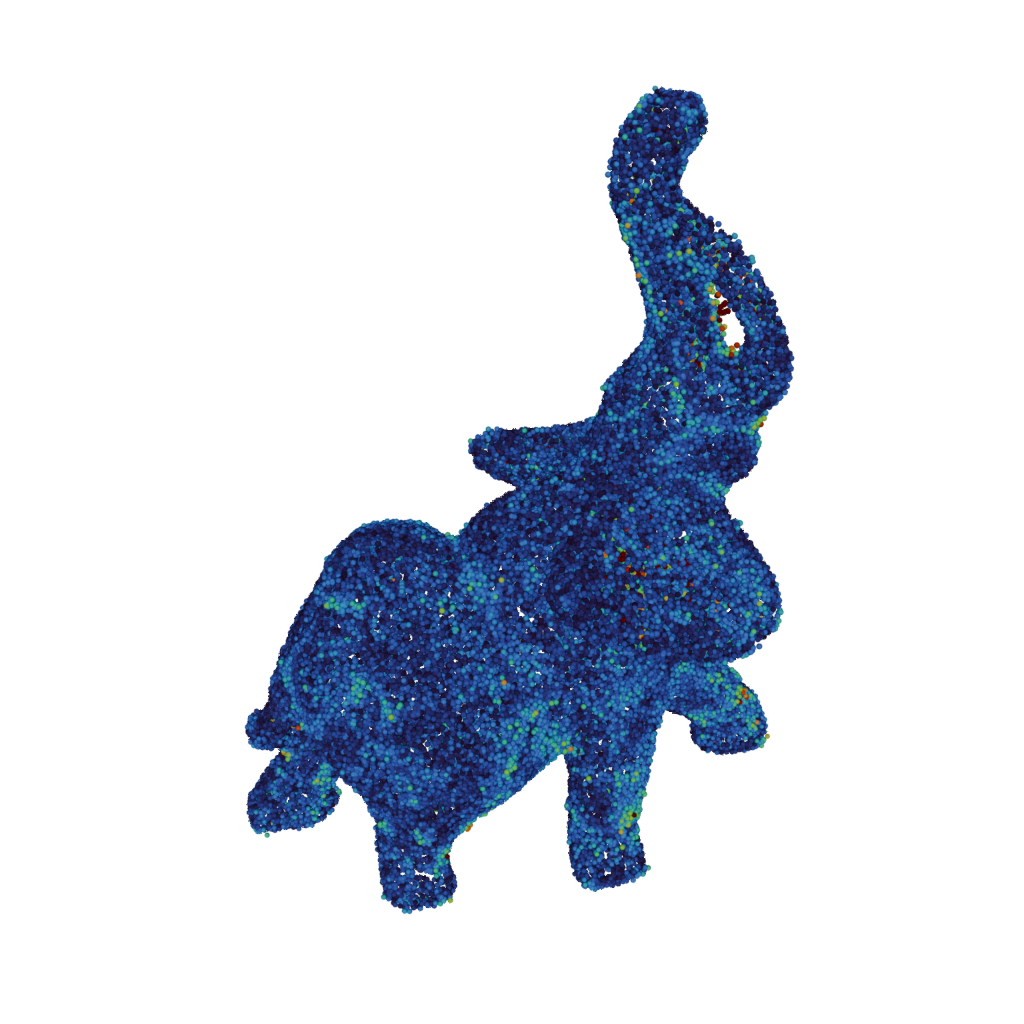}\\[-0.2em]
  \footnotesize DMRDenoise
\end{minipage}\hspace{-23mm}%
\begin{minipage}[b]{0.27\linewidth}\centering
  \includegraphics[width=\linewidth]{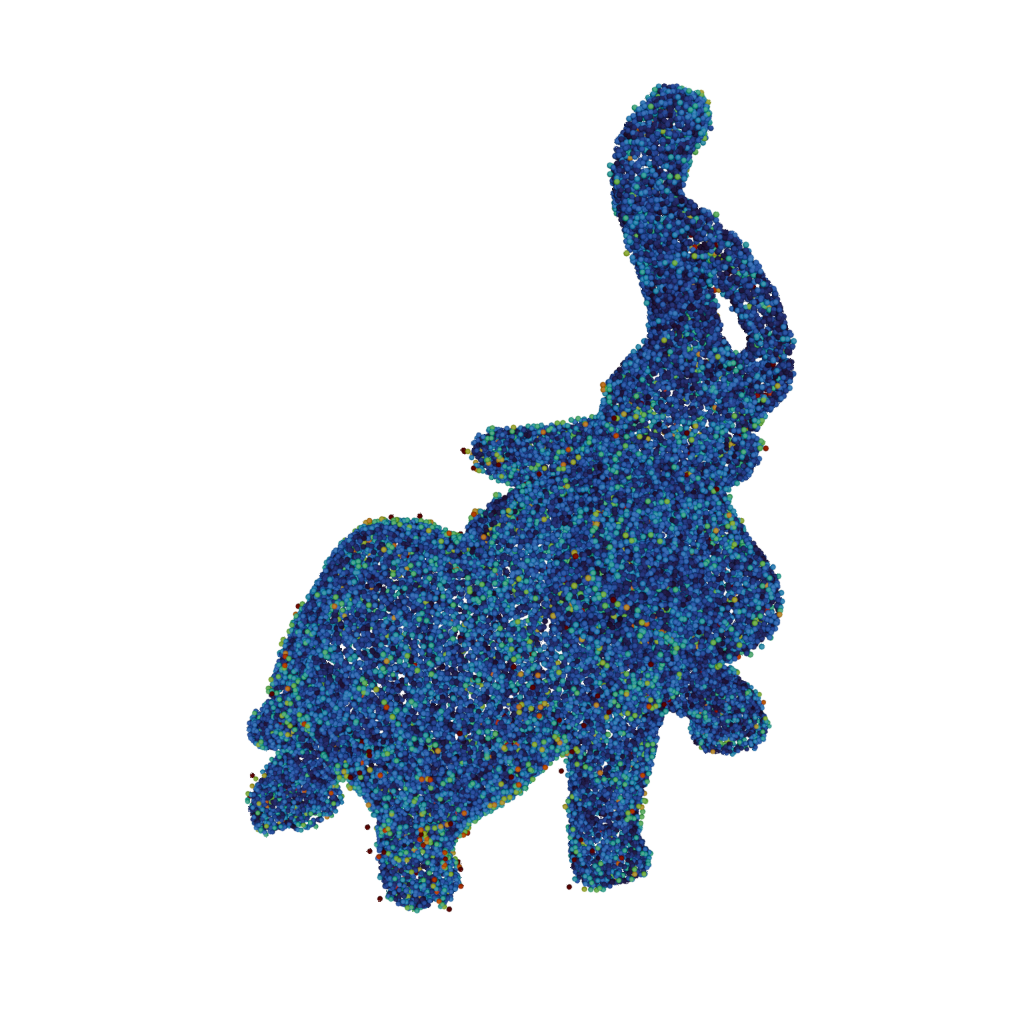}\\[-0.2em]
  \footnotesize GPDNet
\end{minipage}\hspace{-23mm}%
\begin{minipage}[b]{0.27\linewidth}\centering
  \includegraphics[width=\linewidth]{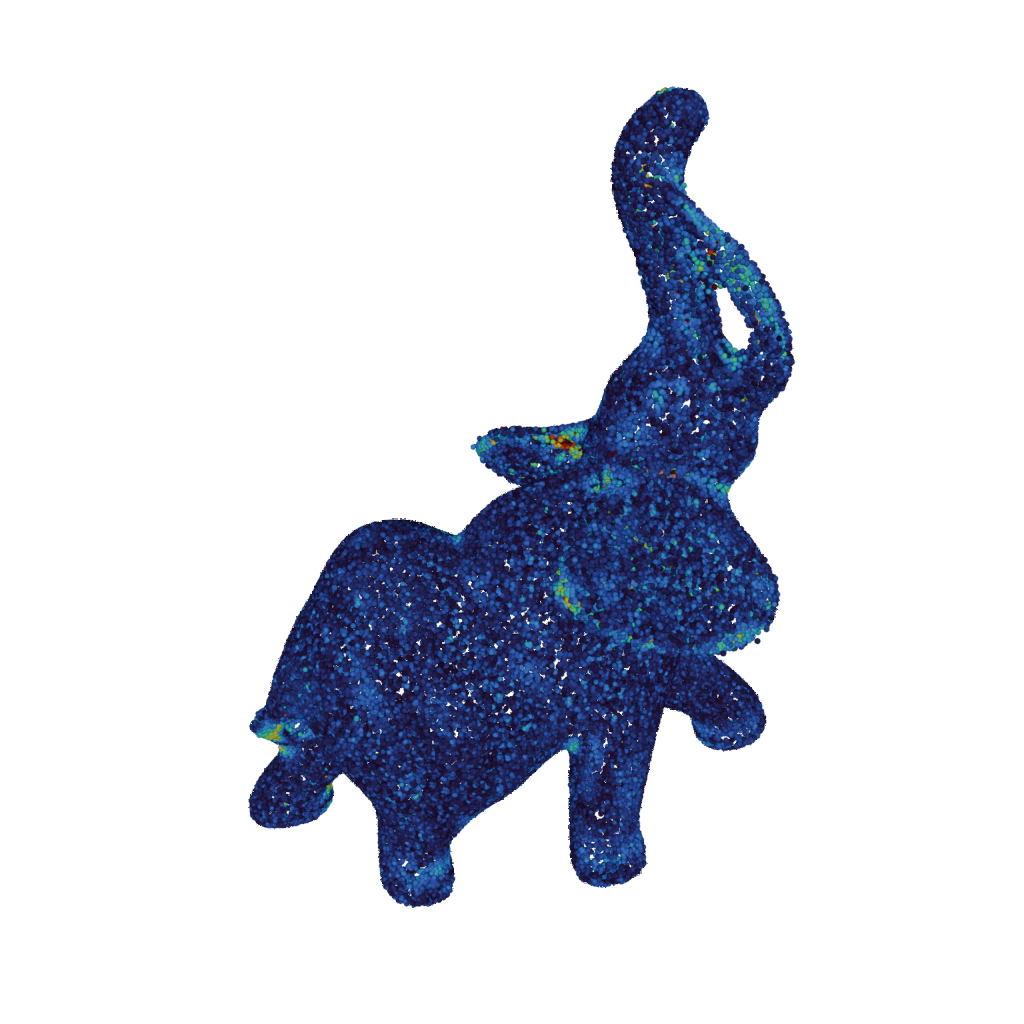}\\[-0.2em]
  \footnotesize Pointfilter
\end{minipage}\hspace{-23mm}%
\begin{minipage}[b]{0.27\linewidth}\centering
  \includegraphics[width=\linewidth]{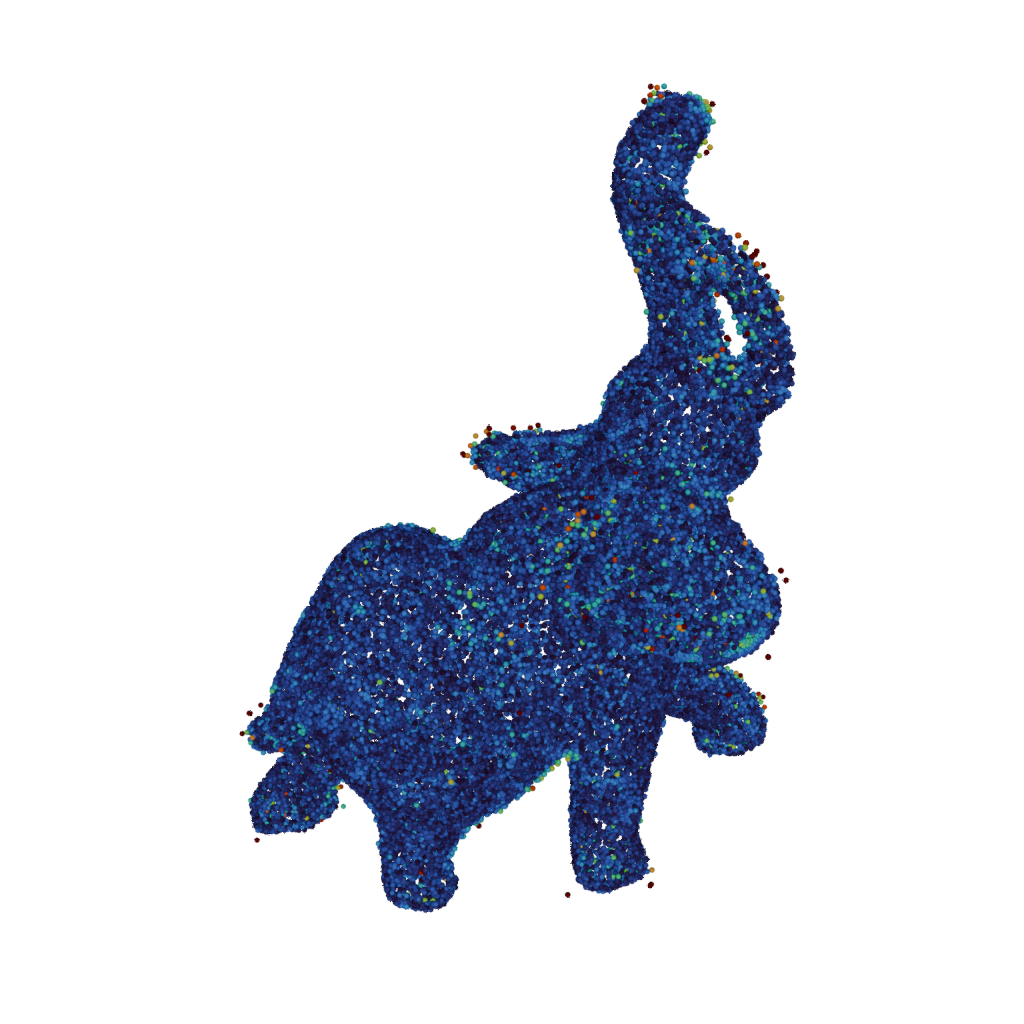}\\[-0.2em]
  \footnotesize PointCleanNet
\end{minipage}%
\end{flushleft}

\vspace{0.3em}
\textbf{Unsupervised}\par\vspace{0.1em}

\begin{flushleft}
\begin{minipage}[b]{0.27\linewidth}\centering
  \includegraphics[width=\linewidth]{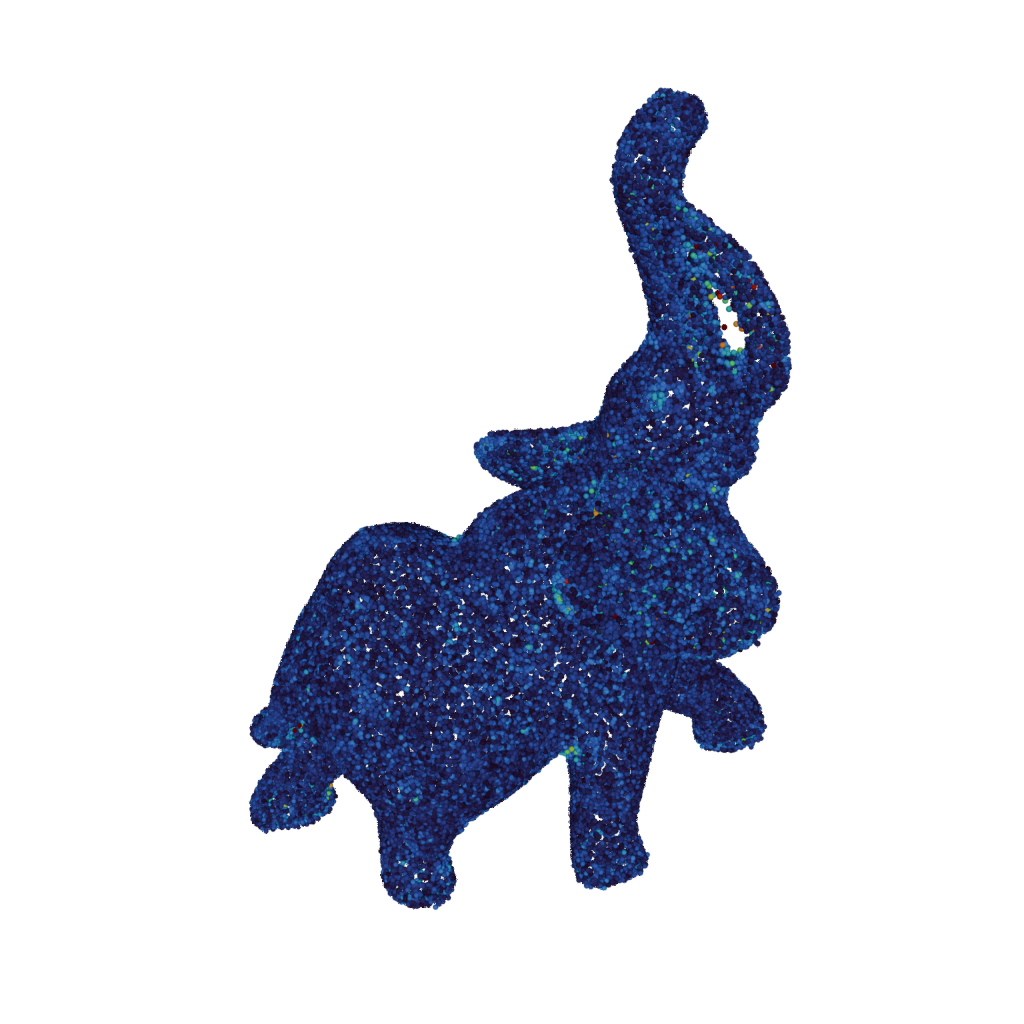}\\[-0.2em]
  \footnotesize Noise4Denoise
\end{minipage}\hspace{-23mm}%
\begin{minipage}[b]{0.27\linewidth}\centering
  \includegraphics[width=\linewidth]{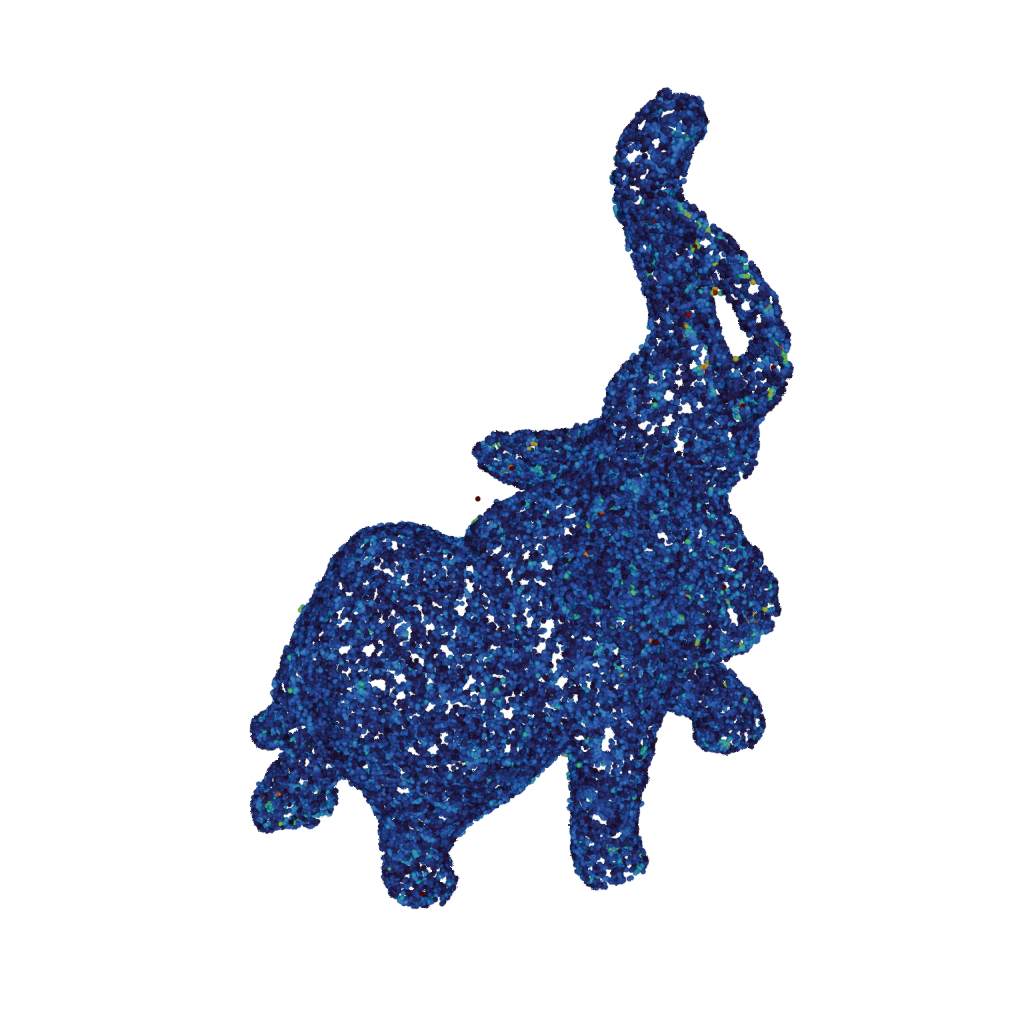}\\[-0.2em]
  \footnotesize Un-Score
\end{minipage}\hspace{-23mm}%
\begin{minipage}[b]{0.27\linewidth}\centering
  \includegraphics[width=\linewidth]{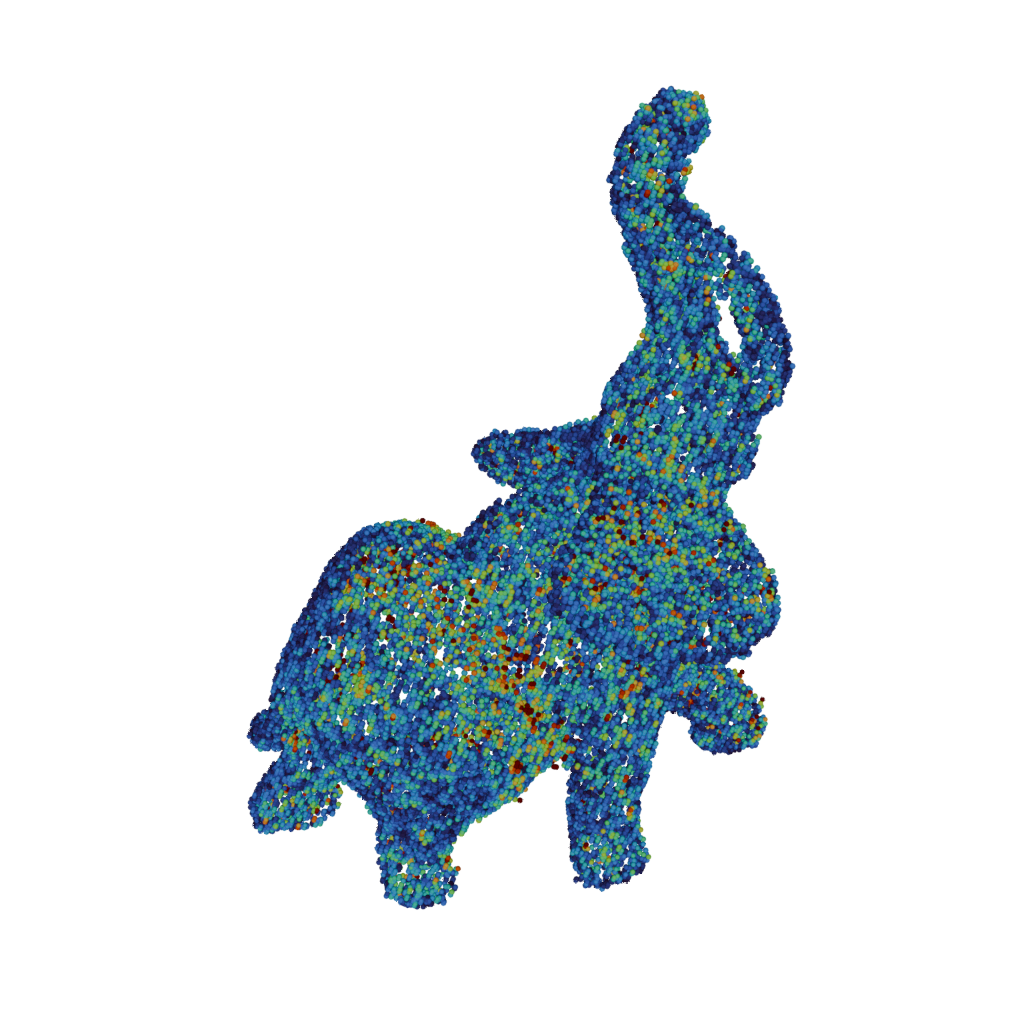}\\[-0.2em]
  \footnotesize Un-DMR
\end{minipage}\hspace{-23mm}%
\begin{minipage}[b]{0.27\linewidth}\centering
  \includegraphics[width=\linewidth]{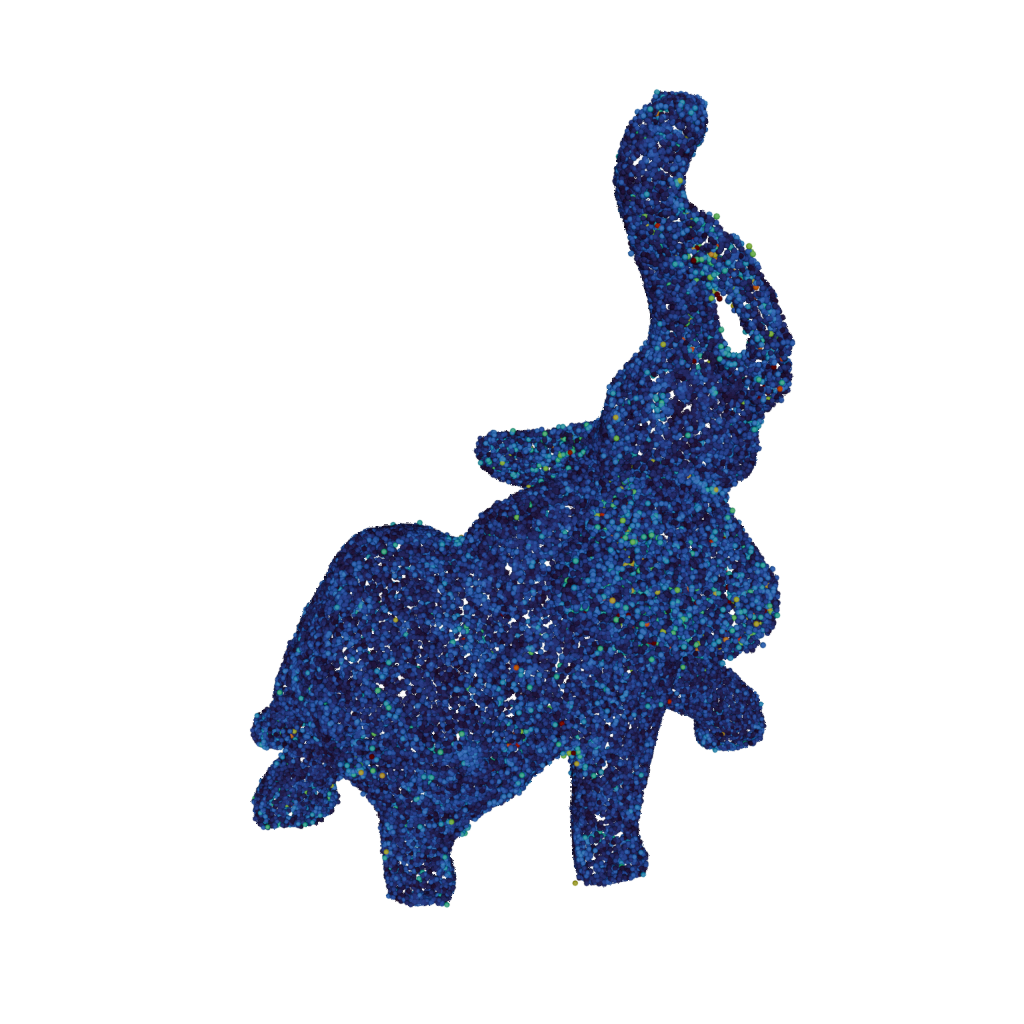}\\[-0.2em]
  \footnotesize TotalDenoising
\end{minipage}\hspace{-23mm}%
\begin{minipage}[b]{0.27\linewidth}\centering
  \includegraphics[width=\linewidth]{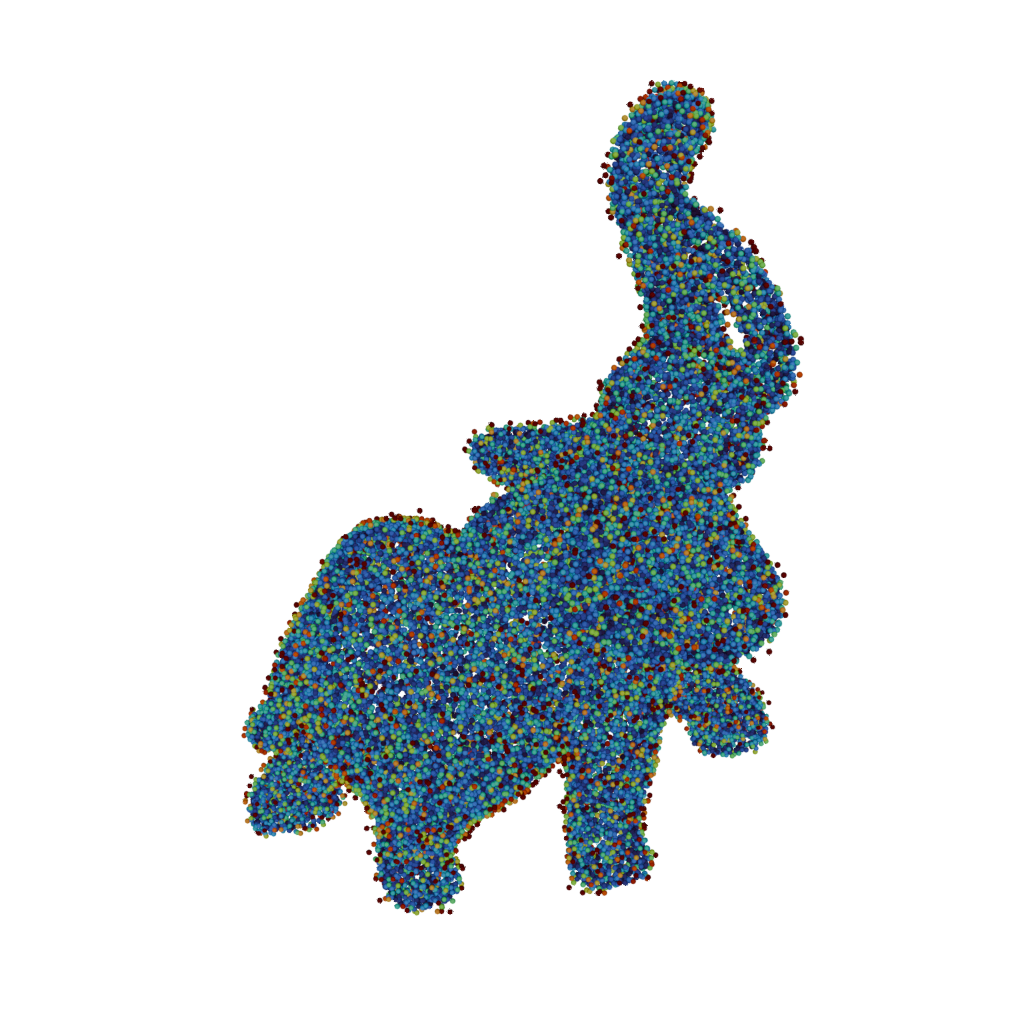}\\[-0.2em]
  \footnotesize Noisy
\end{minipage}\hspace{-23mm}%
\begin{minipage}[b]{0.28\linewidth}\centering
  \includegraphics[height=2.8cm]{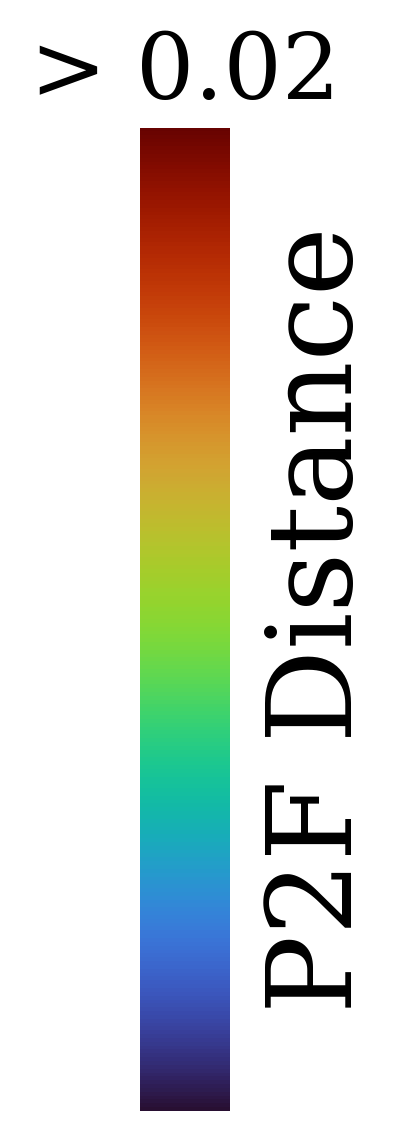}\\[-0.2em]
  \footnotesize
\end{minipage}
\end{flushleft}

\caption{Visual comparison of denoising errors on the \texttt{elephant} shape under Gaussian noise ($\sigma = 0.01$).}
\label{fig:elephant_error_gallery}
\end{figure*}

\begin{figure*}[htbp]
\centering

\begin{flushleft}

\begin{minipage}[t]{0.25\linewidth}\centering
  \begin{tikzpicture}[baseline]
    \node[inner sep=0] (img)
      {\includegraphics[width=\linewidth]{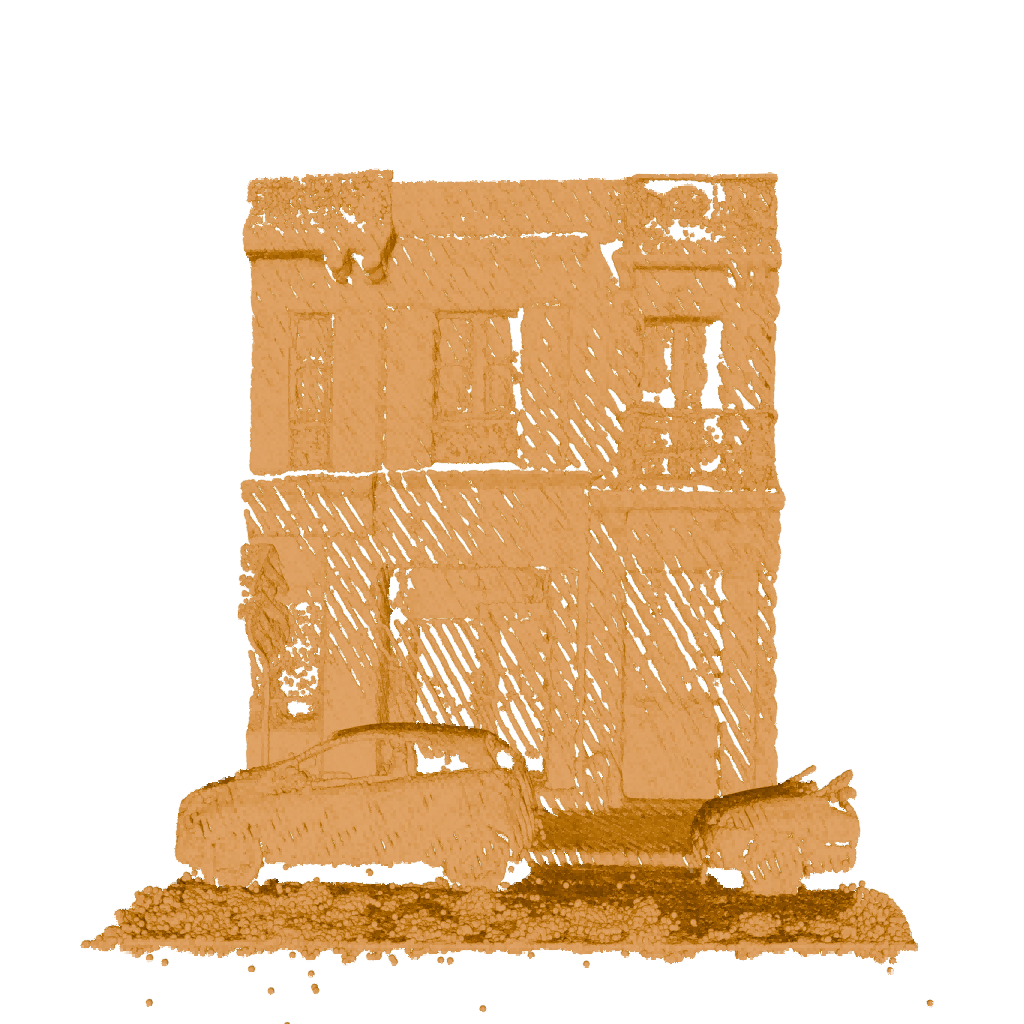}};
    \node[inner sep=0, anchor=north west] (detail)
      at ([xshift=2pt,yshift=-4pt]img.north west)
      {\includegraphics[width=0.50\linewidth]{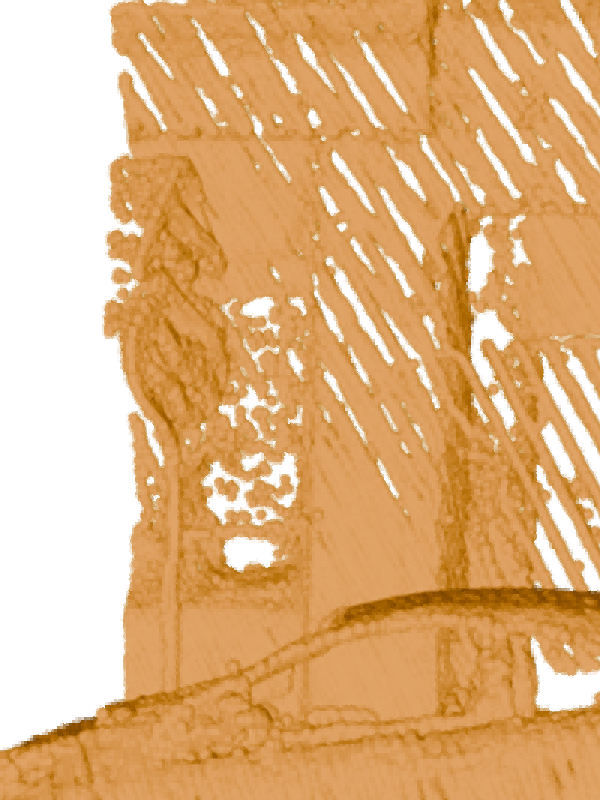}};
    \draw[line width=0.5pt] (detail.north west) rectangle (detail.south east);
  \end{tikzpicture}\\[-0.3em]
  \footnotesize 3DMambaIPF
\end{minipage}\hspace{-11.5mm}%
%
\begin{minipage}[t]{0.25\linewidth}\centering
  \begin{tikzpicture}[baseline]
    \node[inner sep=0] (img)
      {\includegraphics[width=\linewidth]{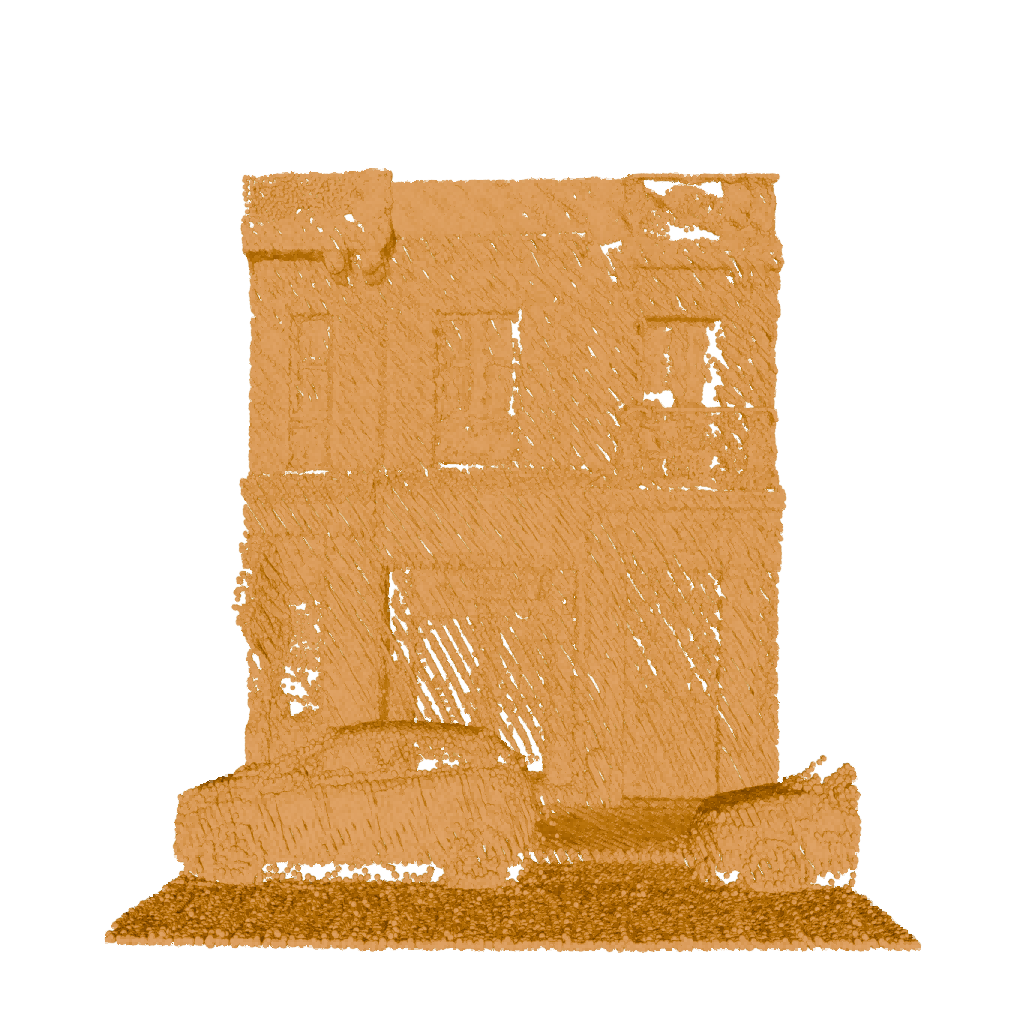}};
    \node[inner sep=0, anchor=north west] (detail)
      at ([xshift=2pt,yshift=-4pt]img.north west)
      {\includegraphics[width=0.50\linewidth]{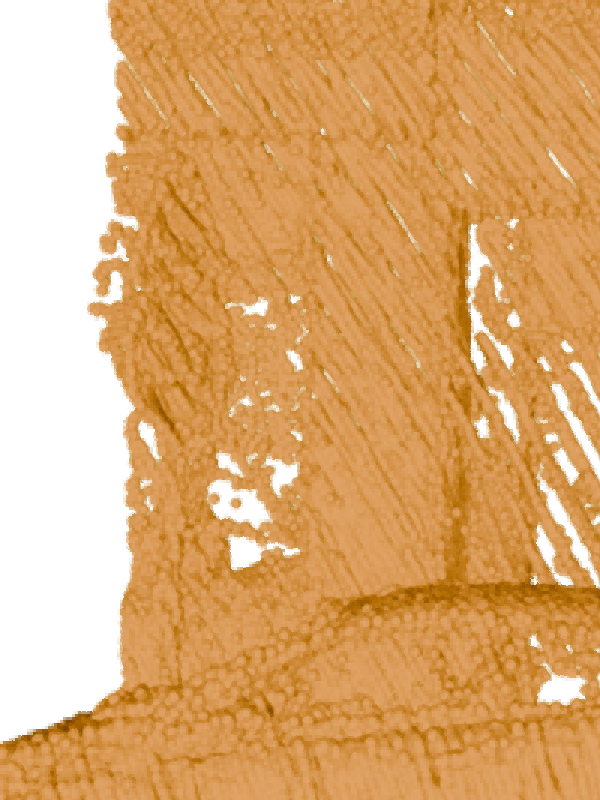}};
    \draw[line width=0.5pt] (detail.north west) rectangle (detail.south east);
  \end{tikzpicture}\\[-0.3em]
  \footnotesize ASDN
\end{minipage}\hspace{-11.5mm}%
%
\begin{minipage}[t]{0.25\linewidth}\centering
  \begin{tikzpicture}[baseline]
    \node[inner sep=0] (img)
      {\includegraphics[width=\linewidth]{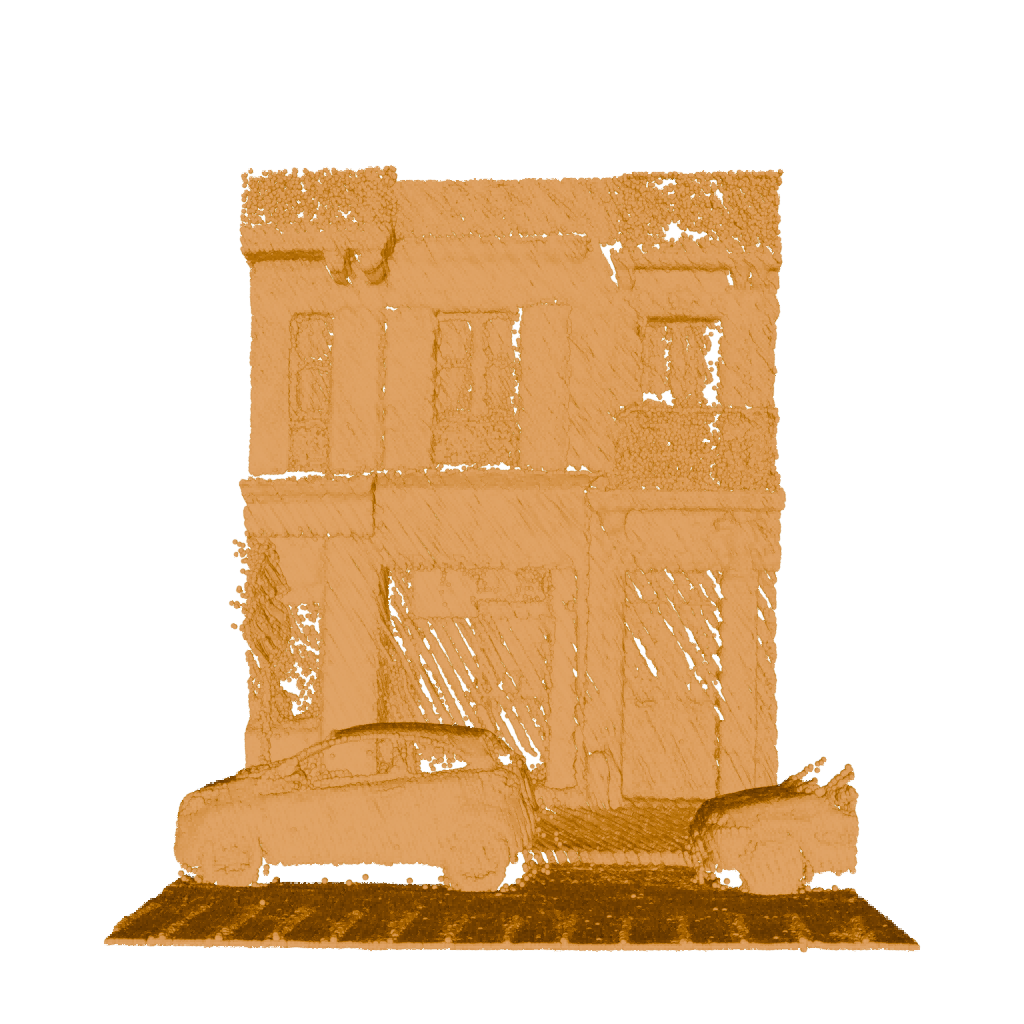}};
    \node[inner sep=0, anchor=north west] (detail)
      at ([xshift=2pt,yshift=-4pt]img.north west)
      {\includegraphics[width=0.50\linewidth]{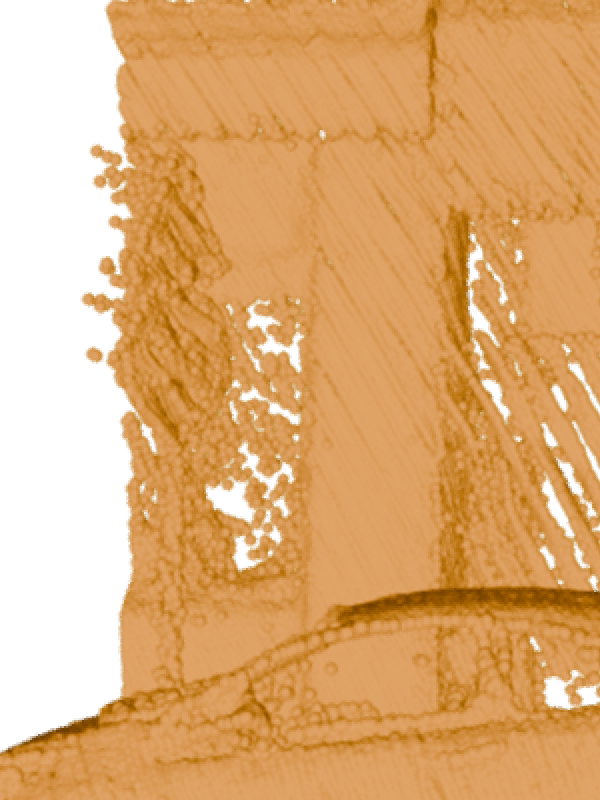}};
    \draw[line width=0.5pt] (detail.north west) rectangle (detail.south east);
  \end{tikzpicture}\\[-0.3em]
  \footnotesize P2P-Bridge
\end{minipage}\hspace{-11.5mm}%
%
\begin{minipage}[t]{0.25\linewidth}\centering
  \begin{tikzpicture}[baseline]
    \node[inner sep=0] (img)
      {\includegraphics[width=\linewidth]{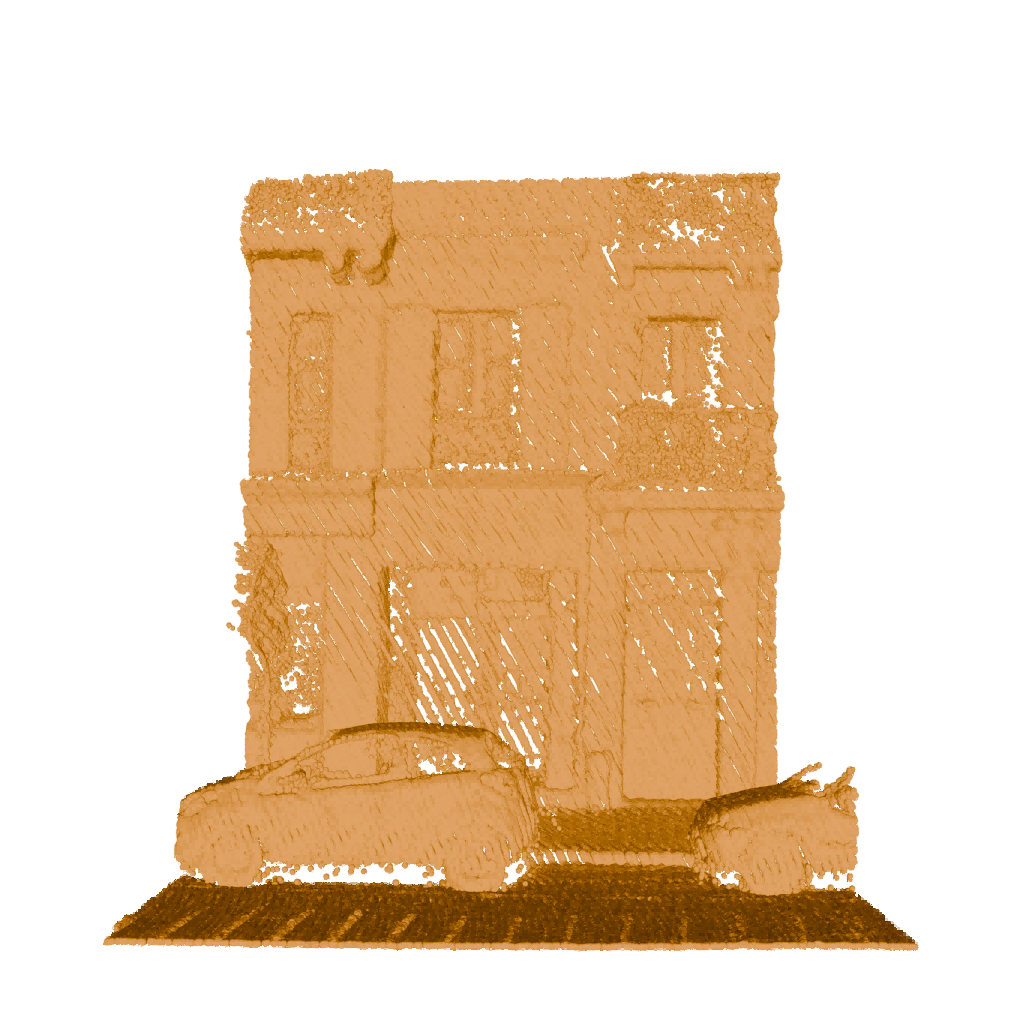}};
    \node[inner sep=0, anchor=north west] (detail)
      at ([xshift=2pt,yshift=-4pt]img.north west)
      {\includegraphics[width=0.50\linewidth]{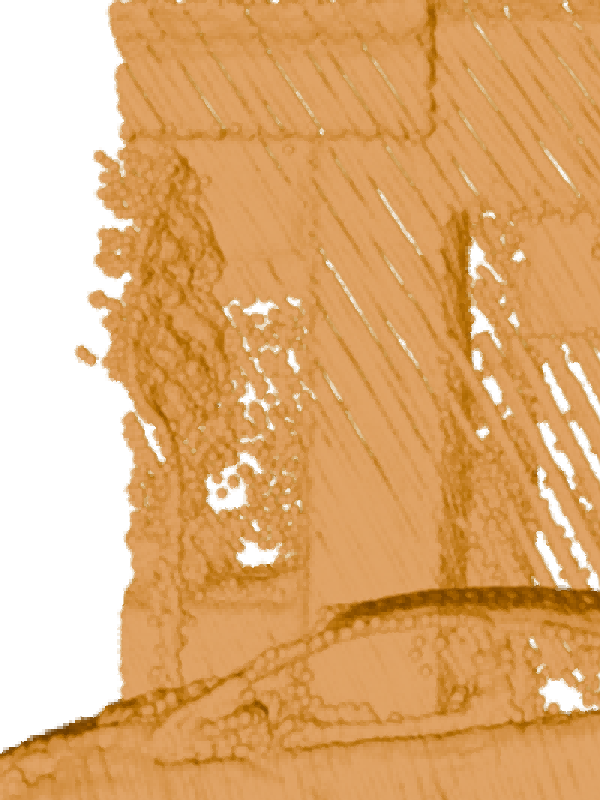}};
    \draw[line width=0.5pt] (detail.north west) rectangle (detail.south east);
  \end{tikzpicture}\\[-0.3em]
  \footnotesize Pathnet
\end{minipage}\hspace{-11.5mm}%
%
\begin{minipage}[t]{0.25\linewidth}\centering
  \begin{tikzpicture}[baseline]
    \node[inner sep=0] (img)
      {\includegraphics[width=\linewidth]{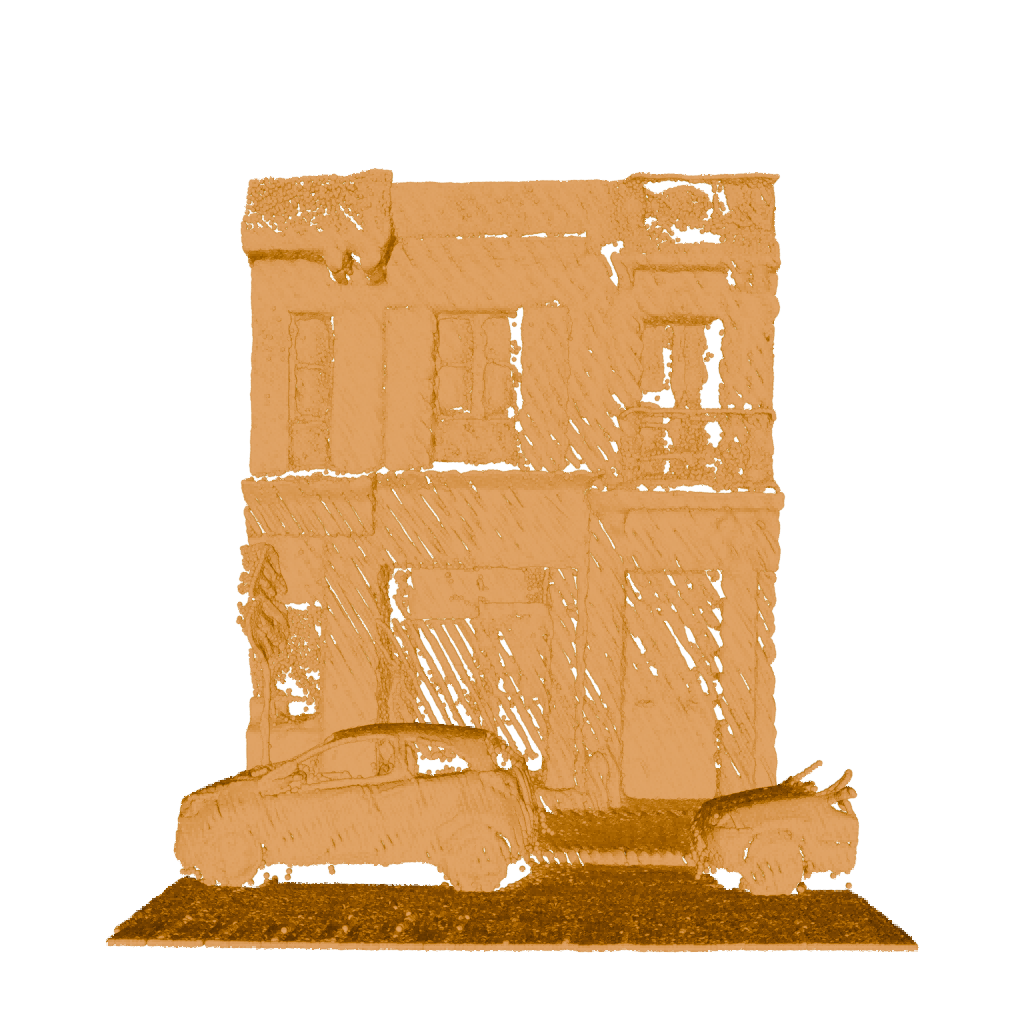}};
    \node[inner sep=0, anchor=north west] (detail)
      at ([xshift=2pt,yshift=-4pt]img.north west)
      {\includegraphics[width=0.50\linewidth]{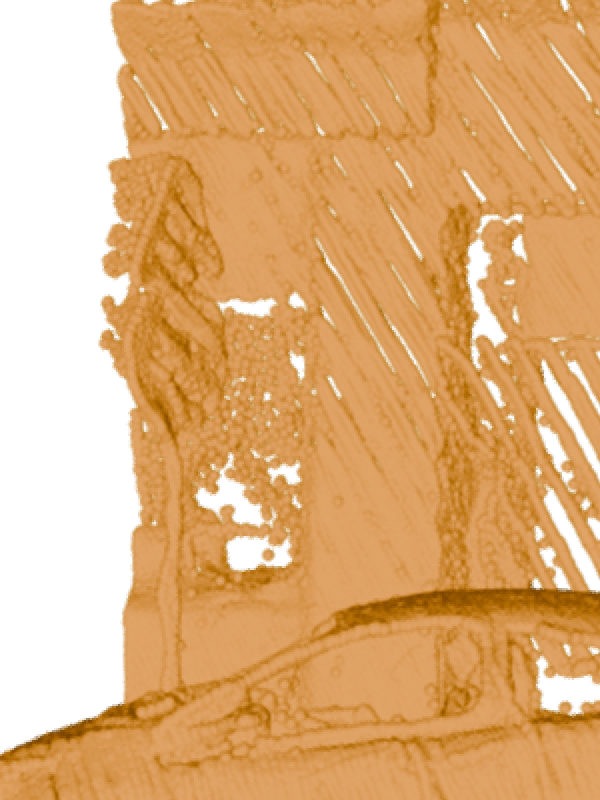}};
    \draw[line width=0.5pt] (detail.north west) rectangle (detail.south east);
  \end{tikzpicture}\\[-0.3em]
  \footnotesize StraightPCF
\end{minipage}%
\end{flushleft}

\vspace{-1.5em}

\begin{flushleft}

\begin{minipage}[t]{0.25\linewidth}\centering
  \begin{tikzpicture}[baseline]
    \node[inner sep=0] (img)
      {\includegraphics[width=\linewidth]{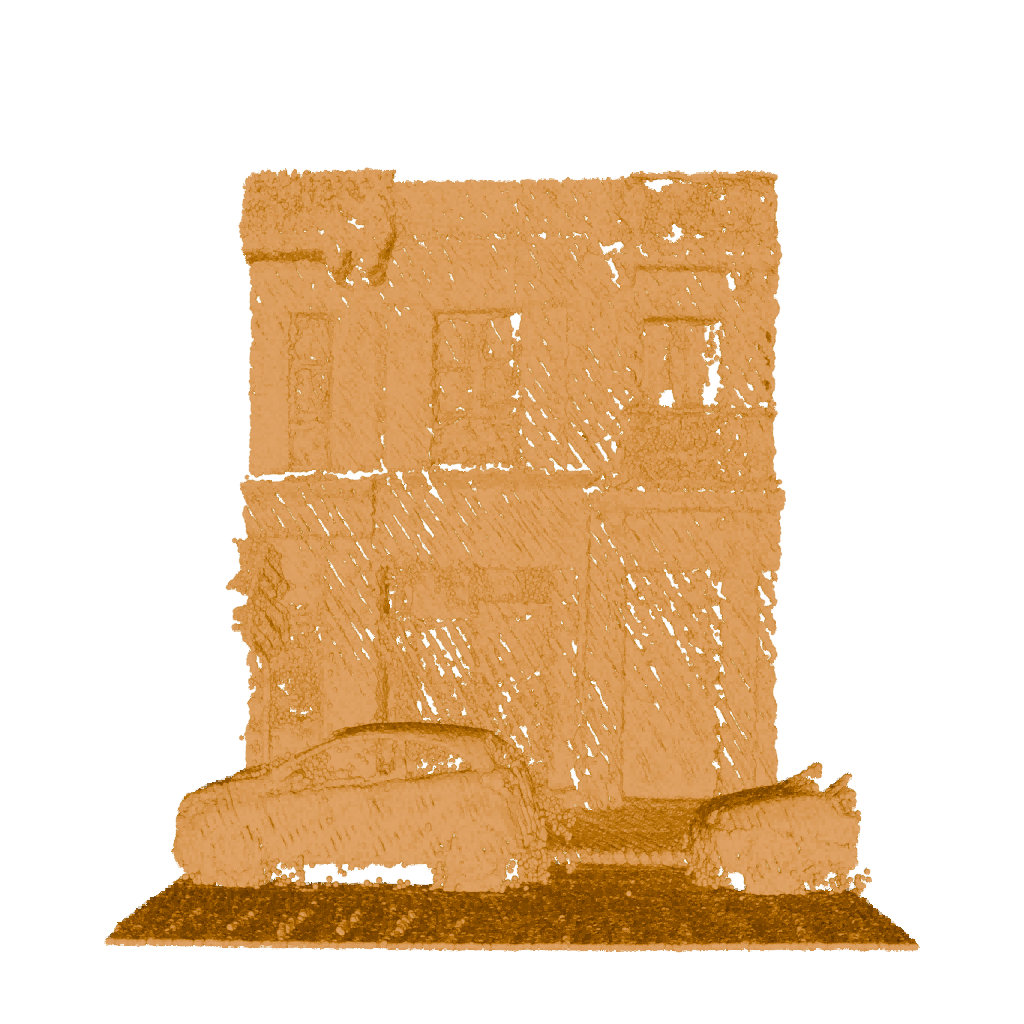}};
    \node[inner sep=0, anchor=north west] (detail)
      at ([xshift=2pt,yshift=-4pt]img.north west)
      {\includegraphics[width=0.50\linewidth]{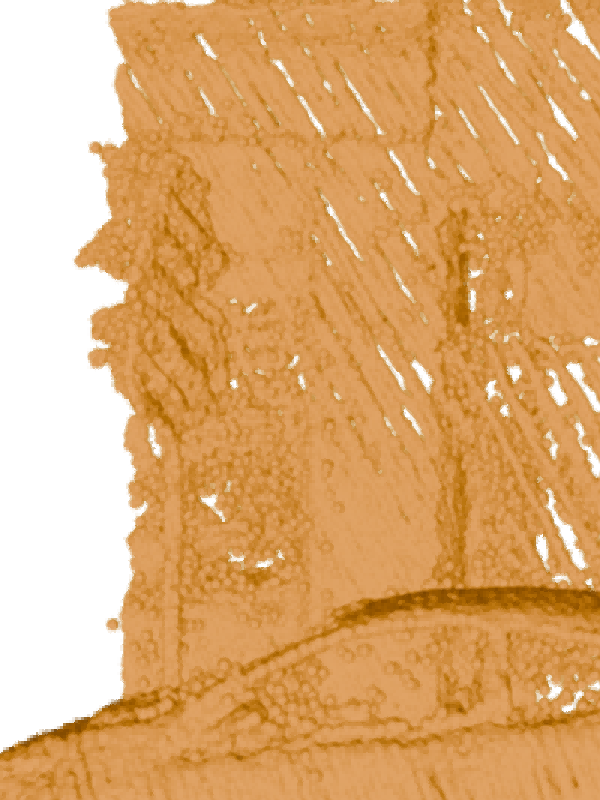}};
    \draw[line width=0.5pt] (detail.north west) rectangle (detail.south east);
  \end{tikzpicture}\\[-0.3em]
  \footnotesize MAG
\end{minipage}\hspace{-11.5mm}%
%
\begin{minipage}[t]{0.25\linewidth}\centering
  \begin{tikzpicture}[baseline]
    \node[inner sep=0] (img)
      {\includegraphics[width=\linewidth]{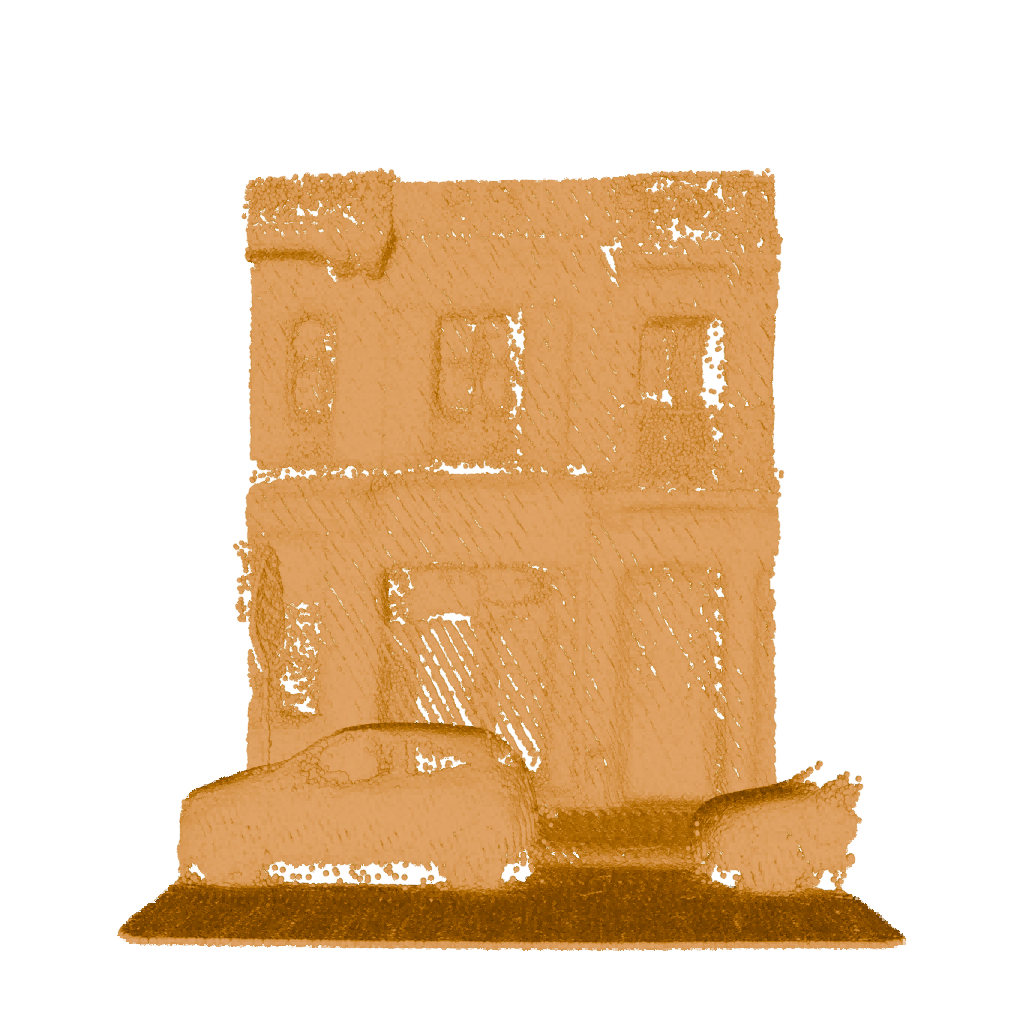}};
    \node[inner sep=0, anchor=north west] (detail)
      at ([xshift=2pt,yshift=-4pt]img.north west)
      {\includegraphics[width=0.50\linewidth]{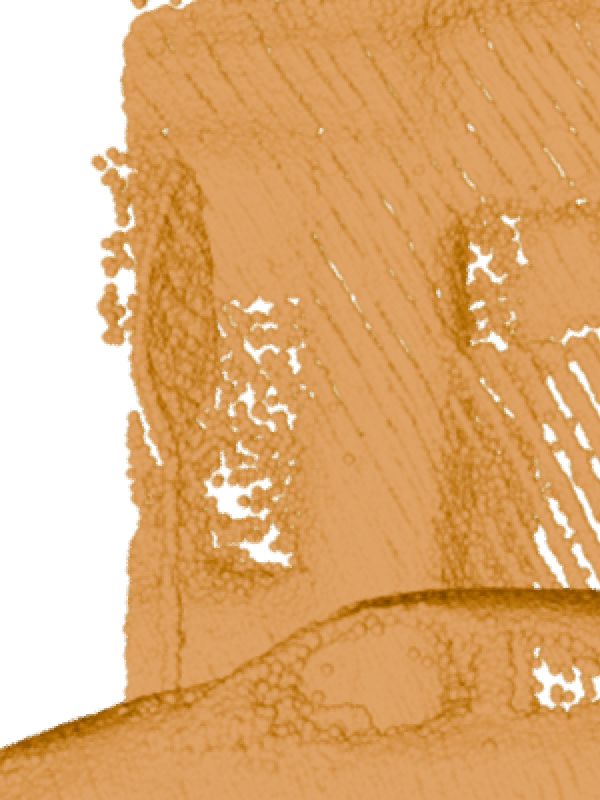}};
    \draw[line width=0.5pt] (detail.north west) rectangle (detail.south east);
  \end{tikzpicture}\\[-0.3em]
  \footnotesize MODNet
\end{minipage}\hspace{-11.5mm}%
%
\begin{minipage}[t]{0.25\linewidth}\centering
  \begin{tikzpicture}[baseline]
    \node[inner sep=0] (img)
      {\includegraphics[width=\linewidth]{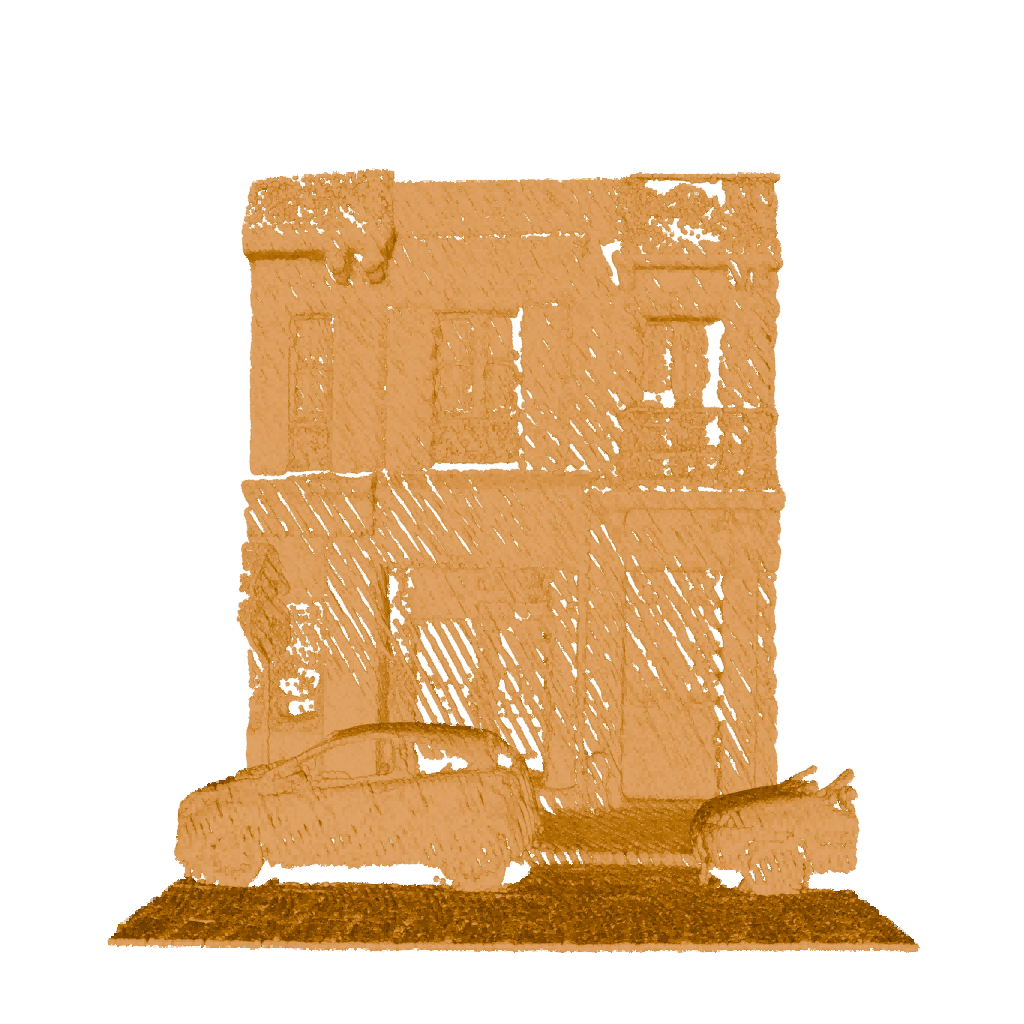}};
    \node[inner sep=0, anchor=north west] (detail)
      at ([xshift=2pt,yshift=-4pt]img.north west)
      {\includegraphics[width=0.50\linewidth]{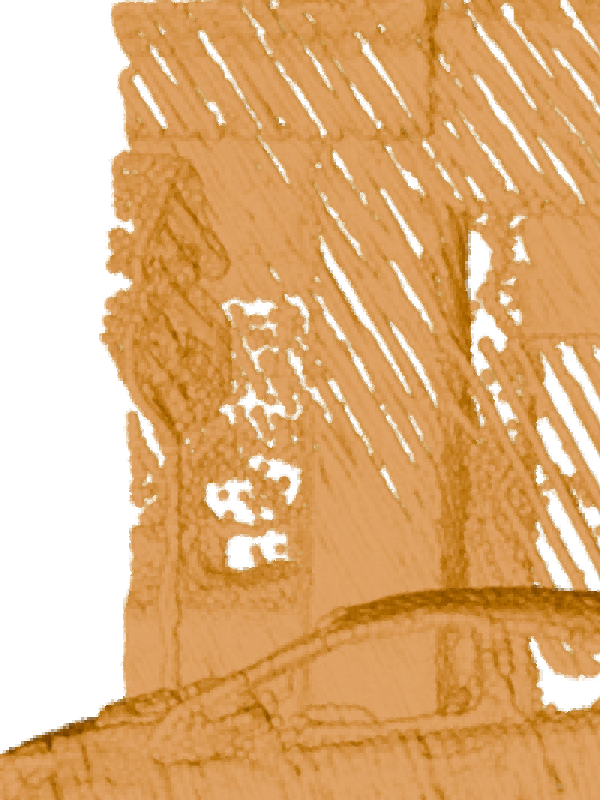}};
    \draw[line width=0.5pt] (detail.north west) rectangle (detail.south east);
  \end{tikzpicture}\\[-0.3em]
  \footnotesize IterativePFN
\end{minipage}\hspace{-11.5mm}%
%
\begin{minipage}[t]{0.25\linewidth}\centering
  \begin{tikzpicture}[baseline]
    \node[inner sep=0] (img)
      {\includegraphics[width=\linewidth]{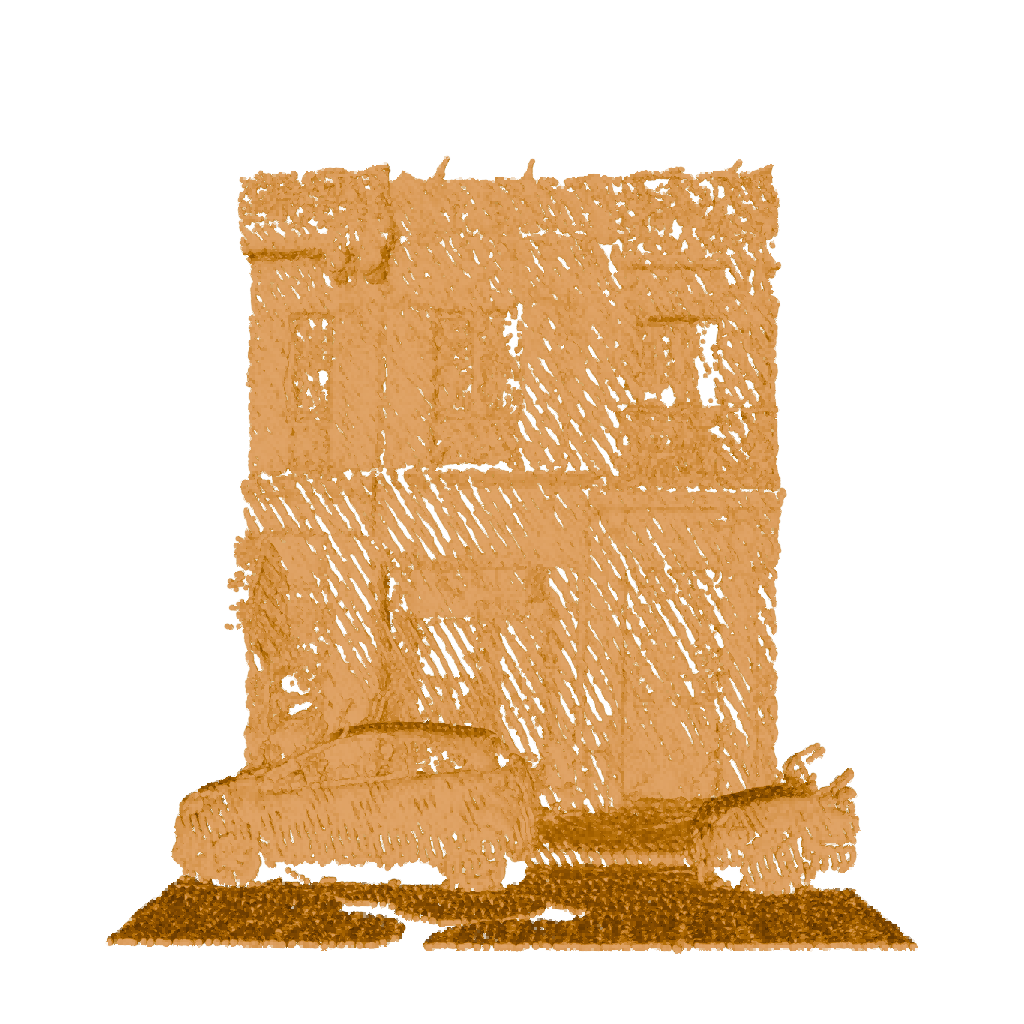}};
    \node[inner sep=0, anchor=north west] (detail)
      at ([xshift=2pt,yshift=-4pt]img.north west)
      {\includegraphics[width=0.50\linewidth]{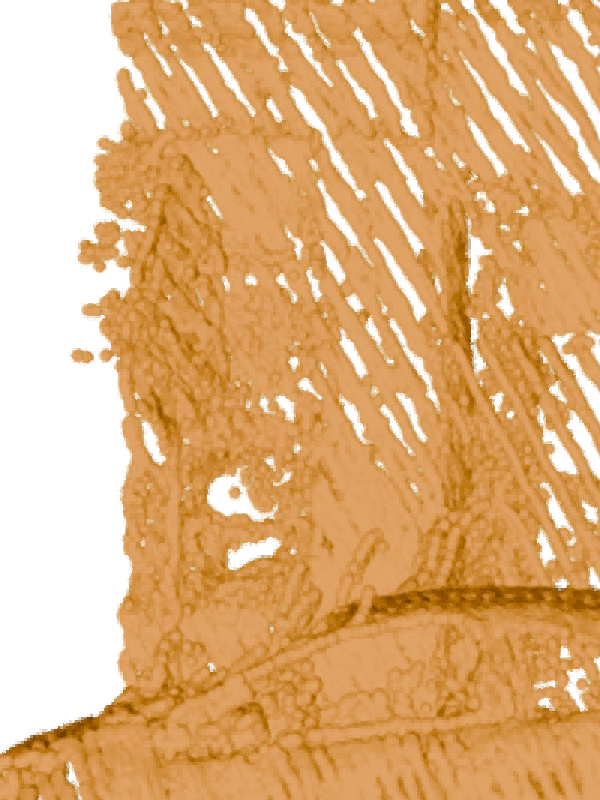}};
    \draw[line width=0.5pt] (detail.north west) rectangle (detail.south east);
  \end{tikzpicture}\\[-0.3em]
  \footnotesize DeepPSR
\end{minipage}\hspace{-11.5mm}%
%
\begin{minipage}[t]{0.25\linewidth}\centering
  \begin{tikzpicture}[baseline]
    \node[inner sep=0] (img)
      {\includegraphics[width=\linewidth]{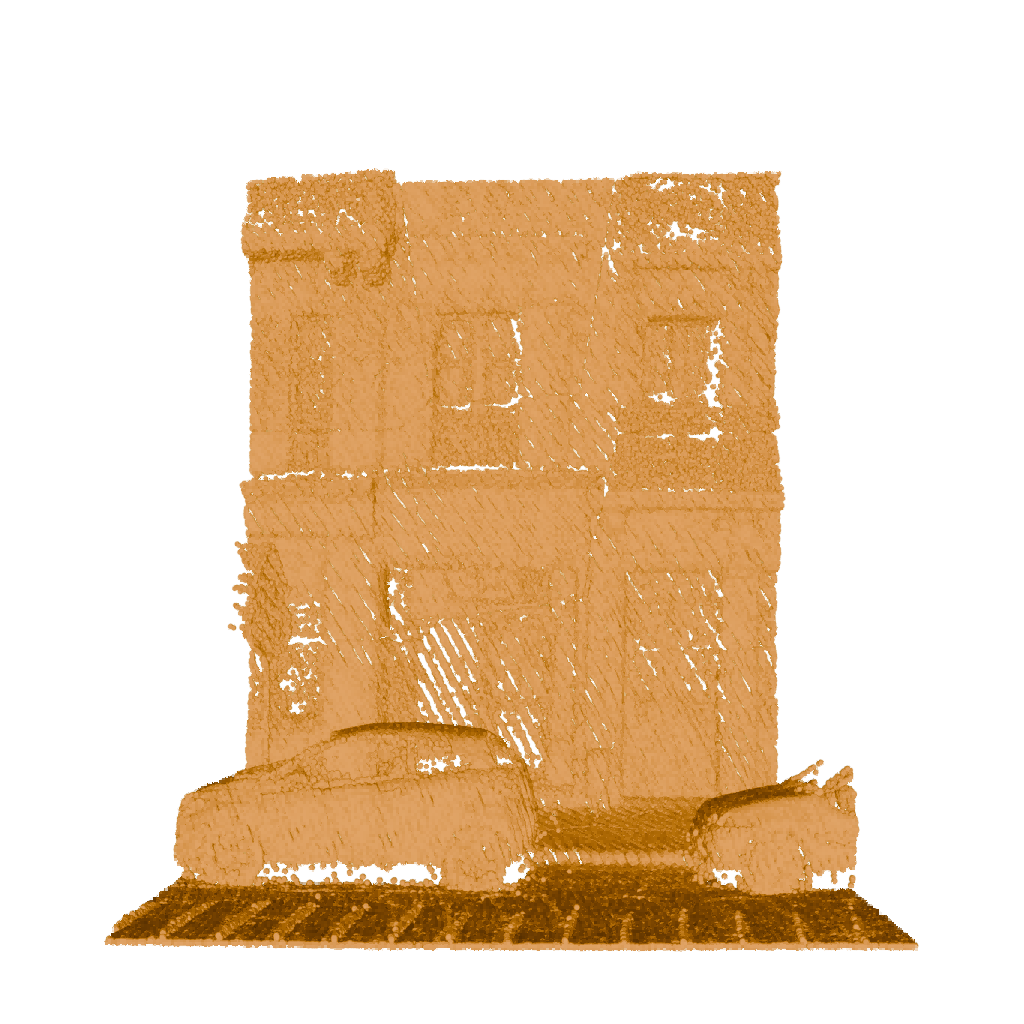}};
    \node[inner sep=0, anchor=north west] (detail)
      at ([xshift=2pt,yshift=-4pt]img.north west)
      {\includegraphics[width=0.50\linewidth]{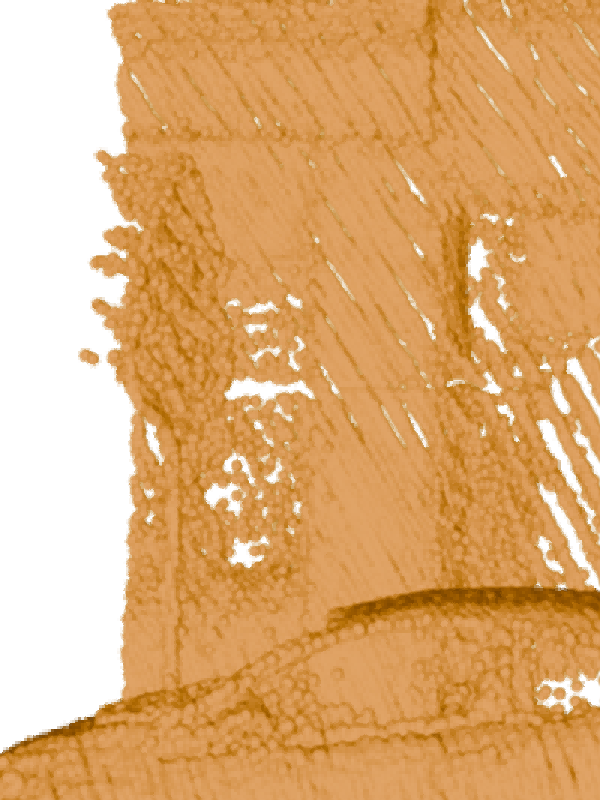}};
    \draw[line width=0.5pt] (detail.north west) rectangle (detail.south east);
  \end{tikzpicture}\\[-0.3em]
  \footnotesize RePCD
\end{minipage}%
\end{flushleft}

\vspace{-1.5em}

\begin{flushleft}

\begin{minipage}[t]{0.25\linewidth}\centering
  \begin{tikzpicture}[baseline]
    \node[inner sep=0] (img)
      {\includegraphics[width=\linewidth]{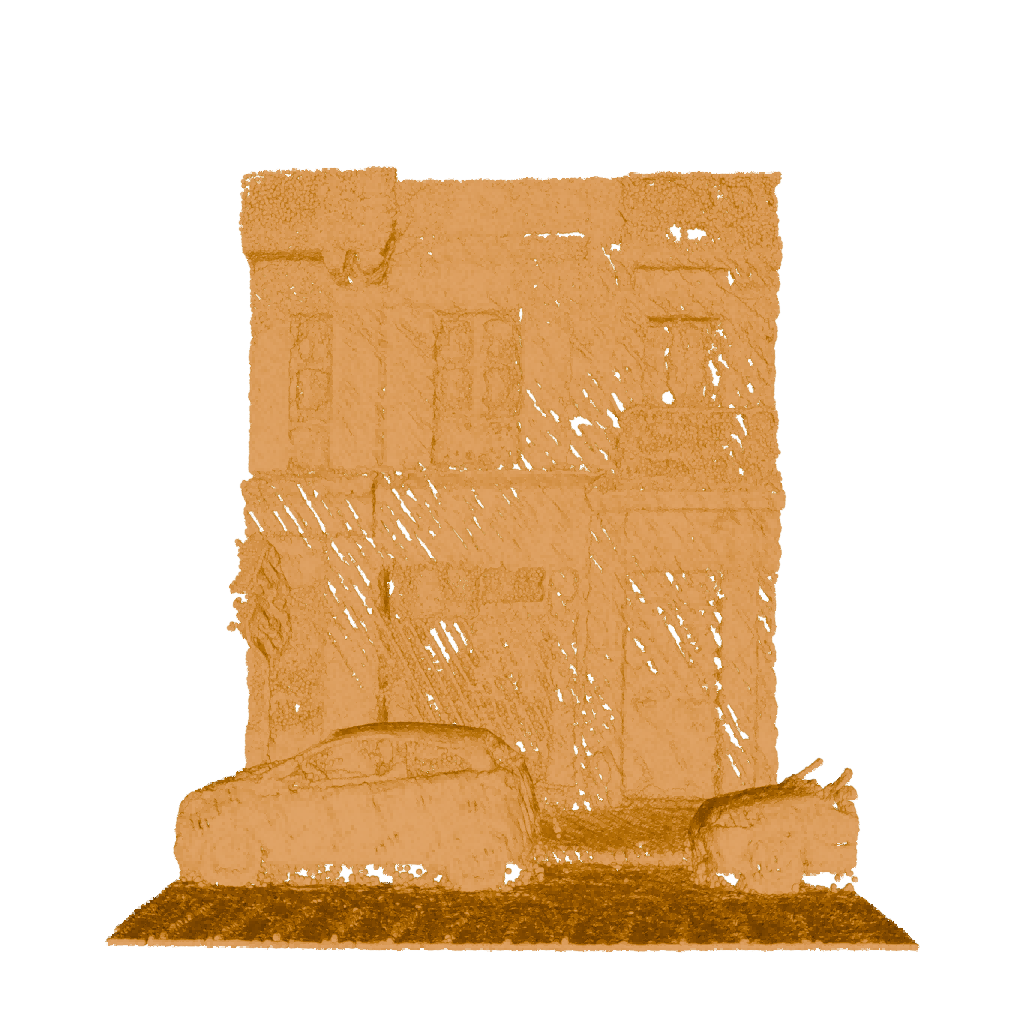}};
    \node[inner sep=0, anchor=north west] (detail)
      at ([xshift=2pt,yshift=-4pt]img.north west)
      {\includegraphics[width=0.50\linewidth]{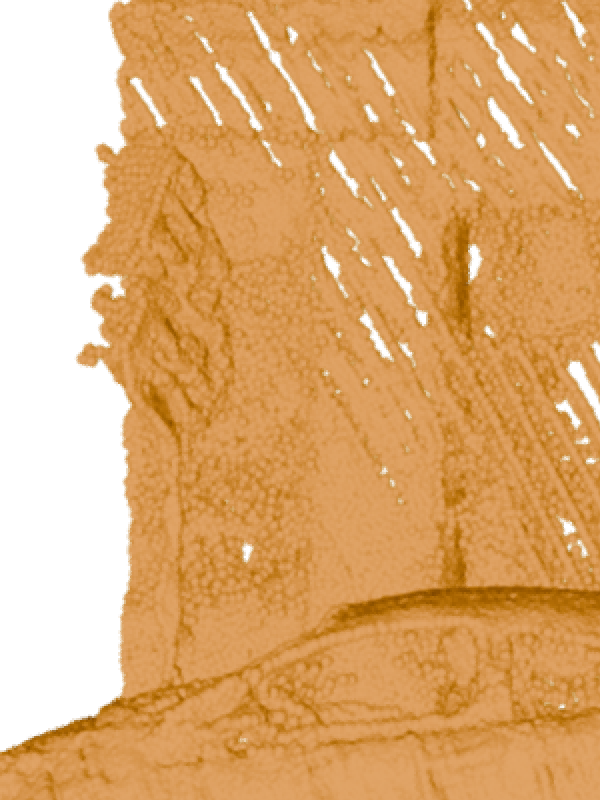}};
    \draw[line width=0.5pt] (detail.north west) rectangle (detail.south east);
  \end{tikzpicture}\\[-0.3em]
  \footnotesize PDFlow
\end{minipage}\hspace{-11.5mm}%
%
\begin{minipage}[t]{0.25\linewidth}\centering
  \begin{tikzpicture}[baseline]
    \node[inner sep=0] (img)
      {\includegraphics[width=\linewidth]{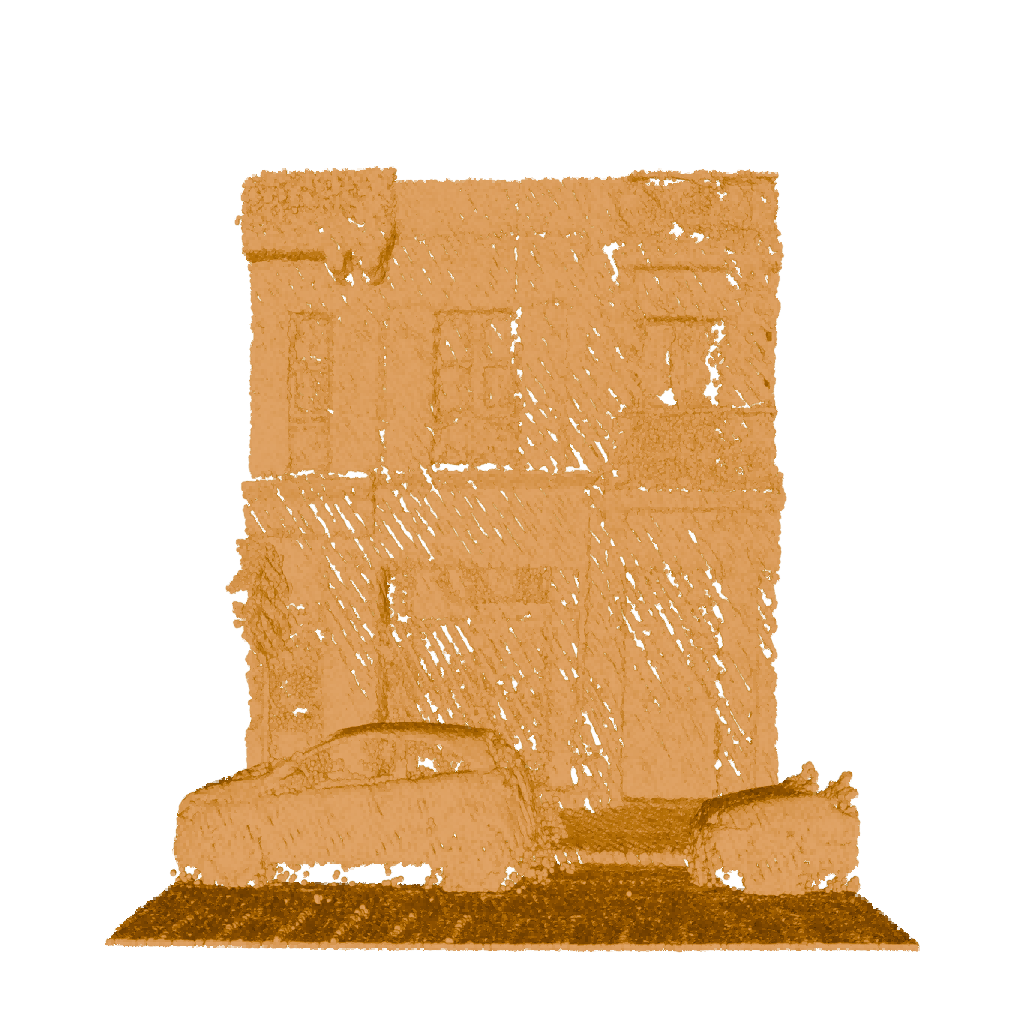}};
    \node[inner sep=0, anchor=north west] (detail)
      at ([xshift=2pt,yshift=-4pt]img.north west)
      {\includegraphics[width=0.50\linewidth]{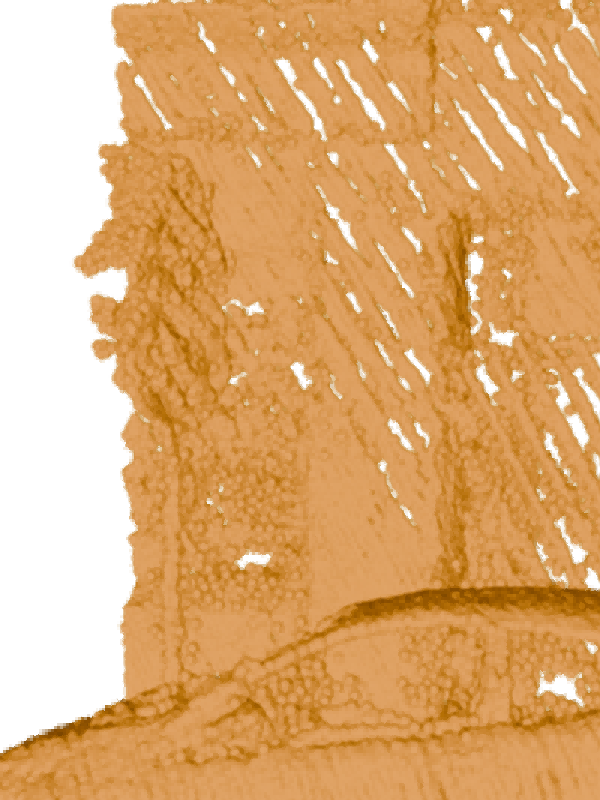}};
    \draw[line width=0.5pt] (detail.north west) rectangle (detail.south east);
  \end{tikzpicture}\\[-0.3em]
  \footnotesize ScoreDenoise
\end{minipage}\hspace{-11.5mm}%
%
\begin{minipage}[t]{0.25\linewidth}\centering
  \begin{tikzpicture}[baseline]
    \node[inner sep=0] (img)
      {\includegraphics[width=\linewidth]{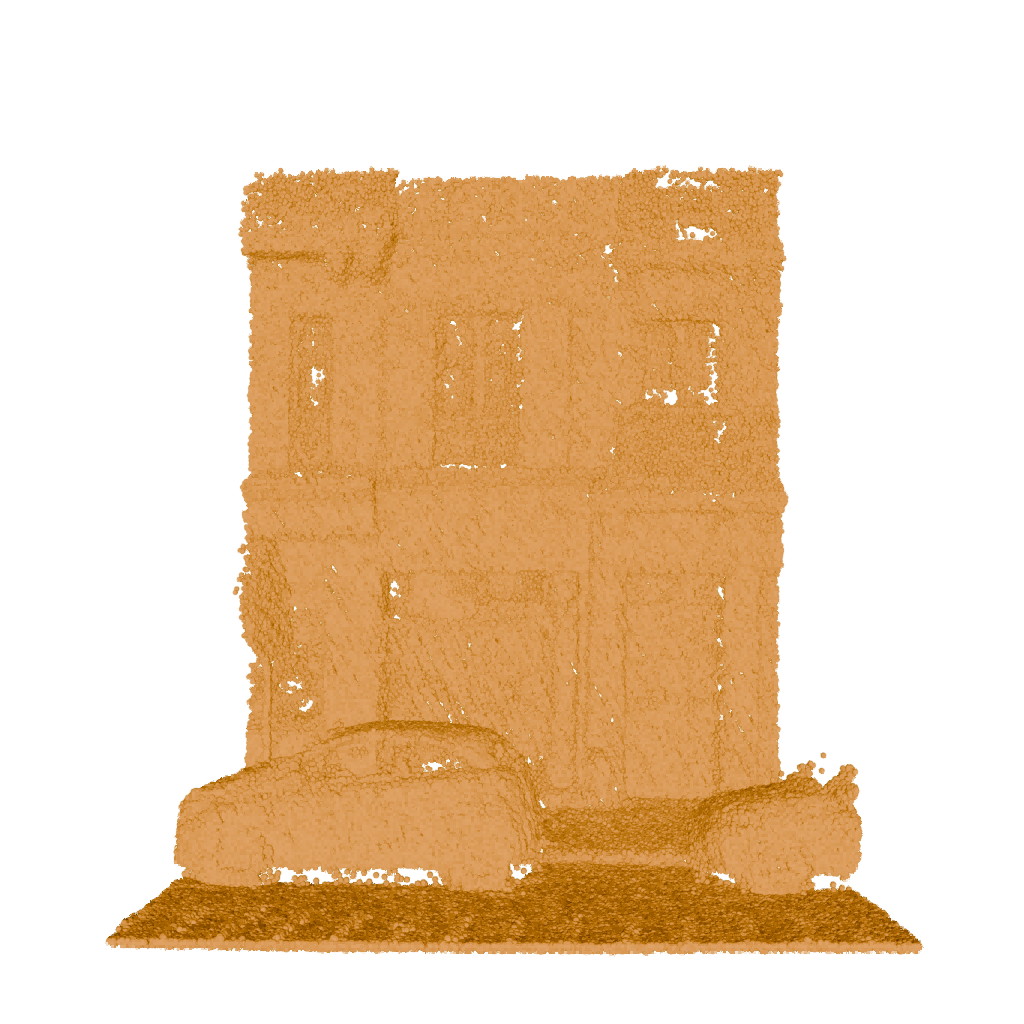}};
    \node[inner sep=0, anchor=north west] (detail)
      at ([xshift=2pt,yshift=-4pt]img.north west)
      {\includegraphics[width=0.50\linewidth]{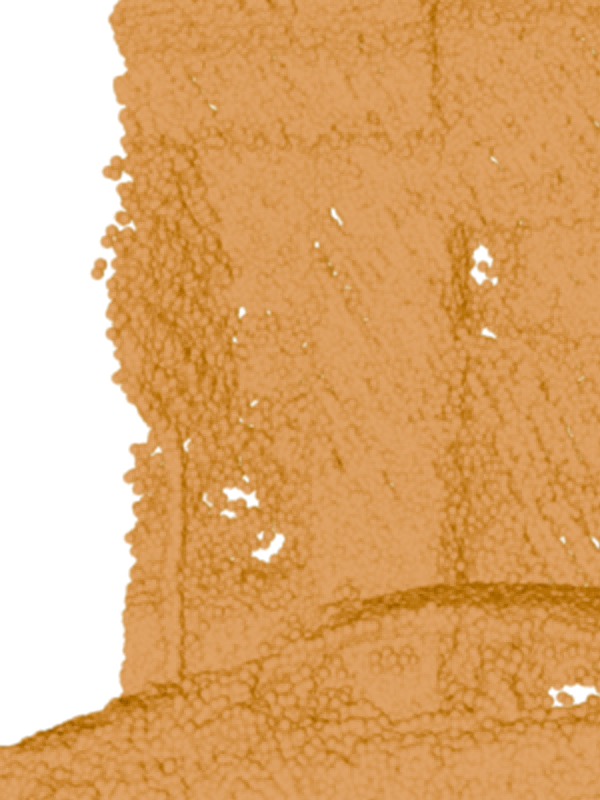}};
    \draw[line width=0.5pt] (detail.north west) rectangle (detail.south east);
  \end{tikzpicture}\\[-0.3em]
  \footnotesize DMRDenoise
\end{minipage}\hspace{-11.5mm}%
%
\begin{minipage}[t]{0.25\linewidth}\centering
  \begin{tikzpicture}[baseline]
    \node[inner sep=0] (img)
      {\includegraphics[width=\linewidth]{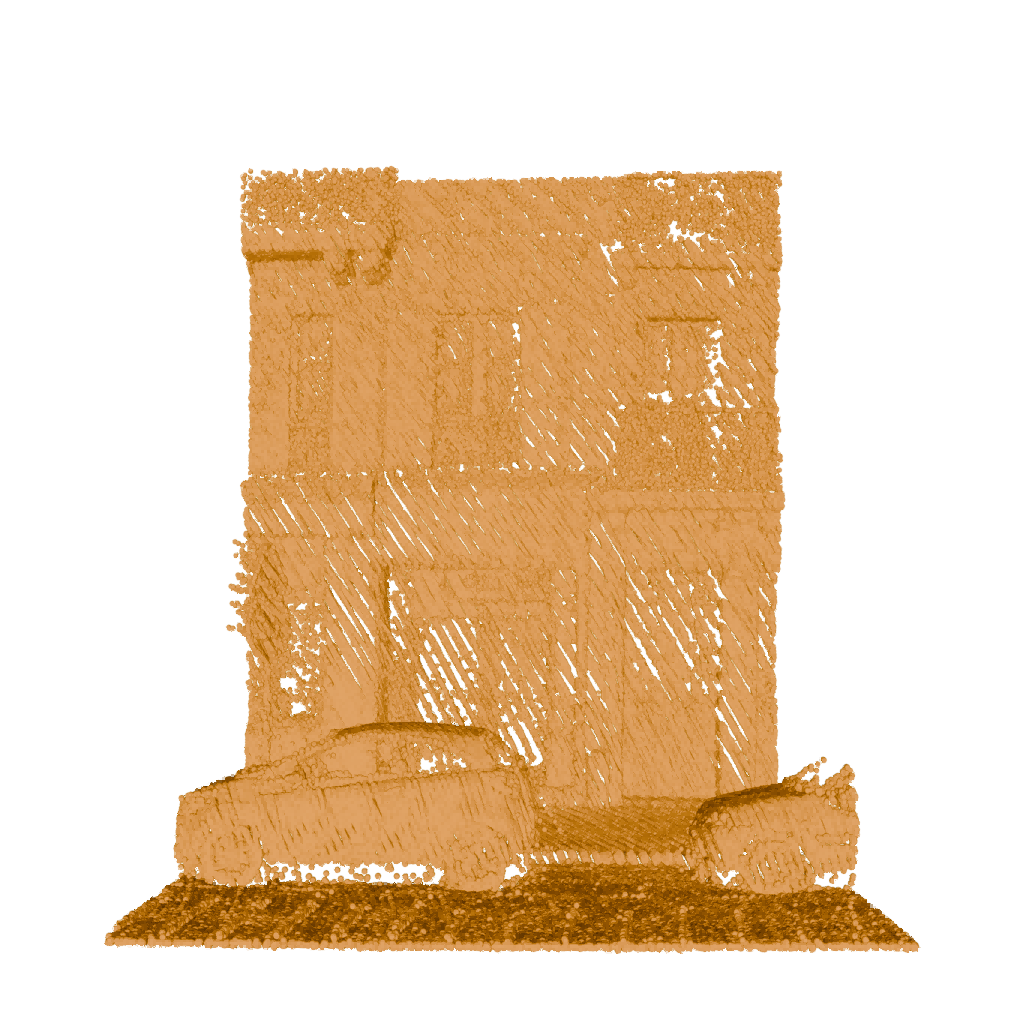}};
    \node[inner sep=0, anchor=north west] (detail)
      at ([xshift=2pt,yshift=-4pt]img.north west)
      {\includegraphics[width=0.50\linewidth]{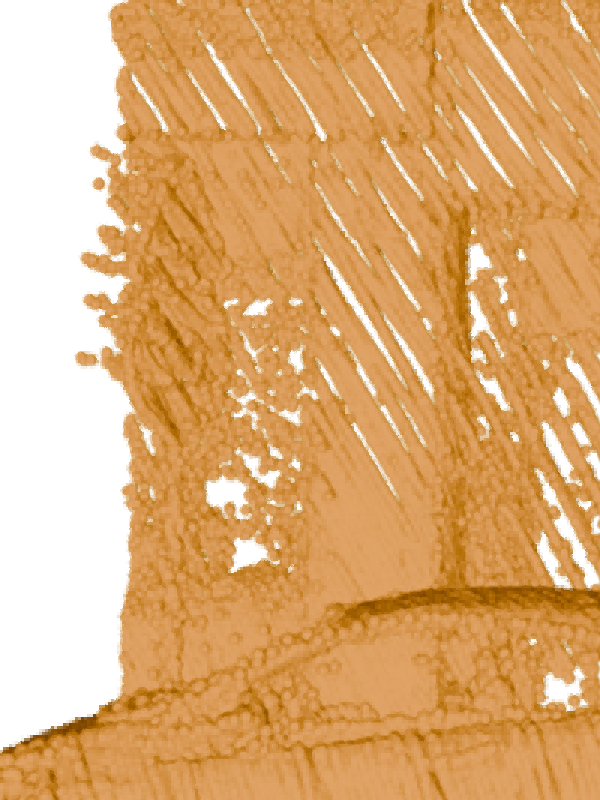}};
    \draw[line width=0.5pt] (detail.north west) rectangle (detail.south east);
  \end{tikzpicture}\\[-0.3em]
  \footnotesize GPDNet
\end{minipage}\hspace{-11.5mm}%
%
\begin{minipage}[t]{0.25\linewidth}\centering
  \begin{tikzpicture}[baseline]
    \node[inner sep=0] (img)
      {\includegraphics[width=\linewidth]{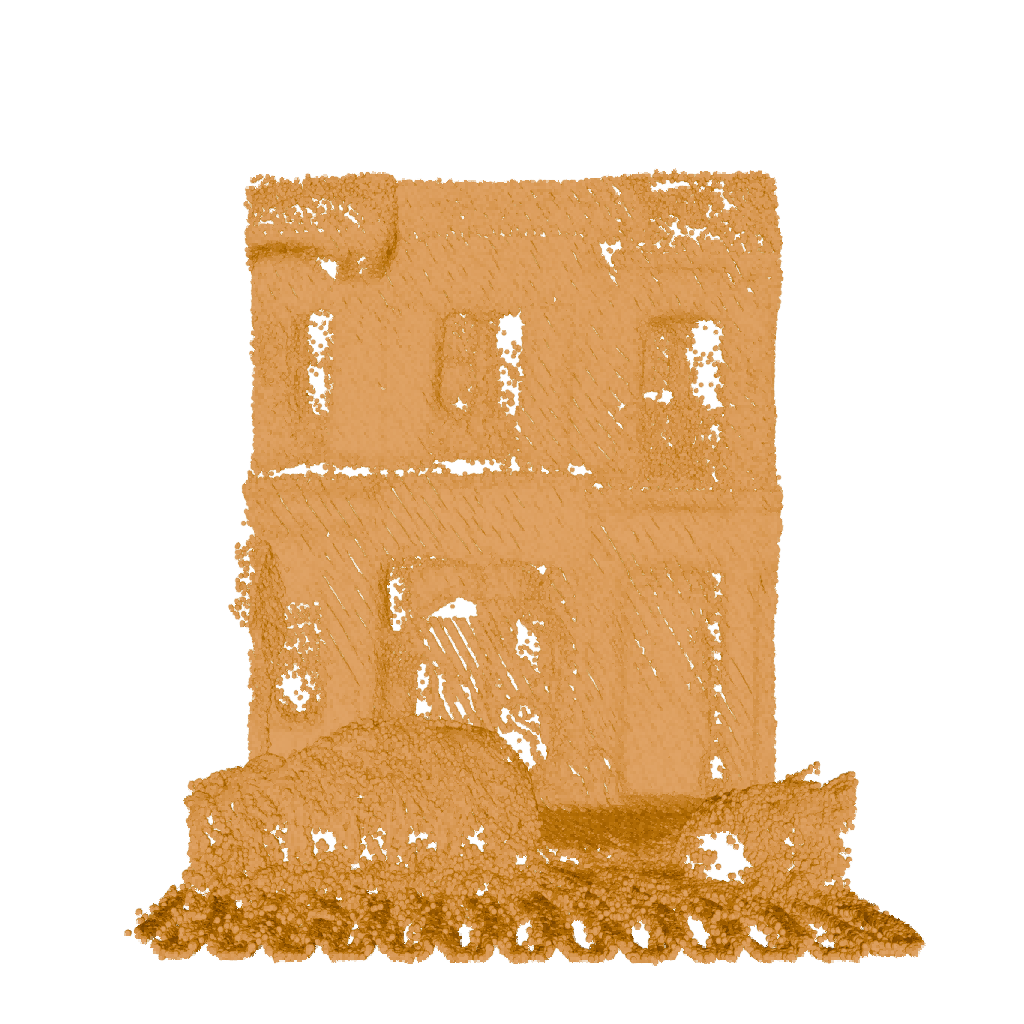}};
    \node[inner sep=0, anchor=north west] (detail)
      at ([xshift=2pt,yshift=-4pt]img.north west)
      {\includegraphics[width=0.50\linewidth]{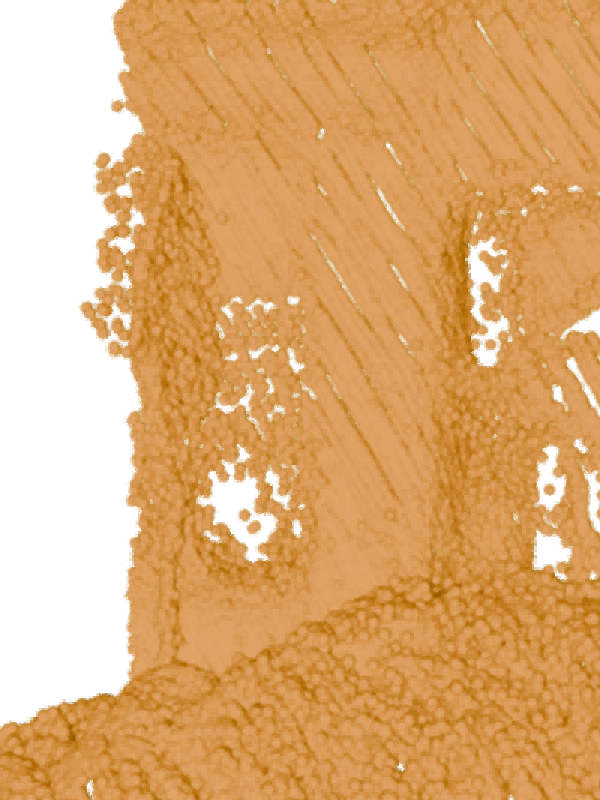}};
    \draw[line width=0.5pt] (detail.north west) rectangle (detail.south east);
  \end{tikzpicture}\\[-0.3em]
  \footnotesize Pointfilter
\end{minipage}%
\end{flushleft}

\vspace{0.5em}
\caption{Visual comparison of denoising results on the \texttt{RueMadame\_3} scan. For each method, a zoom-in detail is embedded in the upper-left corner to highlight denoising performance on fine structures.}
\label{fig:ruemadame3_gallery}
\end{figure*}

We qualitatively evaluate two representative cases: a synthetic shape (\texttt{elephant}) corrupted by Gaussian noise with $\sigma=0.01$, and a real-world scan from the RueMadame dataset. Figs.~\ref{fig:elephant_error_gallery} and~\ref{fig:ruemadame3_gallery} present side-by-side comparisons of all methods, grouped by supervision type (supervised vs.\ unsupervised) and ordered chronologically within each group. Additional results, including \texttt{chair} at $\sigma=0.02$ and \texttt{moai} at $\sigma=0.03$, are provided in the supplementary material.

\textbf{Synthetic noise.}
Fig.~\ref{fig:elephant_error_gallery} shows qualitative results on the synthetic \texttt{elephant} under Gaussian noise with $\sigma=0.01$. Additional shapes and comparisons are provided in the supplementary material.
The visualizations show per-point error maps computed using a unidirectional Point-to-Surface distance to the ground-truth mesh.
The per-point error is visualized using a blue-to-red colormap, where red highlights regions with larger deviations from the ground-truth mesh.
Overall, recent supervised methods such as \textit{3DMambaIPF}, \textit{ASDN}, \textit{PD-LTS}, and \textit{StraightPCF} consistently produce smoother surfaces and better preserve geometric details across all shapes. In contrast, some earlier methods (e.g., \textit{DMRDenoise}, \textit{PointCleanNet}) often result in oversmoothing or leave residual noise, especially under higher noise levels (see \texttt{moai} in the supplementary).

Unsupervised methods are competitive at low noise—for example, \textit{Noise4Denoise} and \textit{Un-Score} recover plausible geometry. However, under higher noise (see \texttt{moai} in the supplementary), most unsupervised approaches struggle to preserve fine structures, underscoring the difficulty of learning without ground-truth supervision.
These visual results align with the quantitative trends, where low CD corresponds to overall smooth recovery, and HD reveals vulnerability to outliers.

\textbf{Real scan noise.}
Fig.~\ref{fig:ruemadame3_gallery} illustrates visual comparisons on a scan from the RueMadame dataset. Unlike synthetic examples, real-world scans contain non-uniform and irregular noise patterns. In this case, methods such as \textit{ASDN}, \textit{StraightPCF}, and \textit{DeepPSR} demonstrate visually coherent reconstructions with well-preserved facade structures. On the other hand, methods like \textit{DMRDenoise} and \textit{Pointfilter} show artifacts or incomplete recovery in fine details such as windows and ledges.
These visual comparisons offer valuable insights into each method’s generalization ability under realistic sensor conditions.


\subsection{Ablation Studies}
We evaluate the most recent point cloud denoising methods (from the past three years), focusing on key aspects such as geometric structure preservation, uniformity of denoised points, downstream applications (e.g., surface reconstruction), and efficiency analysis. These methods represent the current state of the art and embody the latest advances in deep geometric learning.

\textbf{Geometric Edge Preservation.}
We design a quantitative protocol to evaluate the preservation of geometric edge structures after denoising.
First, we select a representative mesh with prominent sharp features from the ABC dataset, which provides high-quality annotations of CAD-derived edge structures. 
The mesh is uniformly resampled (30K points), corrupted with Gaussian noise ($\sigma = 0.01$).
The ground-truth edge points are obtained by sampling along the labeled edge features on the original mesh. To evaluate whether the sharp features are preserved in the denoised output, we select a subset of points from the denoised result by finding the nearest $5$ neighbors (kNN, $k = 5$) for each ground-truth edge point and removing duplicates. This yields a candidate set of denoised edge points.
We then compute the Chamfer Distance (CD) between the denoised subset and the ground-truth edge points. Finally, we visualize the per-point CD as an error map to provide intuitive insight into geometric structure feature preservation quality. 

\begin{figure*}[htbp]
\centering

\hfill
\includegraphics[width=0.2\linewidth]{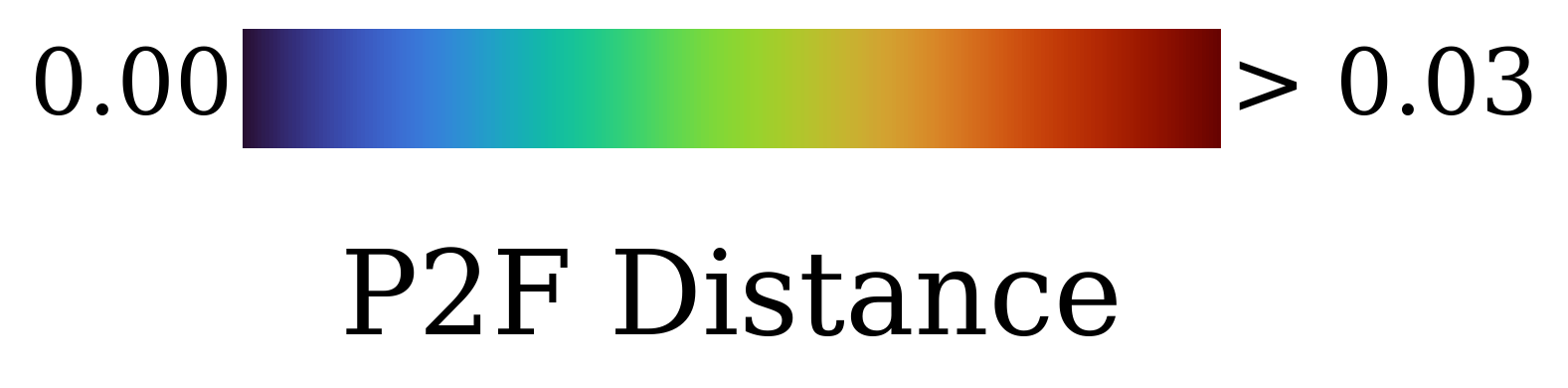}

\vspace{0em}

\begin{minipage}[t]{0.18\linewidth}\centering
  \rotatebox{90}{%
    \includegraphics[height=0.95\linewidth]{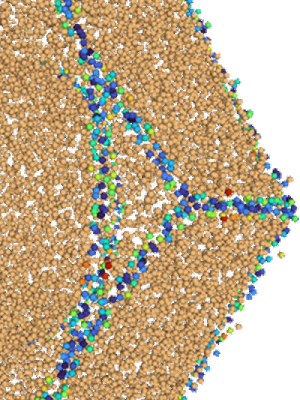}%
  }\\[-0.4em]
  \includegraphics[width=0.495\linewidth]{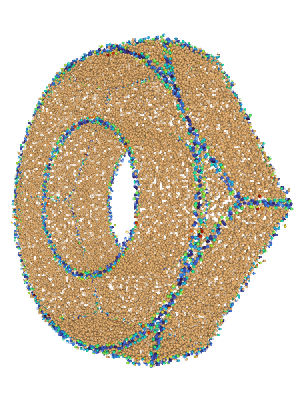}%
  \hfill
  \includegraphics[width=0.495\linewidth]{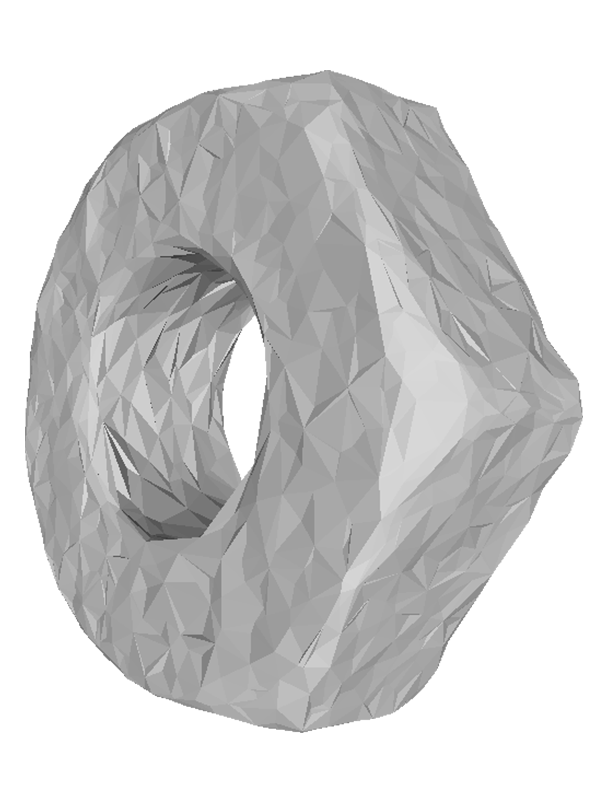}%
  \\[-0.25em]
  \footnotesize Noisy
\end{minipage}\hspace{0.5mm}%
%
\begin{minipage}[t]{0.18\linewidth}\centering
  \rotatebox{90}{%
    \includegraphics[height=0.95\linewidth]{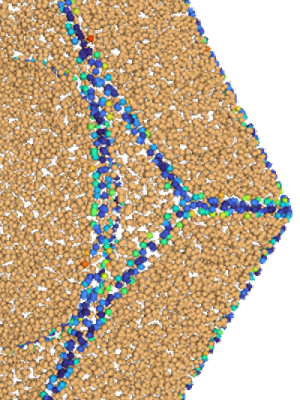}%
  }\\[-0.4em]
  \includegraphics[width=0.495\linewidth]{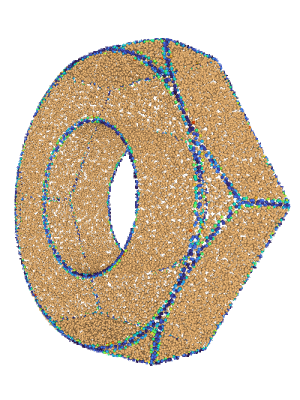}%
  \hfill
  \includegraphics[width=0.495\linewidth]{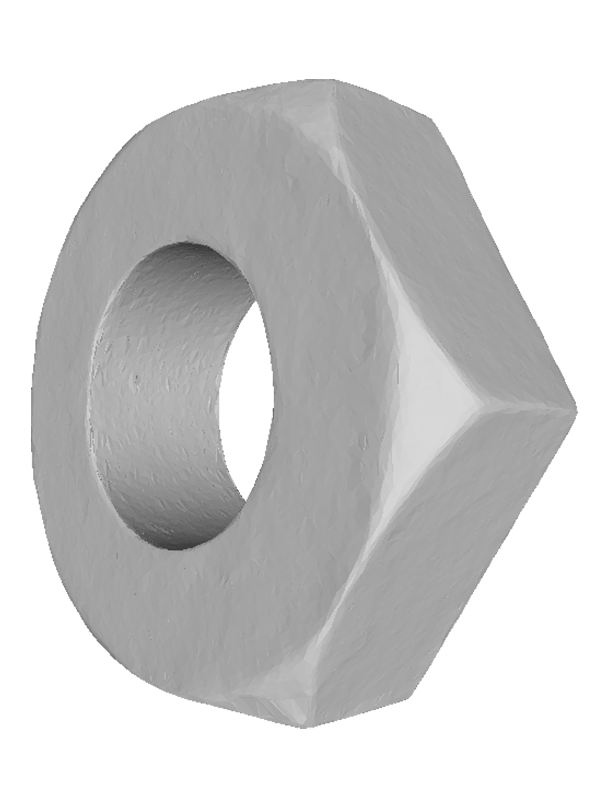}%
  \\[-0.25em]
  \footnotesize 3DMambaIPF
\end{minipage}\hspace{0.5mm}%
%
\begin{minipage}[t]{0.18\linewidth}\centering
  \rotatebox{90}{%
    \includegraphics[height=0.95\linewidth]{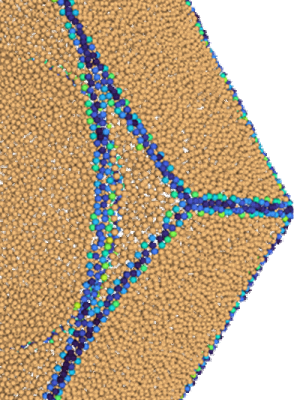}%
  }\\[-0.4em]
  \includegraphics[width=0.495\linewidth]{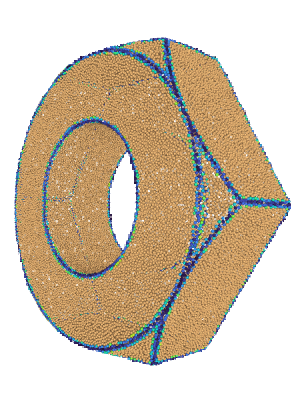}%
  \hfill
  \includegraphics[width=0.495\linewidth]{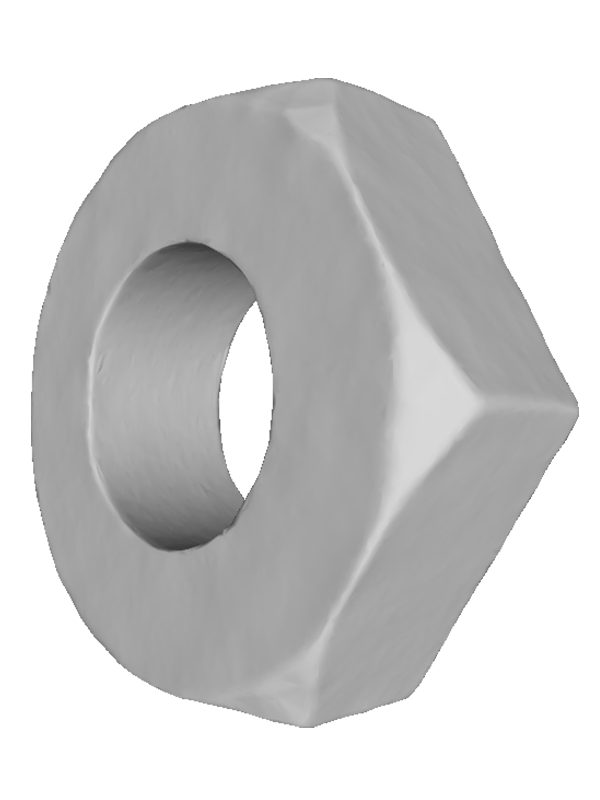}%
  \\[-0.25em]
  \footnotesize ASDN
\end{minipage}\hspace{0.5mm}%
%
\begin{minipage}[t]{0.18\linewidth}\centering
  \rotatebox{90}{%
    \includegraphics[height=0.95\linewidth]{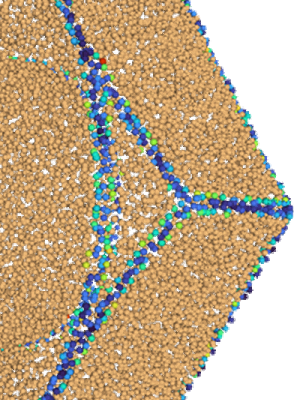}%
  }\\[-0.4em]
  \includegraphics[width=0.495\linewidth]{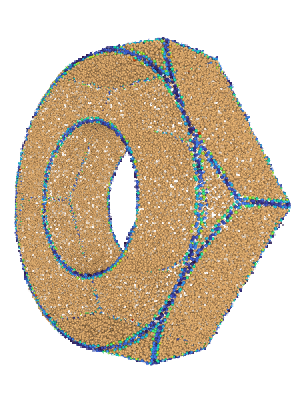}%
  \hfill
  \includegraphics[width=0.495\linewidth]{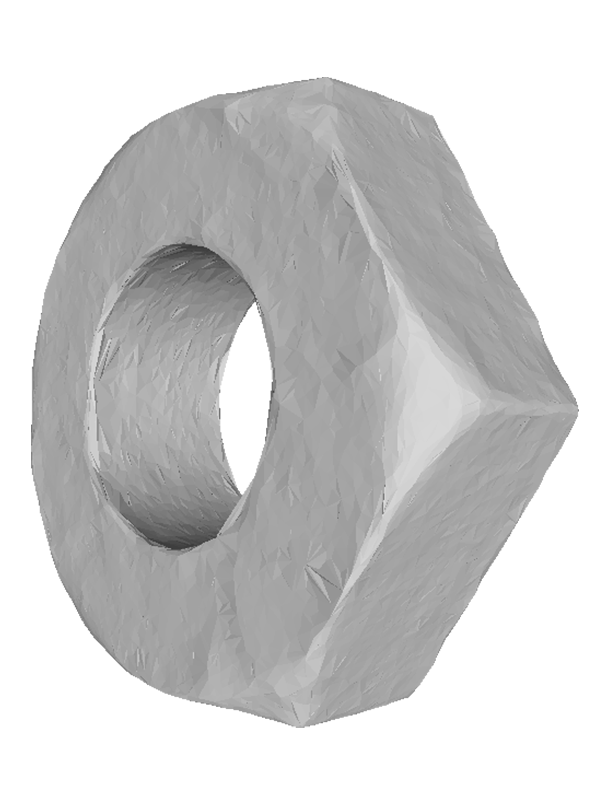}%
  \\[-0.25em]
  \footnotesize P2P\text{-}Bridge
\end{minipage}\hspace{0.5mm}%
%
\begin{minipage}[t]{0.18\linewidth}\centering
  \rotatebox{90}{%
    \includegraphics[height=0.95\linewidth]{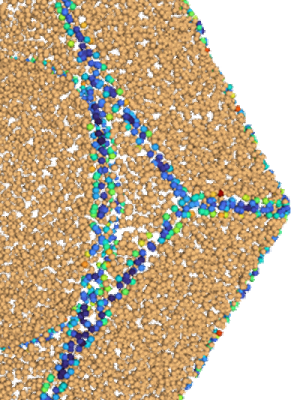}%
  }\\[-0.4em]
  \includegraphics[width=0.495\linewidth]{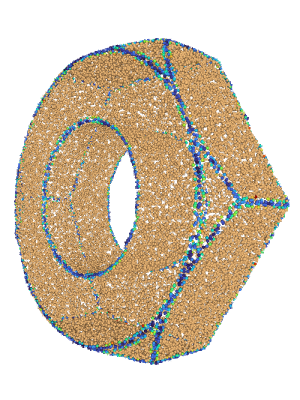}%
  \hfill
  \includegraphics[width=0.495\linewidth]{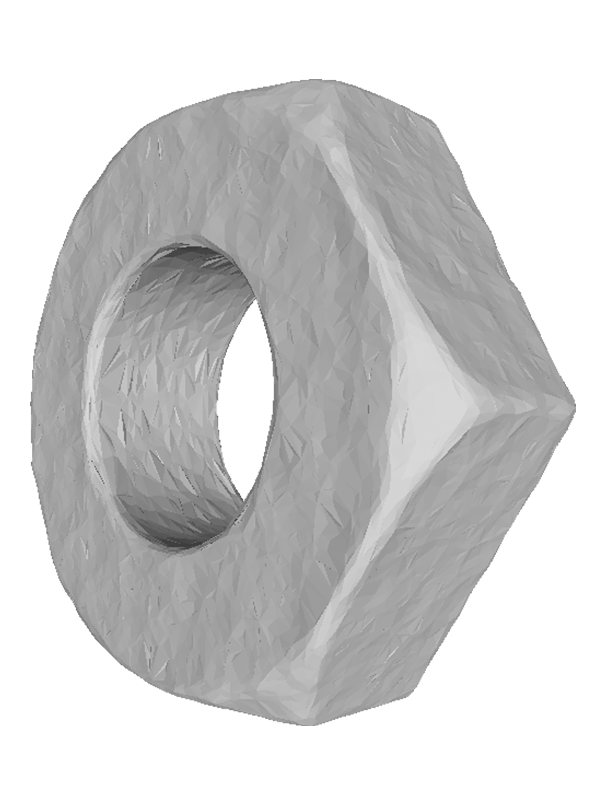}%
  \\[-0.25em]
  \footnotesize Pathnet
\end{minipage}%

\vspace{2mm}

\begin{minipage}[t]{0.18\linewidth}\centering
  \rotatebox{90}{%
    \includegraphics[height=0.95\linewidth]{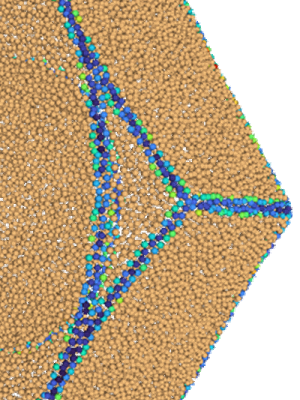}%
  }\\[-0.4em]
  \includegraphics[width=0.495\linewidth]{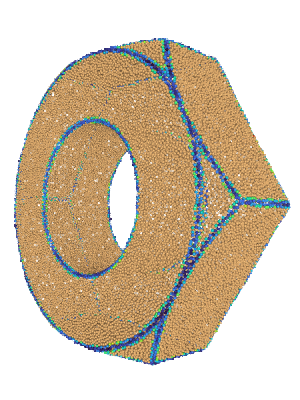}%
  \hfill
  \includegraphics[width=0.495\linewidth]{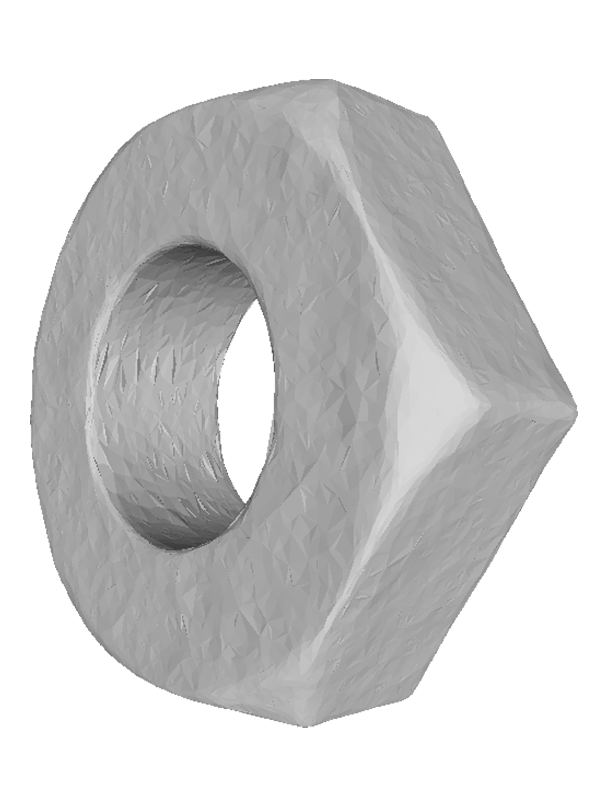}%
  \\[-0.25em]
  \footnotesize StraightPCF
\end{minipage}\hspace{0.5mm}%
%
\begin{minipage}[t]{0.18\linewidth}\centering
  \rotatebox{90}{%
    \includegraphics[height=0.95\linewidth]{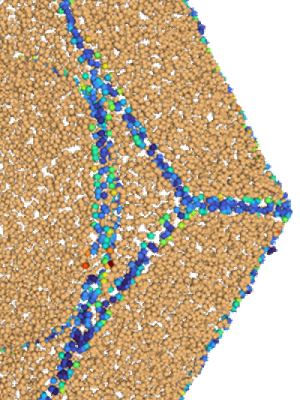}%
  }\\[-0.4em]
  \includegraphics[width=0.495\linewidth]{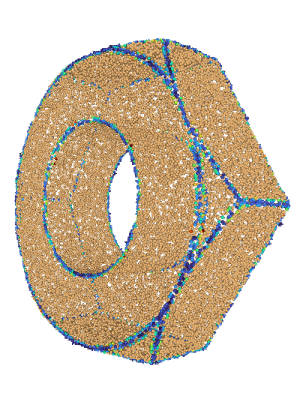}%
  \hfill
  \includegraphics[width=0.495\linewidth]{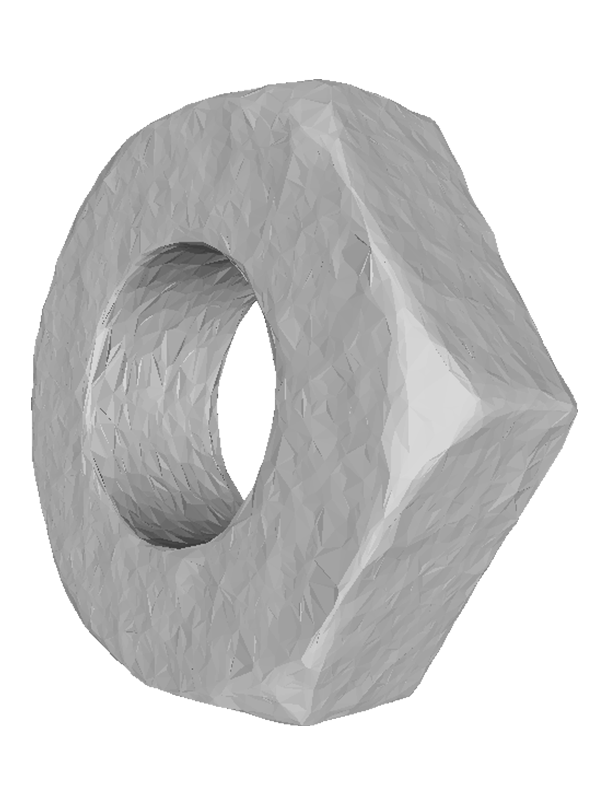}%
  \\[-0.25em]
  \footnotesize MAG
\end{minipage}\hspace{0.5mm}%
%
\begin{minipage}[t]{0.18\linewidth}\centering
  \rotatebox{90}{%
    \includegraphics[height=0.95\linewidth]{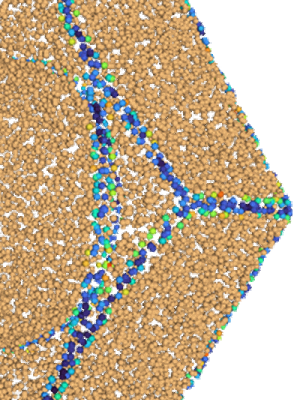}%
  }\\[-0.4em]
  \includegraphics[width=0.495\linewidth]{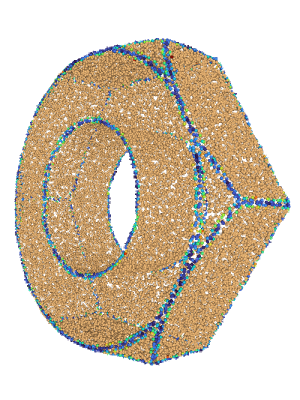}%
  \hfill
  \includegraphics[width=0.495\linewidth]{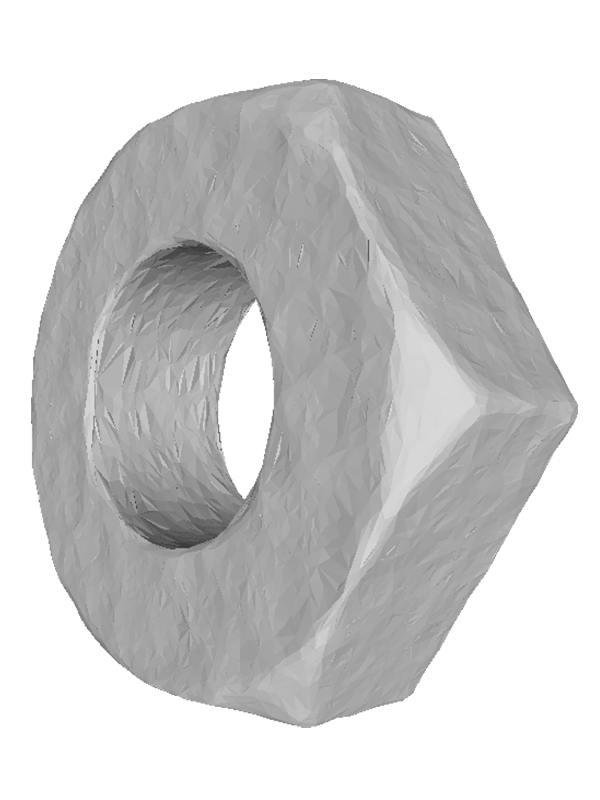}%
  \\[-0.25em]
  \footnotesize MODNet
\end{minipage}\hspace{0.5mm}%
%
\begin{minipage}[t]{0.18\linewidth}\centering
  \rotatebox{90}{%
    \includegraphics[height=0.95\linewidth]{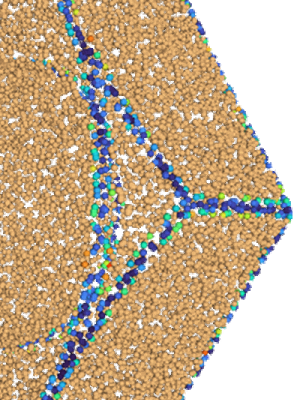}%
  }\\[-0.4em]
  \includegraphics[width=0.495\linewidth]{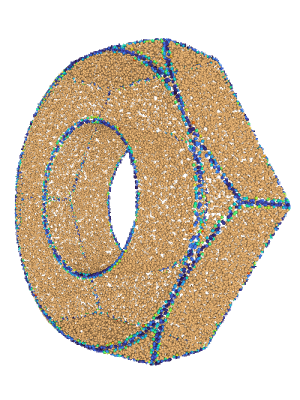}%
  \hfill
  \includegraphics[width=0.495\linewidth]{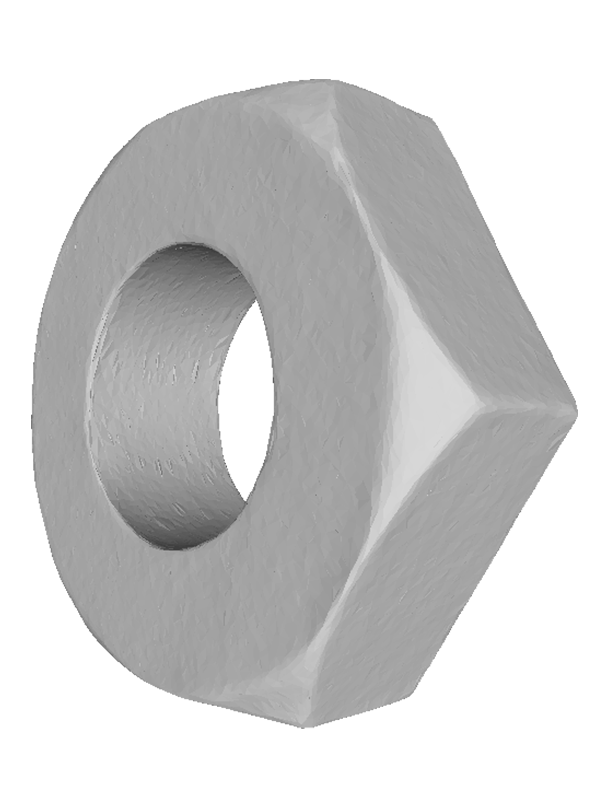}%
  \\[-0.25em]
  \footnotesize IterativePFN
\end{minipage}\hspace{0.5mm}%
%
\begin{minipage}[t]{0.18\linewidth}\centering
  \rotatebox{90}{%
    \includegraphics[height=0.95\linewidth]{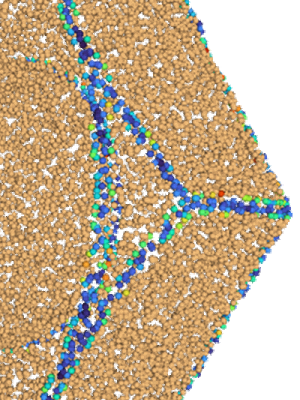}%
  }\\[-0.4em]
  \includegraphics[width=0.495\linewidth]{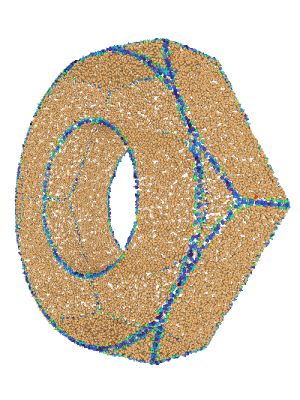}%
  \hfill
  \includegraphics[width=0.495\linewidth]{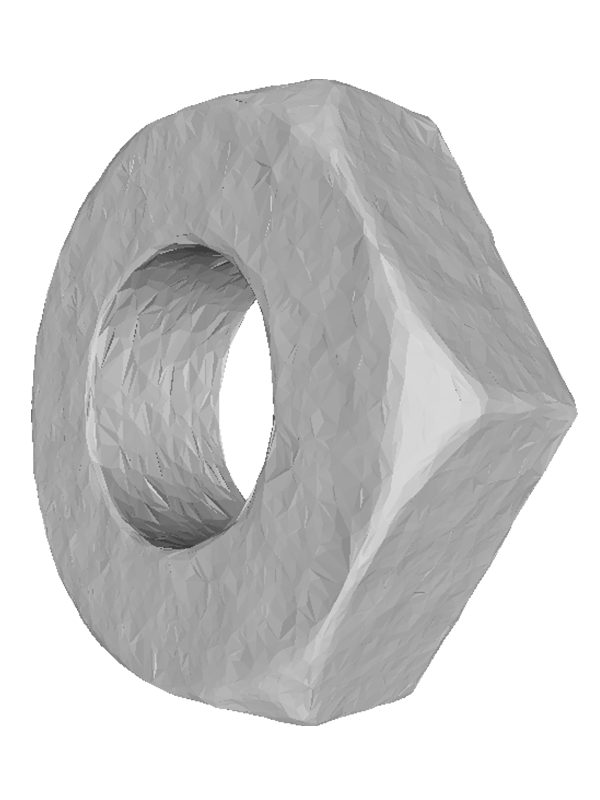}%
  \\[-0.25em]
  \footnotesize PCDNF
\end{minipage}%

\vspace{0.5em}
\caption{Visual comparison of denoising results (two rows). Each column shows one method: the top image is a zoomed-in detail from the point cloud (rotated 90\textdegree), followed by the denoised point cloud and its corresponding reconstructed mesh.}
\label{fig:abla:edge_mesh_detail}
\end{figure*}

Fig.~\ref{fig:abla:edge_mesh_detail} shows the corresponding detail view, denoised point cloud, and reconstructed mesh for each method, highlighting differences in structural preservation.
In the zoomed-in edge regions, \textit{ASDN}, \textit{P2P-Bridge}, and \textit{StraightPCF} show better retention of sharp geometric features, such as creases and corners.

\textbf{Point Distribution.} 
Maintaining a uniform distribution of points is essential for accurate surface reconstruction and downstream geometric processing. To evaluate how well each method promotes uniformity, we visualize local point densities under a consistent setting (50K resolution and Gaussian noise with $\sigma = 0.01$). 
For each point, we compute the kNN density 
$\rho_i = \tfrac{1}{\overline{d}_{k,i}}$ with $k=20$, 
and derive non-uniformity as the robust deviation from the global median. 
The computed densities are normalized to $[0, 1]$ and mapped to colors using the plasma colormap. 

Fig.~\ref{fig:abla:density_detail} shows the colored density visualizations. Compared to the original noisy point cloud, which exhibits a highly non-uniform distribution, all methods demonstrate improved uniformity to varying degrees. From the zoom-in details, it can be observed that \textit{ASDN} and \textit{StraightPCF} achieve relatively uniform point distributions. Additionally, similar conclusions can be drawn from the point density visualizations in Fig.~\ref{fig:abla:edge_mesh_detail}, which are consistent with the observations above.

\textbf{Surface Reconstruction.}
To assess surface quality and geometric fidelity of denoised point clouds, we reconstruct triangle meshes for qualitative comparison. We employ Poisson surface reconstruction for each point cloud, where per-point normals are estimated via principal component analysis (PCA) on local neighborhoods. To ensure global consistency, normal orientations are further aligned using a tangent plane propagation scheme. The mesh resolution is controlled by a depth parameter, which we set to 7 in our experiments.

Fig.~\ref{fig:abla:recons_detail} presents the reconstructed meshes with zoomed-in views for detailed comparison. We observe that smoother denoised point clouds generally lead to smoother reconstructed surfaces. For instance, \textit{3DMambaIPF} produces high-quality denoising results and yields notably smoother mesh surfaces. Another interesting observation is that a more uniform point distribution tends to improve reconstruction quality. Under the same reconstruction parameters, methods like StraightPCF produce watertight and well-formed surfaces due to their even point sampling.

\textbf{Efficiency Analysis.}
To enable a more comprehensive and fair comparison of different denoising methods, we report the number of parameters, FLOPs, and inference time for each model based on a 10K-resolution input point cloud. 
The parameter count is obtained directly from the models used in our quantitative evaluations, ensuring consistency across metrics.
For FLOPs calculation, we adopt two different strategies based on the denoising design of each method. Specifically, we categorize all methods into two types: Patch-to-Patch and Patch-to-Point. Patch-to-Patch methods predict the denoised coordinates of all points within a local patch, while Patch-to-Point methods (denoted with $^\ast$) only predict the denoised position of a single center point given the entire patch as input.
To ensure fair comparison, we extract the same 1000-point local patch from the input shape and use it across all methods.
For Patch-to-Patch methods, we directly feed this patch into the network and compute the floating-point operations required for a full forward pass.
For Patch-to-Point methods, we use the same patch as a source, extract a subset of input points according to the method’s original design, compute the FLOPs for predicting one center point, and multiply the result by 1000 to normalize the cost under the same prediction volume (i.e., 1000 output points).
Inference time is measured by running each model once with a batch size of 1, recording only the forward pass time while excluding any data loading, preprocessing, or postprocessing. The reported runtime reflects the average time required to denoise 1000 points. All methods are evaluated in a single-pass (non-iterative) setting.

From the results in Table~\ref{tab:complexity}, we observe that the relationship between FLOPs and runtime is not strictly linear. 
For instance, despite having relatively low FLOPs, Patch-to-Point methods like \textit{MODNet} and \textit{PCDNF} exhibit significantly higher inference times. 
This is likely due to their inherently sequential architectures, where each patch produces only one output point, making them less parallelizable than Patch-to-Patch counterparts that denoise all points in a patch simultaneously.
In contrast, methods like \textit{P2P-Bridge} and MAG achieve efficient runtimes despite higher theoretical FLOPs, benefiting from streamlined feed-forward designs. Among all methods, P2P-Bridge is the fastest (17.8 ms per 1000 points) but also the most parameter-heavy (498M). \textit{MAG} and \textit{ASDN} present a better trade-off between model size, computation, and runtime, suggesting their suitability for practical deployment.

\begin{table}[!t]
\centering
\caption{Comparison of network complexity. Methods marked with $^\ast$ belong to the Patch-to-Point category. FLOPs and Time are measured per 1000-point patch. “Params” indicates the total number of model parameters. M = $10^6$, G = $10^9$, ms = milliseconds.}
\begin{tabular}{lccc}
\toprule
\textbf{Method} & \textbf{Params (M)} & \textbf{FLOPs (G)} & \textbf{Time (ms)} \\
                & (whole model)      & (per patch)    & (per patch) \\
\midrule
3DMambaIPF          & 138.309         & 75.072          & 128.601         \\
ASDN                & 6.605           & 2.566           & 32.085          \\
P2P-Bridge          & 498.762         & 116.411         & 17.844          \\
Pathnet$^\ast$      & 77.858          & 90.000          & 636.595         \\
StraightPCF         & 3.702           & 17.348          & 63.106          \\
MAG                 & 1.115           & 21.005          & 28.901          \\
MODNet$^\ast$       & 16.602          & 528.000         & 512.620         \\
IterativePFN        & 21.056          & 67.266          & 107.017         \\
PCDNF$^\ast$        & 5.450           & 378.000         & 966.674         \\
\bottomrule
\end{tabular}
\label{tab:complexity}
\end{table}

\newcommand{\colgap}{0.5mm}








\begin{figure*}[!t]
\centering

\hfill
\includegraphics[width=0.2\linewidth]{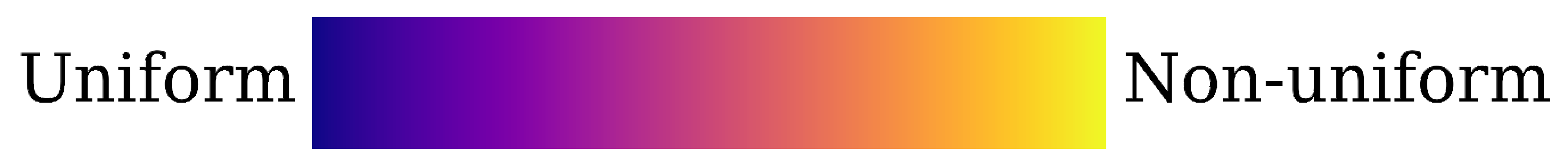}

\vspace{1em}

\begin{minipage}[t]{0.18\linewidth}\centering
  \begin{tikzpicture}[baseline=(base)]
    \node[inner sep=0pt] (img) at (0,0) {\includegraphics[width=0.65\linewidth]{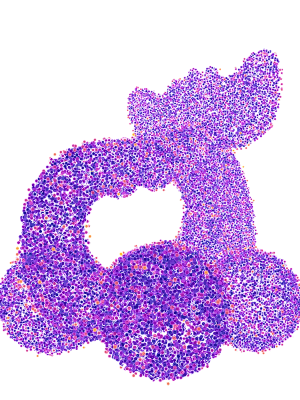}};
    \path (img.south) coordinate (base);
    \def\rectW{0.09\linewidth}
    \def\rectH{0.24\linewidth}
    \draw[line width=0.4pt, color=black]
      ([xshift=11pt,yshift=3pt]img.west) ++(0,0.5*\rectH)
      rectangle ++(\rectW, -\rectH);
  \end{tikzpicture}%
  \hfill
  \setlength\fboxsep{0pt}%
  \fbox{\includegraphics[width=0.33\linewidth]{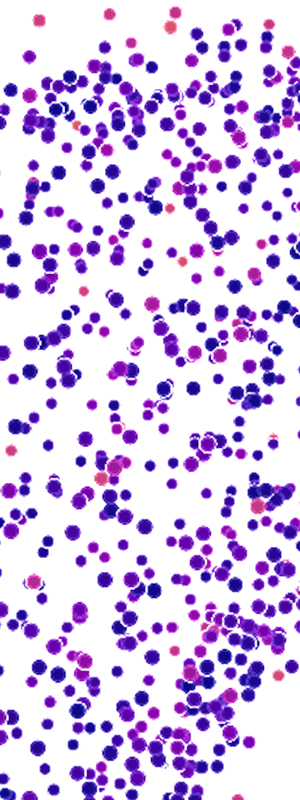}}%
  \\[0.1em]
  \footnotesize Noisy
\end{minipage}\hspace{0.5mm}%
%
\begin{minipage}[t]{0.18\linewidth}\centering
  \begin{tikzpicture}[baseline=(base)]
    \node[inner sep=0pt] (img) at (0,0) {\includegraphics[width=0.65\linewidth]{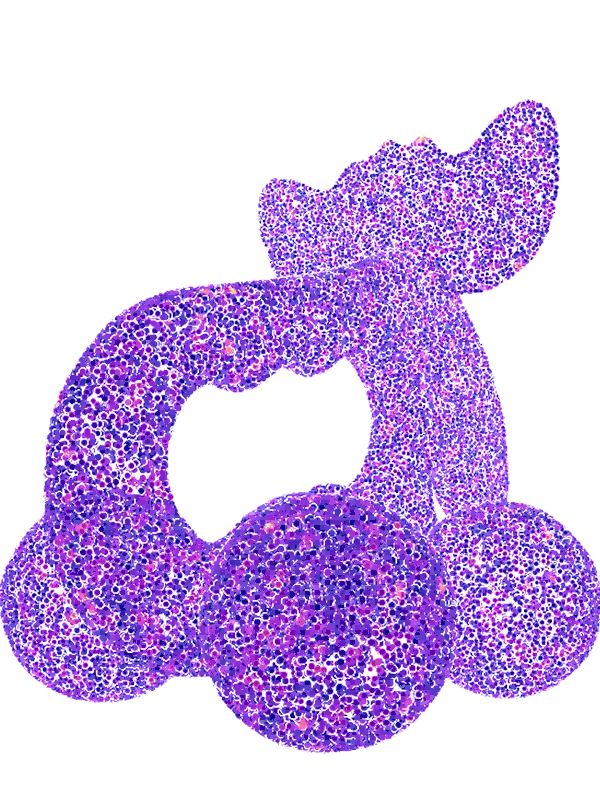}};
    \path (img.south) coordinate (base);
    \def\rectW{0.09\linewidth}
    \def\rectH{0.24\linewidth}
    \draw[line width=0.4pt, color=black]
      ([xshift=11pt,yshift=3pt]img.west) ++(0,0.5*\rectH)
      rectangle ++(\rectW, -\rectH);
  \end{tikzpicture}%
  \hfill
  \setlength\fboxsep{0pt}%
  \fbox{\includegraphics[width=0.33\linewidth]{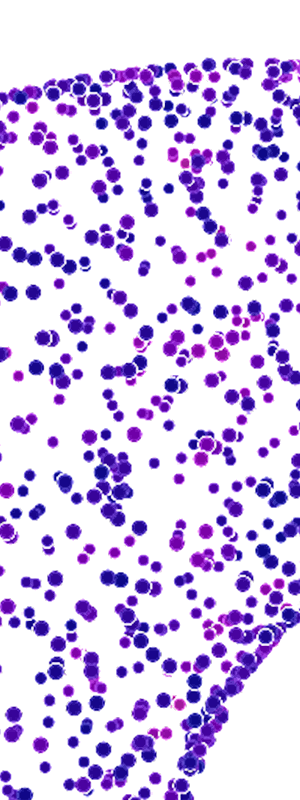}}%
  \\[0.1em]
  \footnotesize 3DMambaIPF
\end{minipage}\hspace{0.5mm}%
%
\begin{minipage}[t]{0.18\linewidth}\centering
  \begin{tikzpicture}[baseline=(base)]
    \node[inner sep=0pt] (img) at (0,0) {\includegraphics[width=0.65\linewidth]{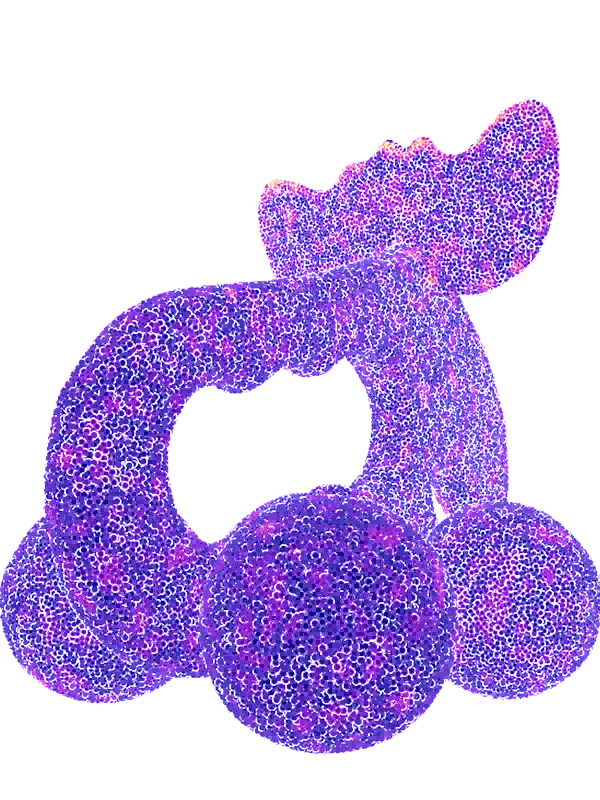}};
    \path (img.south) coordinate (base);
    \def\rectW{0.09\linewidth}
    \def\rectH{0.24\linewidth}
    \draw[line width=0.4pt, color=black]
      ([xshift=11pt,yshift=3pt]img.west) ++(0,0.5*\rectH)
      rectangle ++(\rectW, -\rectH);
  \end{tikzpicture}%
  \hfill
  \setlength\fboxsep{0pt}%
  \fbox{\includegraphics[width=0.33\linewidth]{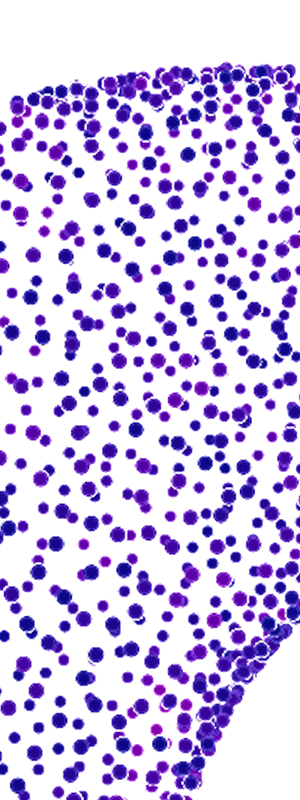}}%
  \\[0.1em]
  \footnotesize ASDN
\end{minipage}\hspace{0.5mm}%
%
\begin{minipage}[t]{0.18\linewidth}\centering
  \begin{tikzpicture}[baseline=(base)]
    \node[inner sep=0pt] (img) at (0,0) {\includegraphics[width=0.65\linewidth]{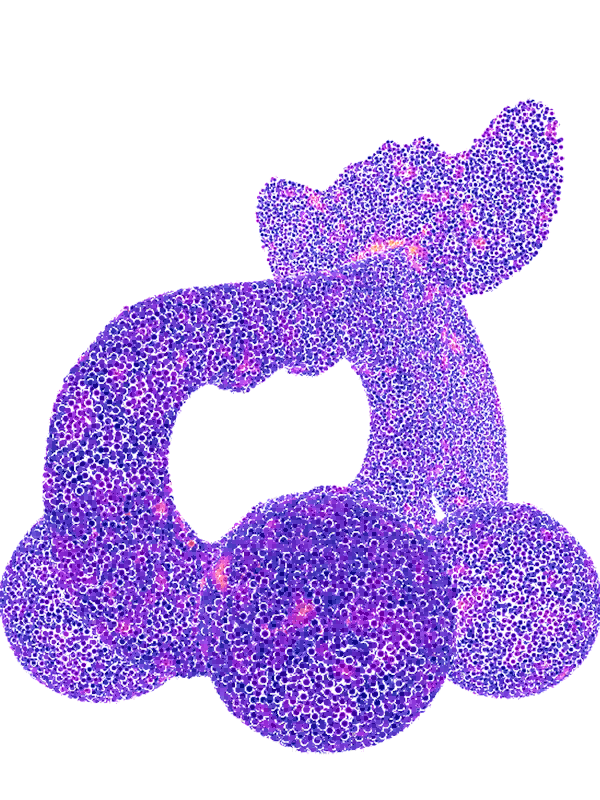}};
    \path (img.south) coordinate (base);
    \def\rectW{0.09\linewidth}
    \def\rectH{0.24\linewidth}
    \draw[line width=0.4pt, color=black]
      ([xshift=11pt,yshift=3pt]img.west) ++(0,0.5*\rectH)
      rectangle ++(\rectW, -\rectH);
  \end{tikzpicture}%
  \hfill
  \setlength\fboxsep{0pt}%
  \fbox{\includegraphics[width=0.33\linewidth]{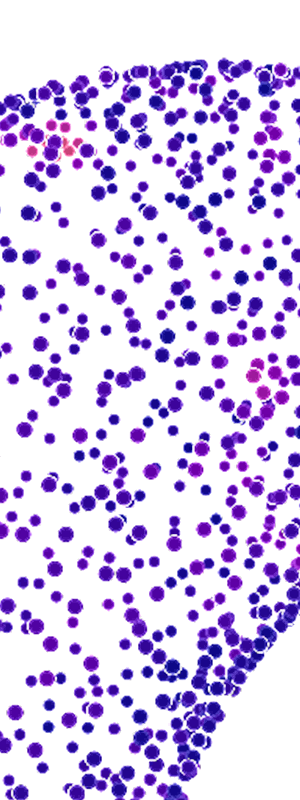}}%
  \\[0.1em]
  \footnotesize P2P\text{-}Bridge
\end{minipage}\hspace{0.5mm}%
%
\begin{minipage}[t]{0.18\linewidth}\centering
  \begin{tikzpicture}[baseline=(base)]
    \node[inner sep=0pt] (img) at (0,0) {\includegraphics[width=0.65\linewidth]{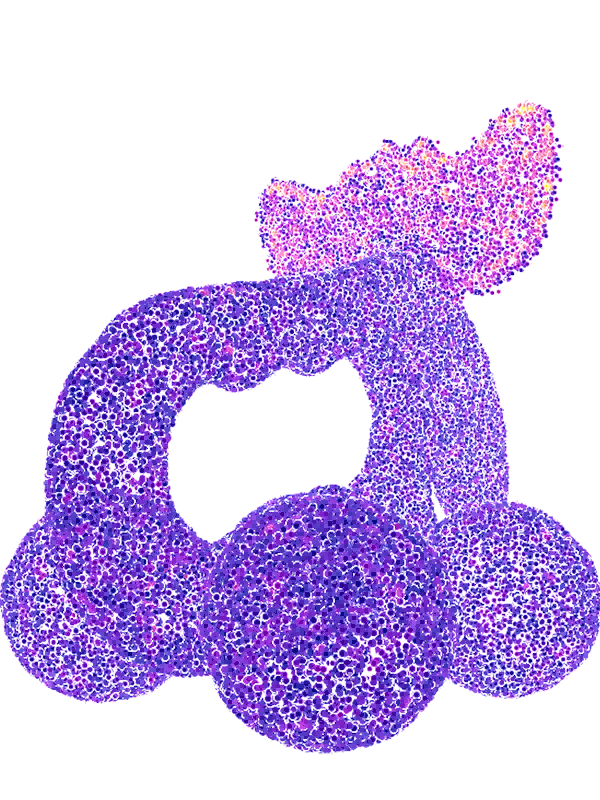}};
    \path (img.south) coordinate (base);
    \def\rectW{0.09\linewidth}
    \def\rectH{0.24\linewidth}
    \draw[line width=0.4pt, color=black]
      ([xshift=11pt,yshift=3pt]img.west) ++(0,0.5*\rectH)
      rectangle ++(\rectW, -\rectH);
  \end{tikzpicture}%
  \hfill
  \setlength\fboxsep{0pt}%
  \fbox{\includegraphics[width=0.33\linewidth]{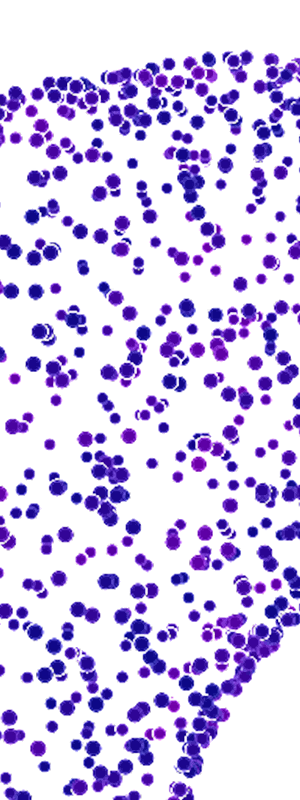}}%
  \\[0.1em]
  \footnotesize Pathnet
\end{minipage}%

\vspace{2mm}

\begin{minipage}[t]{0.18\linewidth}\centering
  \begin{tikzpicture}[baseline=(base)]
    \node[inner sep=0pt] (img) at (0,0) {\includegraphics[width=0.65\linewidth]{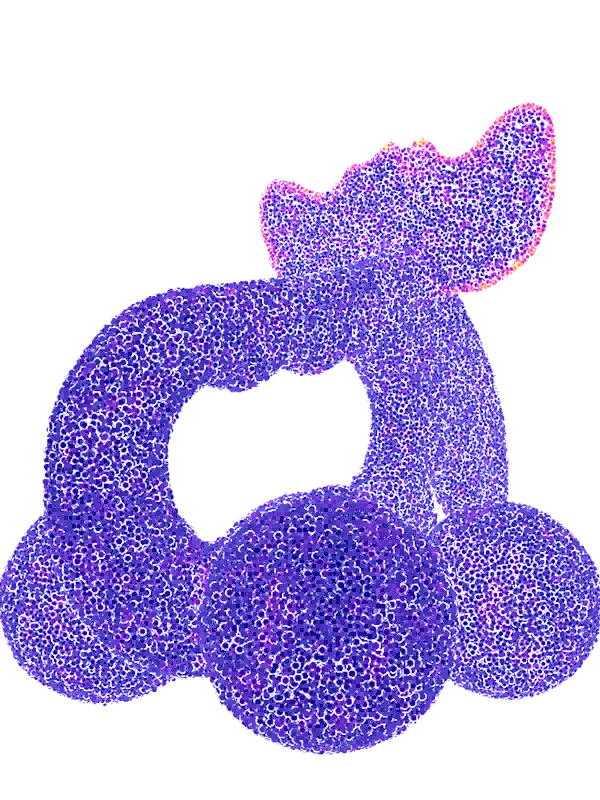}};
    \path (img.south) coordinate (base);
    \def\rectW{0.09\linewidth}
    \def\rectH{0.24\linewidth}
    \draw[line width=0.4pt, color=black]
      ([xshift=11pt,yshift=3pt]img.west) ++(0,0.5*\rectH)
      rectangle ++(\rectW, -\rectH);
  \end{tikzpicture}%
  \hfill
  \setlength\fboxsep{0pt}%
  \fbox{\includegraphics[width=0.33\linewidth]{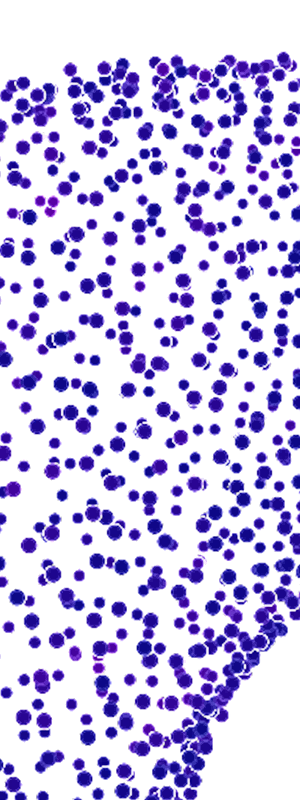}}%
  \\[0.1em]
  \footnotesize StraightPCF
\end{minipage}\hspace{0.5mm}%
%
\begin{minipage}[t]{0.18\linewidth}\centering
  \begin{tikzpicture}[baseline=(base)]
    \node[inner sep=0pt] (img) at (0,0) {\includegraphics[width=0.65\linewidth]{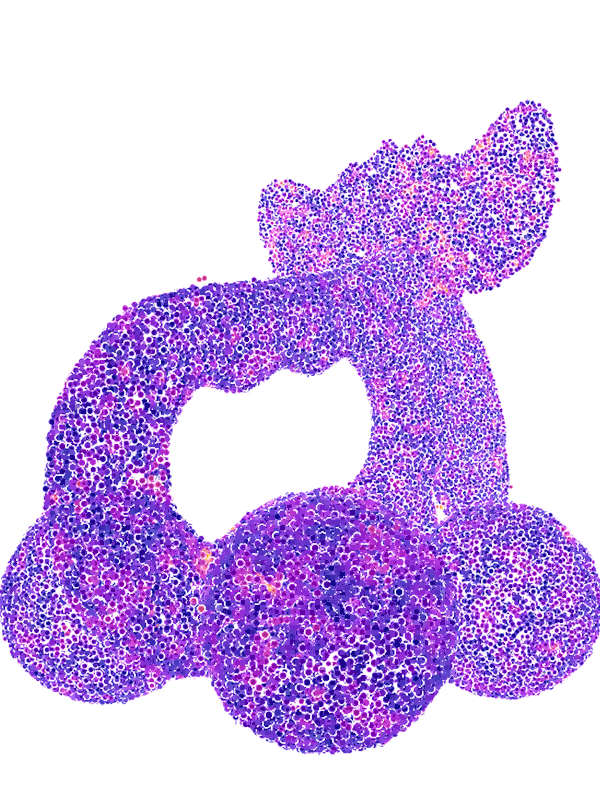}};
    \path (img.south) coordinate (base);
    \def\rectW{0.09\linewidth}
    \def\rectH{0.24\linewidth}
    \draw[line width=0.4pt, color=black]
      ([xshift=11pt,yshift=3pt]img.west) ++(0,0.5*\rectH)
      rectangle ++(\rectW, -\rectH);
  \end{tikzpicture}%
  \hfill
  \setlength\fboxsep{0pt}%
  \fbox{\includegraphics[width=0.33\linewidth]{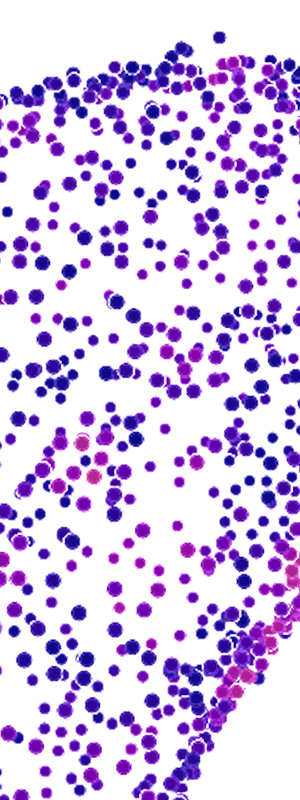}}%
  \\[0.1em]
  \footnotesize MAG
\end{minipage}\hspace{0.5mm}%
%
\begin{minipage}[t]{0.18\linewidth}\centering
  \begin{tikzpicture}[baseline=(base)]
    \node[inner sep=0pt] (img) at (0,0) {\includegraphics[width=0.65\linewidth]{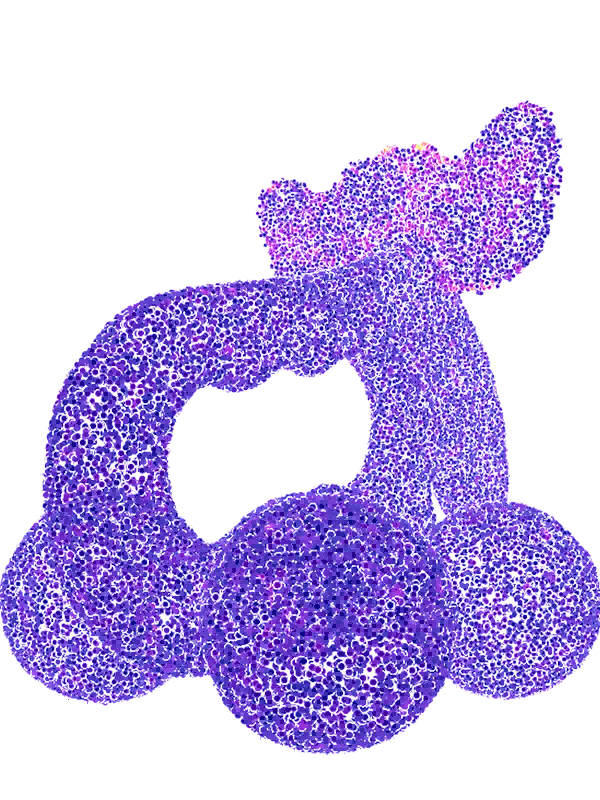}};
    \path (img.south) coordinate (base);
    \def\rectW{0.09\linewidth}
    \def\rectH{0.24\linewidth}
    \draw[line width=0.4pt, color=black]
      ([xshift=11pt,yshift=3pt]img.west) ++(0,0.5*\rectH)
      rectangle ++(\rectW, -\rectH);
  \end{tikzpicture}%
  \hfill
  \setlength\fboxsep{0pt}%
  \fbox{\includegraphics[width=0.33\linewidth]{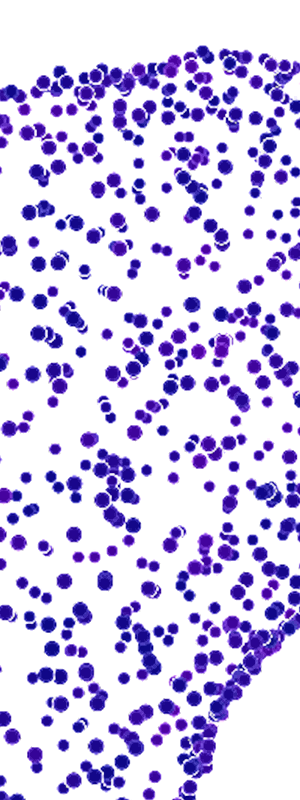}}%
  \\[0.1em]
  \footnotesize MODNet
\end{minipage}\hspace{0.5mm}%
%
\begin{minipage}[t]{0.18\linewidth}\centering
  \begin{tikzpicture}[baseline=(base)]
    \node[inner sep=0pt] (img) at (0,0) {\includegraphics[width=0.65\linewidth]{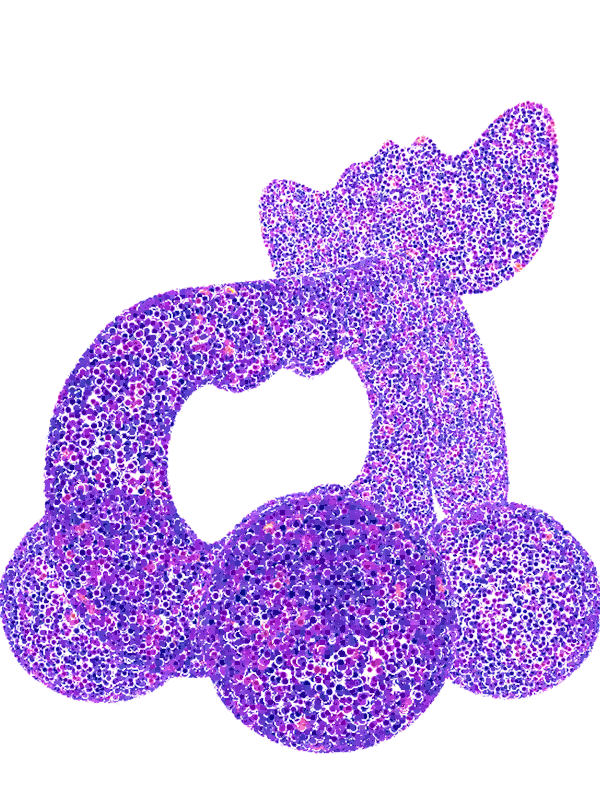}};
    \path (img.south) coordinate (base);
    \def\rectW{0.09\linewidth}
    \def\rectH{0.24\linewidth}
    \draw[line width=0.4pt, color=black]
      ([xshift=11pt,yshift=3pt]img.west) ++(0,0.5*\rectH)
      rectangle ++(\rectW, -\rectH);
  \end{tikzpicture}%
  \hfill
  \setlength\fboxsep{0pt}%
  \fbox{\includegraphics[width=0.33\linewidth]{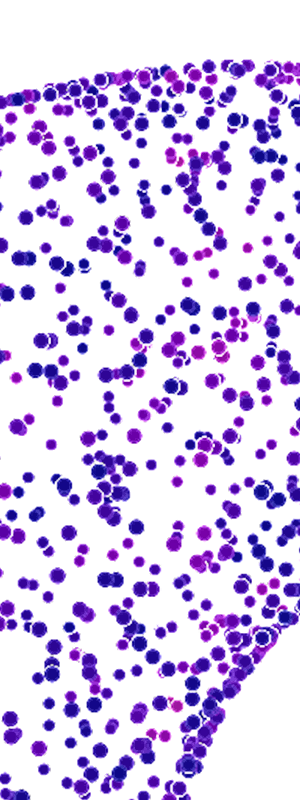}}%
  \\[0.1em]
  \footnotesize IterativePFN
\end{minipage}\hspace{0.5mm}%
%
\begin{minipage}[t]{0.18\linewidth}\centering
  \begin{tikzpicture}[baseline=(base)]
    \node[inner sep=0pt] (img) at (0,0) {\includegraphics[width=0.65\linewidth]{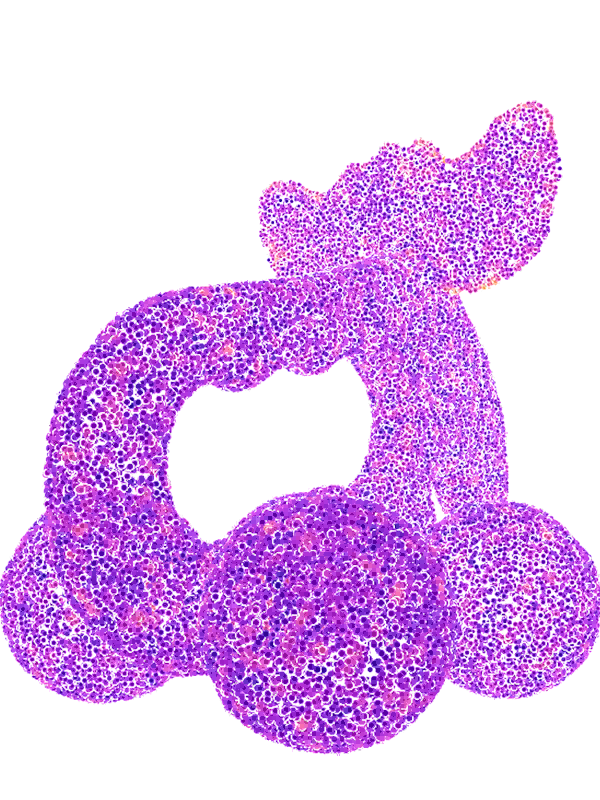}};
    \path (img.south) coordinate (base);
    \def\rectW{0.09\linewidth}
    \def\rectH{0.24\linewidth}
    \draw[line width=0.4pt, color=black]
      ([xshift=11pt,yshift=3pt]img.west) ++(0,0.5*\rectH)
      rectangle ++(\rectW, -\rectH);
  \end{tikzpicture}%
  \hfill
  \setlength\fboxsep{0pt}%
  \fbox{\includegraphics[width=0.33\linewidth]{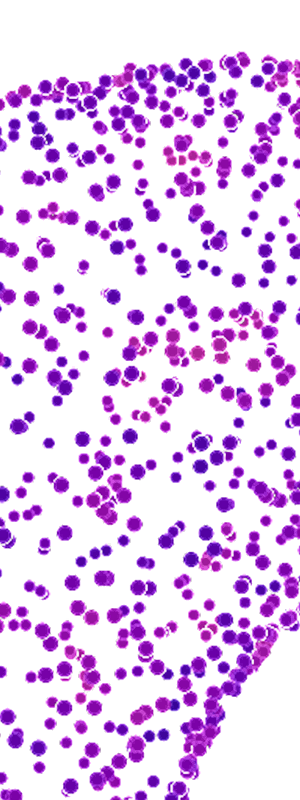}}%
  \\[0.1em]
  \footnotesize PCDNF
\end{minipage}%

\vspace{0.5em}
\caption{Visual comparison of denoising results (two rows). Each column corresponds to one method. The main image shows the denoised point cloud, overlaid with a black box indicating a zoomed-in region. The detail on the right presents the magnified view of that region.}
\label{fig:abla:density_detail}
\end{figure*}
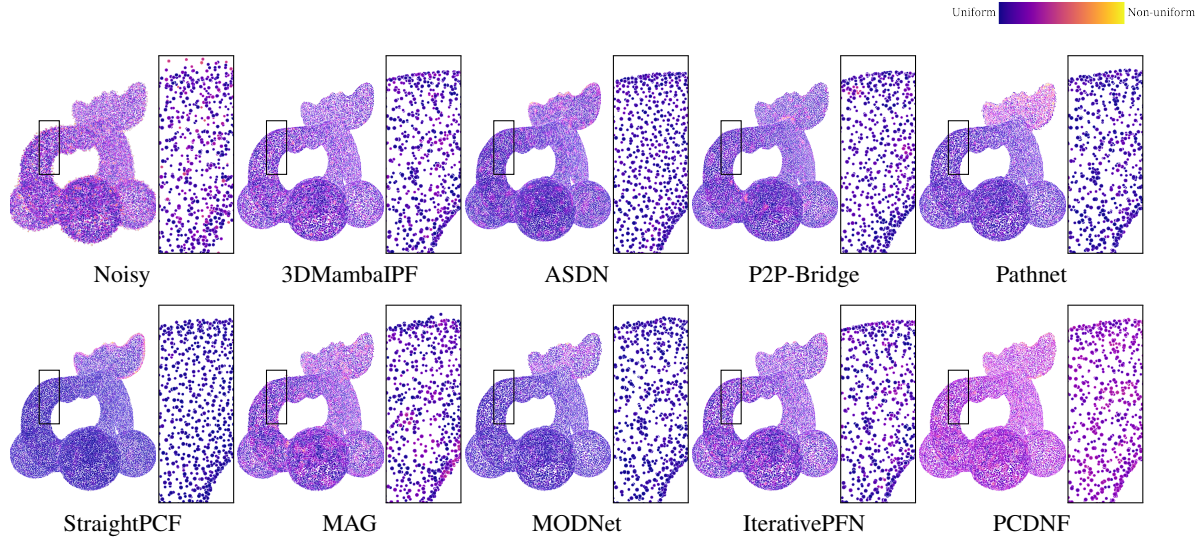
\begin{figure*}[htbp]
\centering

\begin{minipage}[t]{0.18\linewidth}\centering
  \begin{tikzpicture}[baseline=(base)]
    \node[inner sep=0pt] (img) at (0,0) {\includegraphics[width=0.65\linewidth]{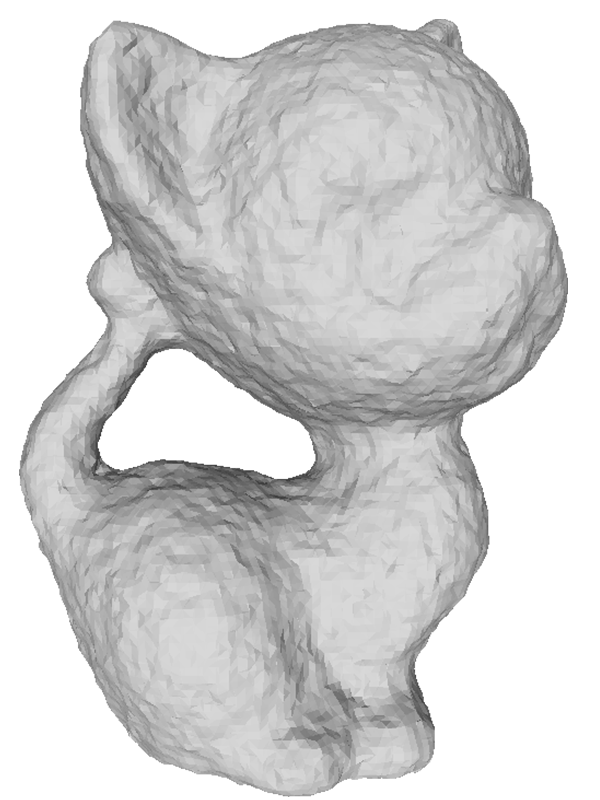}};
    \path (img.south) coordinate (base);
    \def\rectW{0.09\linewidth}
    \def\rectH{0.24\linewidth}
    \draw[line width=0.4pt, color=black]
      ([xshift=5pt,yshift=25pt]img.west) ++(0,0.5*\rectH)
      rectangle ++(\rectW, -\rectH);
  \end{tikzpicture}%
  \hfill
  \setlength\fboxsep{0pt}%
  \fbox{\includegraphics[width=0.33\linewidth]{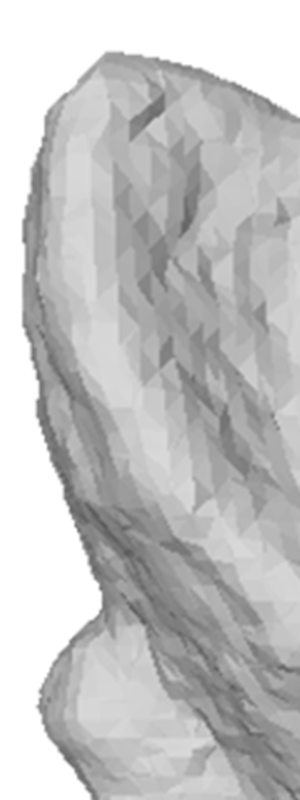}}%
  \\[0.1em]
  \footnotesize Noisy
\end{minipage}\hspace{0.5mm}%
%
\begin{minipage}[t]{0.18\linewidth}\centering
  \begin{tikzpicture}[baseline=(base)]
    \node[inner sep=0pt] (img) at (0,0) {\includegraphics[width=0.65\linewidth]{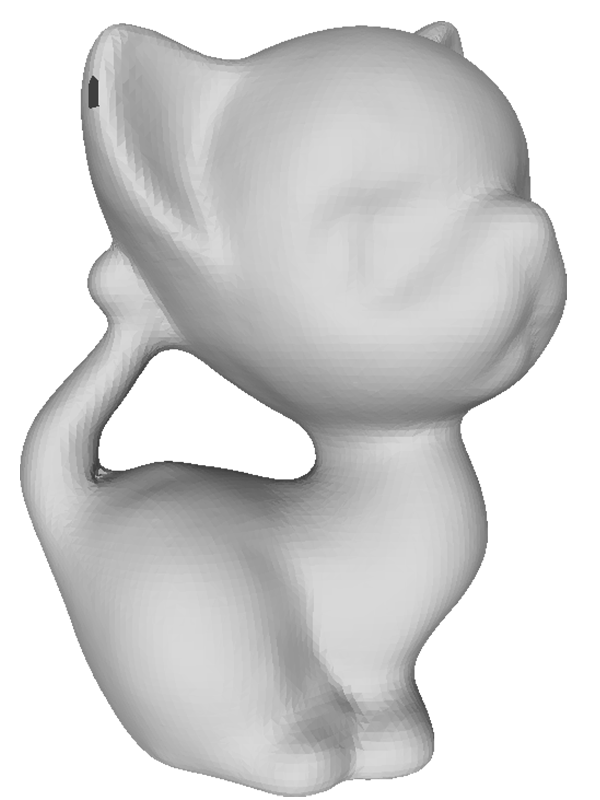}};
    \path (img.south) coordinate (base);
    \def\rectW{0.09\linewidth}
    \def\rectH{0.24\linewidth}
    \draw[line width=0.4pt, color=black]
      ([xshift=5pt,yshift=25pt]img.west) ++(0,0.5*\rectH)
      rectangle ++(\rectW, -\rectH);
  \end{tikzpicture}%
  \hfill
  \setlength\fboxsep{0pt}%
  \fbox{\includegraphics[width=0.33\linewidth]{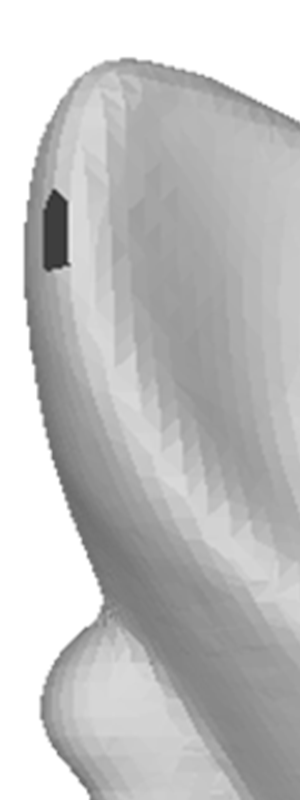}}%
  \\[0.1em]
  \footnotesize 3DMambaIPF
\end{minipage}\hspace{0.5mm}%
%
\begin{minipage}[t]{0.18\linewidth}\centering
  \begin{tikzpicture}[baseline=(base)]
    \node[inner sep=0pt] (img) at (0,0) {\includegraphics[width=0.65\linewidth]{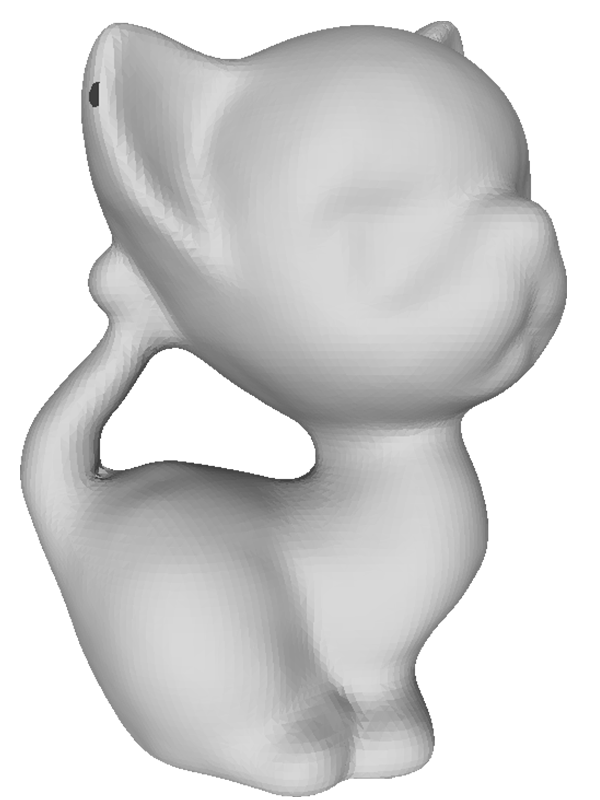}};
    \path (img.south) coordinate (base);
    \def\rectW{0.09\linewidth}
    \def\rectH{0.24\linewidth}
    \draw[line width=0.4pt, color=black]
      ([xshift=5pt,yshift=25pt]img.west) ++(0,0.5*\rectH)
      rectangle ++(\rectW, -\rectH);
  \end{tikzpicture}%
  \hfill
  \setlength\fboxsep{0pt}%
  \fbox{\includegraphics[width=0.33\linewidth]{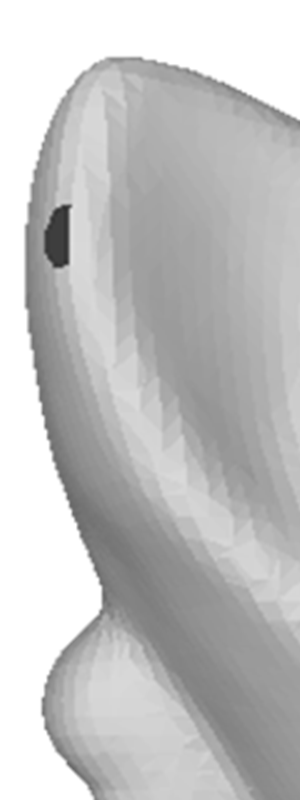}}%
  \\[0.1em]
  \footnotesize ASDN
\end{minipage}\hspace{0.5mm}%
%
\begin{minipage}[t]{0.18\linewidth}\centering
  \begin{tikzpicture}[baseline=(base)]
    \node[inner sep=0pt] (img) at (0,0) {\includegraphics[width=0.65\linewidth]{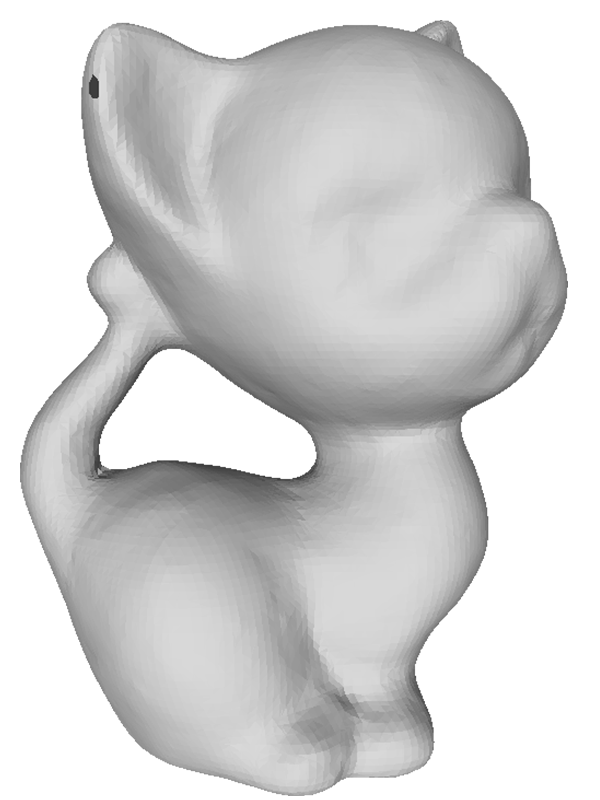}};
    \path (img.south) coordinate (base);
    \def\rectW{0.09\linewidth}
    \def\rectH{0.24\linewidth}
    \draw[line width=0.4pt, color=black]
      ([xshift=5pt,yshift=25pt]img.west) ++(0,0.5*\rectH)
      rectangle ++(\rectW, -\rectH);
  \end{tikzpicture}%
  \hfill
  \setlength\fboxsep{0pt}%
  \fbox{\includegraphics[width=0.33\linewidth]{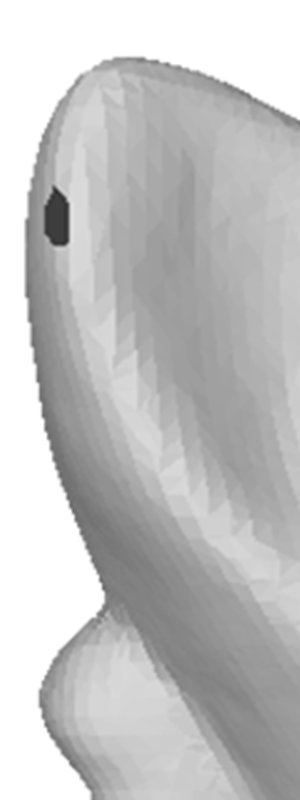}}%
  \\[0.1em]
  \footnotesize P2P\text{-}Bridge
\end{minipage}\hspace{0.5mm}%
%
\begin{minipage}[t]{0.18\linewidth}\centering
  \begin{tikzpicture}[baseline=(base)]
    \node[inner sep=0pt] (img) at (0,0) {\includegraphics[width=0.65\linewidth]{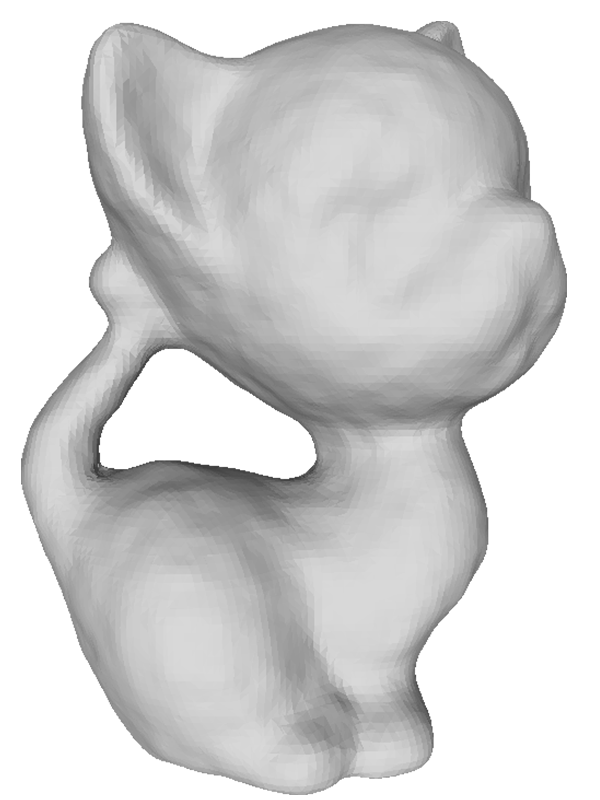}};
    \path (img.south) coordinate (base);
    \def\rectW{0.09\linewidth}
    \def\rectH{0.24\linewidth}
    \draw[line width=0.4pt, color=black]
      ([xshift=5pt,yshift=25pt]img.west) ++(0,0.5*\rectH)
      rectangle ++(\rectW, -\rectH);
  \end{tikzpicture}%
  \hfill
  \setlength\fboxsep{0pt}%
  \fbox{\includegraphics[width=0.33\linewidth]{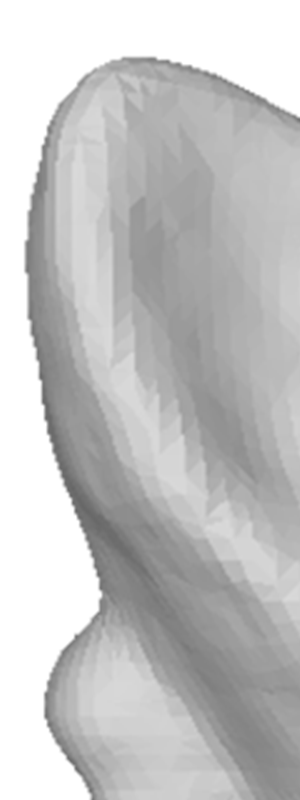}}%
  \\[0.1em]
  \footnotesize Pathnet
\end{minipage}%

\vspace{2mm}

\begin{minipage}[t]{0.18\linewidth}\centering
  \begin{tikzpicture}[baseline=(base)]
    \node[inner sep=0pt] (img) at (0,0) {\includegraphics[width=0.65\linewidth]{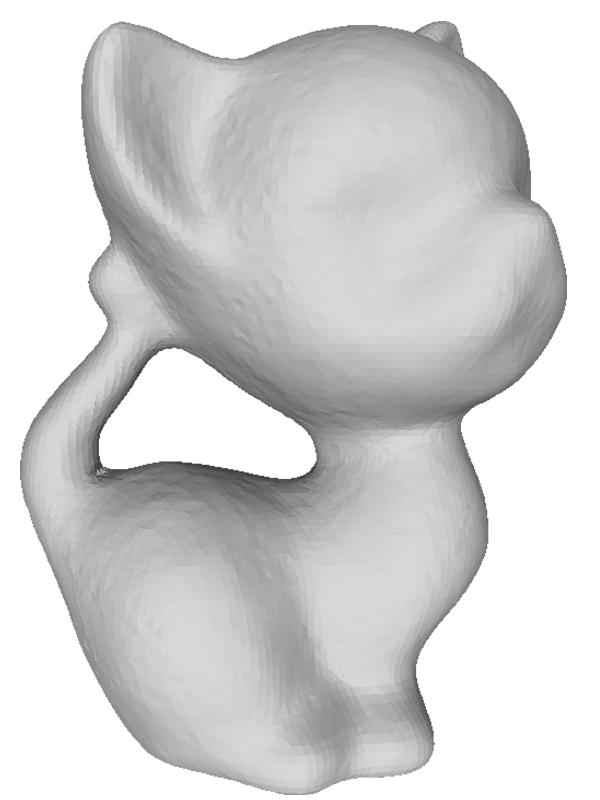}};
    \path (img.south) coordinate (base);
    \def\rectW{0.09\linewidth}
    \def\rectH{0.24\linewidth}
    \draw[line width=0.4pt, color=black]
      ([xshift=5pt,yshift=25pt]img.west) ++(0,0.5*\rectH)
      rectangle ++(\rectW, -\rectH);
  \end{tikzpicture}%
  \hfill
  \setlength\fboxsep{0pt}%
  \fbox{\includegraphics[width=0.33\linewidth]{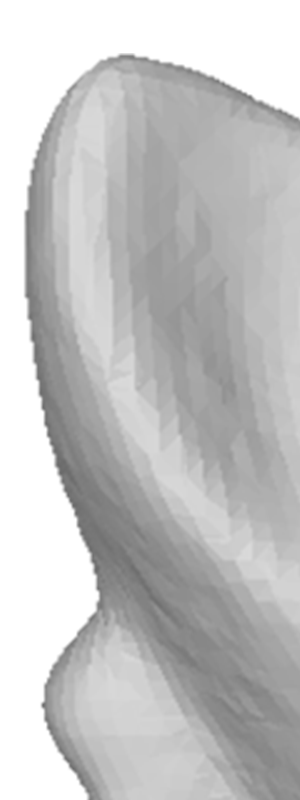}}%
  \\[0.1em]
  \footnotesize StraightPCF
\end{minipage}\hspace{0.5mm}%
%
\begin{minipage}[t]{0.18\linewidth}\centering
  \begin{tikzpicture}[baseline=(base)]
    \node[inner sep=0pt] (img) at (0,0) {\includegraphics[width=0.65\linewidth]{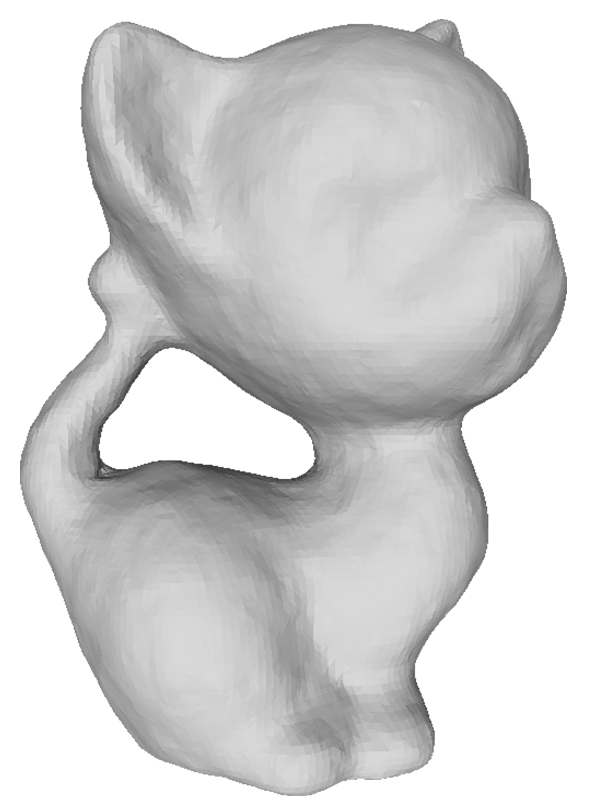}};
    \path (img.south) coordinate (base);
    \def\rectW{0.09\linewidth}
    \def\rectH{0.24\linewidth}
    \draw[line width=0.4pt, color=black]
      ([xshift=5pt,yshift=25pt]img.west) ++(0,0.5*\rectH)
      rectangle ++(\rectW, -\rectH);
  \end{tikzpicture}%
  \hfill
  \setlength\fboxsep{0pt}%
  \fbox{\includegraphics[width=0.33\linewidth]{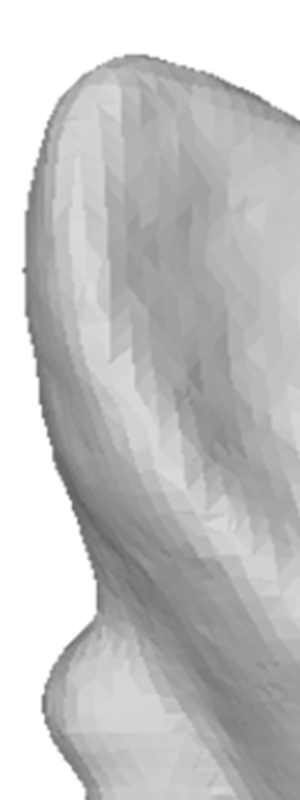}}%
  \\[0.1em]
  \footnotesize MAG
\end{minipage}\hspace{0.5mm}%
%
\begin{minipage}[t]{0.18\linewidth}\centering
  \begin{tikzpicture}[baseline=(base)]
    \node[inner sep=0pt] (img) at (0,0) {\includegraphics[width=0.65\linewidth]{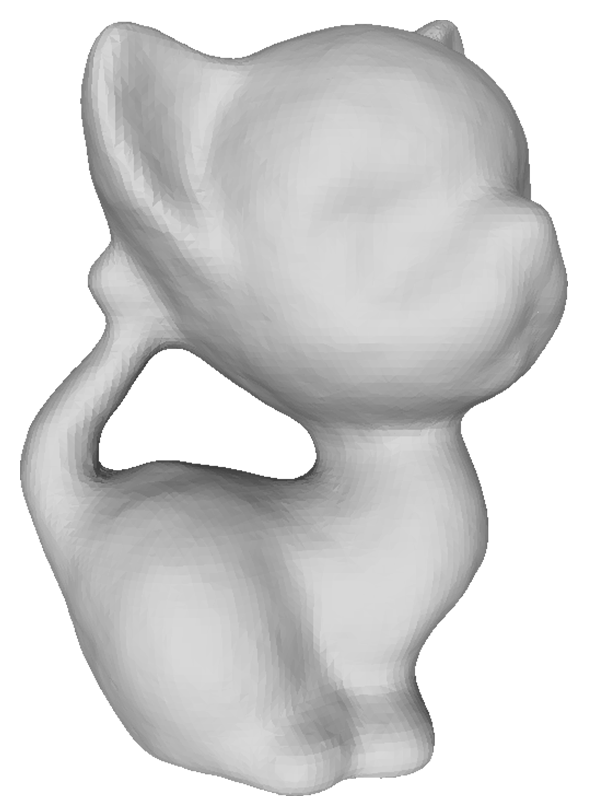}};
    \path (img.south) coordinate (base);
    \def\rectW{0.09\linewidth}
    \def\rectH{0.24\linewidth}
    \draw[line width=0.4pt, color=black]
      ([xshift=5pt,yshift=25pt]img.west) ++(0,0.5*\rectH)
      rectangle ++(\rectW, -\rectH);
  \end{tikzpicture}%
  \hfill
  \setlength\fboxsep{0pt}%
  \fbox{\includegraphics[width=0.33\linewidth]{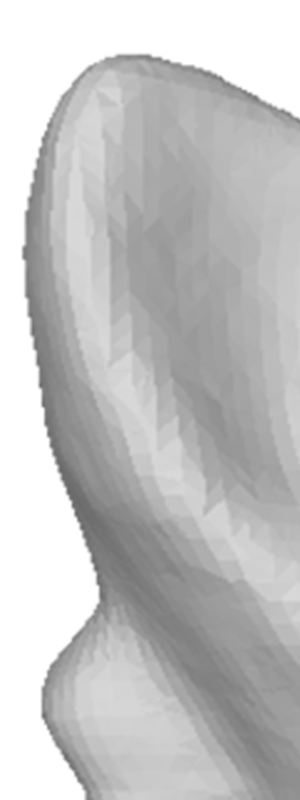}}%
  \\[0.1em]
  \footnotesize MODNet
\end{minipage}\hspace{0.5mm}%
%
\begin{minipage}[t]{0.18\linewidth}\centering
  \begin{tikzpicture}[baseline=(base)]
    \node[inner sep=0pt] (img) at (0,0) {\includegraphics[width=0.65\linewidth]{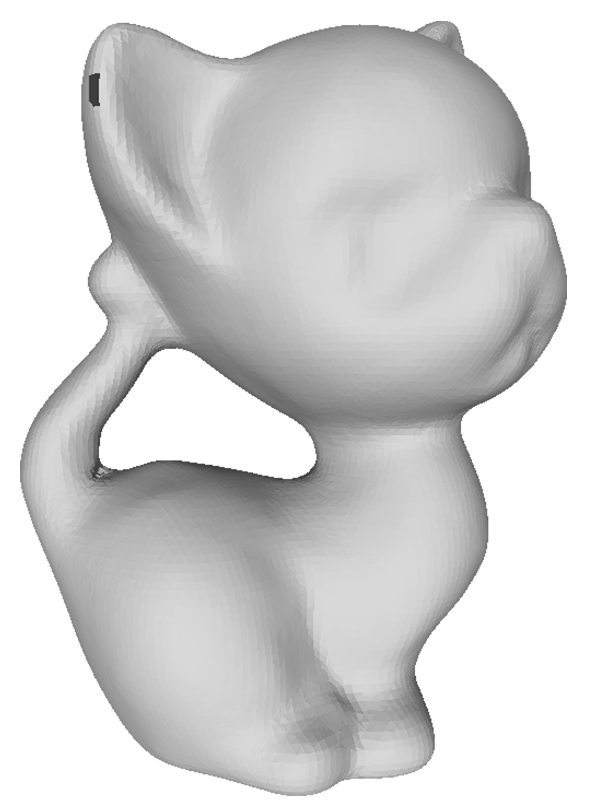}};
    \path (img.south) coordinate (base);
    \def\rectW{0.09\linewidth}
    \def\rectH{0.24\linewidth}
    \draw[line width=0.4pt, color=black]
      ([xshift=5pt,yshift=25pt]img.west) ++(0,0.5*\rectH)
      rectangle ++(\rectW, -\rectH);
  \end{tikzpicture}%
  \hfill
  \setlength\fboxsep{0pt}%
  \fbox{\includegraphics[width=0.33\linewidth]{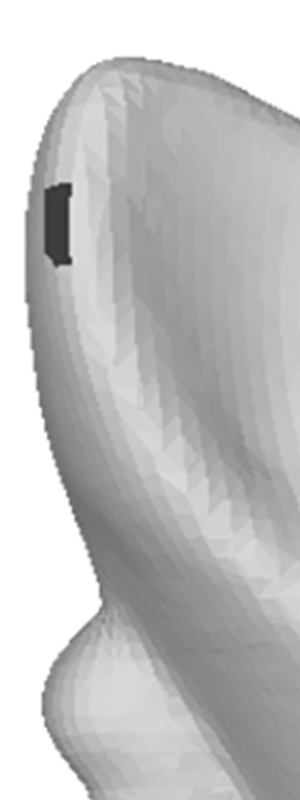}}%
  \\[0.1em]
  \footnotesize IterativePFN
\end{minipage}\hspace{0.5mm}%
%
\begin{minipage}[t]{0.18\linewidth}\centering
  \begin{tikzpicture}[baseline=(base)]
    \node[inner sep=0pt] (img) at (0,0) {\includegraphics[width=0.65\linewidth]{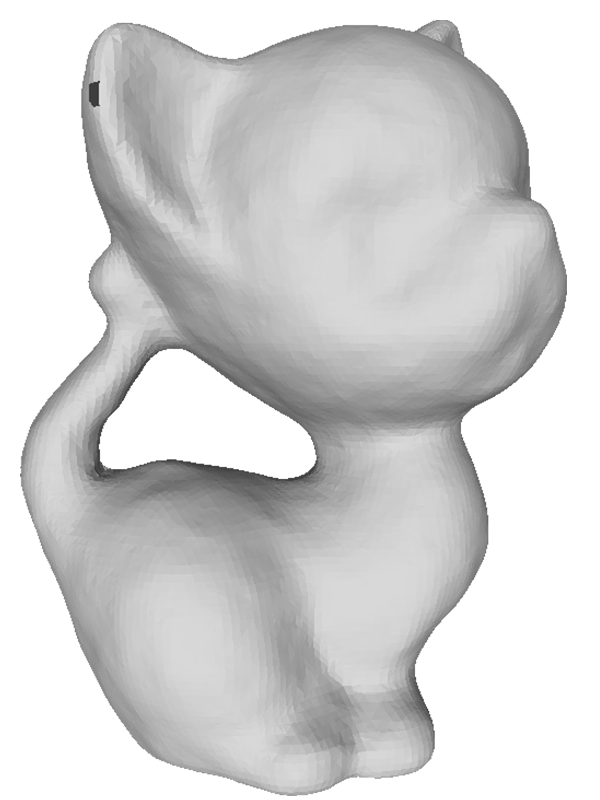}};
    \path (img.south) coordinate (base);
    \def\rectW{0.09\linewidth}
    \def\rectH{0.24\linewidth}
    \draw[line width=0.4pt, color=black]
      ([xshift=5pt,yshift=25pt]img.west) ++(0,0.5*\rectH)
      rectangle ++(\rectW, -\rectH);
  \end{tikzpicture}%
  \hfill
  \setlength\fboxsep{0pt}%
  \fbox{\includegraphics[width=0.33\linewidth]{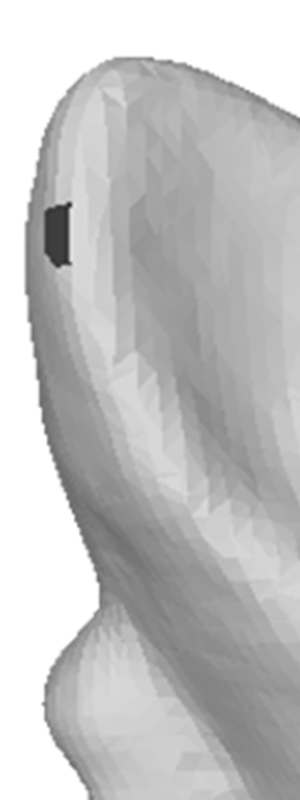}}%
  \\[0.1em]
  \footnotesize PCDNF
\end{minipage}%

\vspace{0.5em}
\caption{Visual comparison of reconstructed meshes (two rows). Each column corresponds to one method. The main panel shows the reconstructed mesh, with a black box marking the region of interest at the top-left corner. The zoomed-in view on the right presents a magnified crop of this area for detailed comparison.}
\label{fig:abla:recons_detail}
\end{figure*}
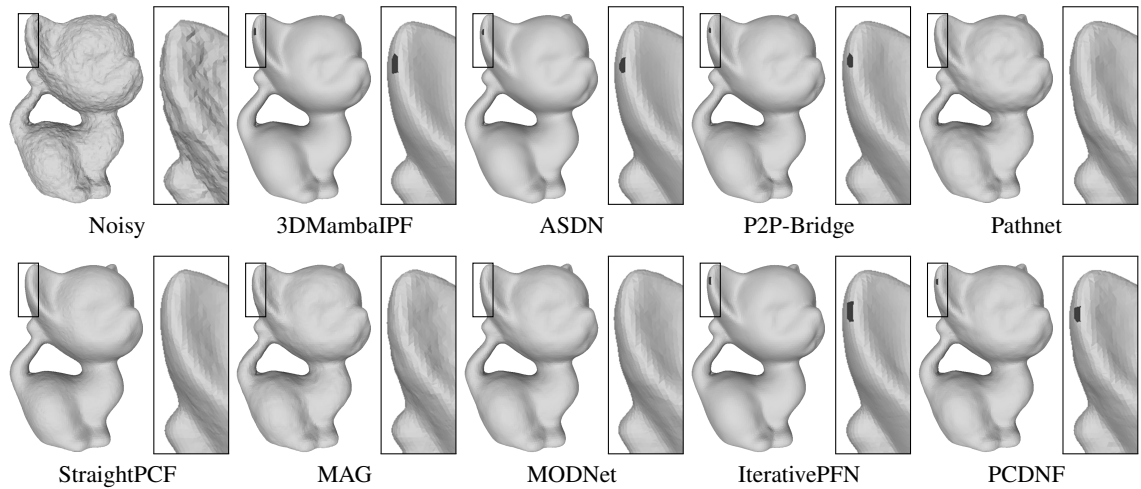

\section{Discussions and Future Directions}
\label{sec:Discussion}
Looking forward, point cloud denoising research may benefit from improvements across two key dimensions:
enhancing practical usability, and 
exploring emerging trends and innovations. Below, we outline several promising directions under each theme.

\subsection{Enhancing Practical Usability}
\begin{itemize}
    \item \textbf{Lightweight and efficient models.} High computational costs limit the deployment of many models. Efficient architectures (e.g., Mamba, SNNs) and model compression techniques (e.g., pruning, quantization) offer potential solutions.
    
    \item \textbf{Task-aware denoising.} Rather than pursuing generic denoising, future models could consider downstream tasks like reconstruction, registration, or segmentation, jointly optimizing for both denoising fidelity and task performance.
    
    \item \textbf{Cloud-level denoising beyond patches.} 
    Most existing methods operate on local patches to reduce memory and computational costs, but this can lead to inconsistent global structures or patch boundary artifacts. With increasing model capacity and hardware advances, directly denoising the entire point cloud (cloud-based processing) could be feasible. This opens up opportunities for learning more globally consistent and semantically aware denoising models.
\end{itemize}

\subsection{Exploring Emerging Models and Trends}
\begin{itemize}
    \item \textbf{Generative and diffusion-based modeling.} Emerging methods based on diffusion models, generative networks, or flow-based approaches have shown strong denoising potential. These models can learn flexible noise distributions and support high-quality restoration.
    
    \item \textbf{Reducing supervision dependency.} Most current methods rely heavily on paired noisy-clean data, which is often costly or impractical to obtain. Exploring unsupervised, weakly supervised, or self-supervised approaches remains a critical direction.

    \item \textbf{Generalization to complex and unknown noise.} Existing methods often assume idealized or synthetic noise patterns. Real-world scans exhibit diverse and mixed noise types, necessitating models with stronger robustness and generalization capacity.
\end{itemize}

\section{Conclusion}
\label{sec:conclusion}
This paper has presented a comprehensive survey of learning-based point cloud denoising up to August 2025. We organized the literature from two complementary perspectives: (i) supervision level (supervised vs.\ unsupervised), and (ii) modeling perspective, where we proposed a functional taxonomy that unifies diverse designs by their underlying denoising principles. We further analyzed architectural trends both structurally and chronologically, and established a unified benchmark that evaluates representative methods under consistent training and testing protocols, covering denoising quality, surface fidelity, point distribution, and computational efficiency. Together, these components provide a consolidated view of where the field stands and what design choices matter in practice.
We hope that the taxonomy, analysis, and benchmark in this survey serve as a practical guide for researchers and practitioners in the community.

\bibliographystyle{unsrt}
\bibliography{ref}

\vfill

\end{document}